\documentclass[5p,longtitle,sort&compress,times,numafflabel]{elsarticle}

\usepackage{amssymb}
\usepackage{amsmath}
\usepackage{lineno}
\usepackage{float}
\usepackage{cuted}
\usepackage{adjustbox}
\usepackage[utf8]{inputenc}
\usepackage[T1]{fontenc}
\usepackage{multirow}

\newcommand\pt{p_\text{T}}
\newcommand\Qs{\ensuremath{Q^{2}}}
\journal{Physics Letters B}

\begin{document}
\begin{frontmatter}



\date{DESY-23-034, August 2023}
\title{Unbinned Deep Learning  Jet Substructure Measurement in High \Qs{} $ep$ collisions at HERA}


\author[44]{V.~Andreev}
\author[29]{M.~Arratia}
\author[40]{A.~Baghdasaryan}
\author[16]{A.~Baty}
\author[34]{K.~Begzsuren}
\author[14]{A.~Bolz}
\author[25]{V.~Boudry}
\author[13]{G.~Brandt}
\author[22]{D.~Britzger}
\author[7]{A.~Buniatyan}
% ...
\affiliation[7]{organisation={School of Physics and Astronomy, University of Birmingham, Birmingham, United Kingdom
$^{c}$}}
% ...
\affiliation[13]{organisation={II. Physikalisches Institut, Universität Göttingen, Göttingen, Germany
}}
\affiliation[14]{organisation={Deutsches Elektronen-Synchrotron DESY, Hamburg and Zeuthen, Germany
}}
% ...
\affiliation[16]{organisation={Rice University, Houston, TX 77005-1827, USA
}}
% ...
\affiliation[22]{organisation={Max-Planck-Institut für Physik, München, Germany
}}
% ...
\affiliation[25]{organisation={LLR, Ecole Polytechnique, CNRS/IN2P3, Palaiseau, France
}}
% ...
\affiliation[29]{organisation={University of California, Riverside, CA 92521, USA
}}
% ...
\affiliation[34]{organisation={Institute of Physics and Technology of the Mongolian Academy of Sciences, Ulaanbaatar, Mongolia
}}
% ...
\affiliation[40]{organisation={Yerevan Physics Institute, Yerevan, Armenia
}}
% ...
\affiliation[44]{organisation={Affiliated with an institute covered by a current or former collaboration agreement with DESY
}}

% \affiliation[inst1]{organization={Department One},%Department and Organization
%             addressline={Address One}, 
%             city={City One},
%             postcode={00000}, 
%             state={State One},
%             country={Country One}}

% \author[inst2]{Author Two}
% \author[inst1,inst2]{Author Three}

% \affiliation[inst2]{organization={Department Two},%Department and Organization
%             addressline={Address Two}, 
%             city={City Two},
%             postcode={22222}, 
%             state={State Two},
%             country={Country Two}}

\begin{abstract}
The radiation pattern within high energy quark- and gluon-initiated jets (jet substructure) is used extensively as a precision probe of the strong force as well as an environment for optimizing event generators with numerous applications in high energy particle and nuclear physics. Looking at electron-proton collisions is of particular interest as many of the complications present at hadron colliders are absent. A detailed study of  modern jet substructure observables, jet angularities, in electron-proton collisions is presented using data recorded using the H1 detector at HERA. The measurement is unbinned and multi-dimensional, using  machine learning to correct for detector effects. All of the available reconstructed object information of the respective jets is interpreted by a graph neural network, achieving superior precision on a selected set of jet angularities. Training these networks was enabled by the use of a large number of GPUs in the Perlmutter supercomputer at Berkeley Lab.  The particle jets are reconstructed in the laboratory frame, using the \unboldmath{$k_{\text{T}}$} jet clustering algorithm. Results are reported at high transverse momentum transfer $Q^2>150~\text{GeV}^2$, and inelasticity $0.2 < y < 0.7$. The analysis is also performed in sub-regions of $Q^2$, thus probing scale dependencies of the substructure variables. The data are compared with a variety of predictions and point towards possible improvements of such models.
\end{abstract}


% \begin{highlights}
% \end{highlights}

% \begin{keyword}
% %% keywords here, in the form: keyword \sep keyword
% keyword one \sep keyword two
% %% PACS codes here, in the form: \PACS code \sep code
% %% MSC codes here, in the form: \MSC code \sep code
% %% or \MSC[2008] code \sep code (2000 is the default)
% \end{keyword}

\end{frontmatter}


Interactions between quarks and gluons (partons) are described by the theory of Quantum Chromodynamics (QCD)~\cite{Gross:2022hyw}. At high energy particle colliders, outgoing partons produce collimated sprays of particles known as jets. 
The radiation pattern inside jets (jet substructure) provides insight into the emergent properties of QCD at high energies.
Electron-proton collisions are ideal for studies of strong interaction processes. They are
induced by an intermediate electroweak gauge particle and provide clean initial conditions. This is in contrast to studies at hadron colliders where the subprocesses are to be studied in the complex context of strong interactions including underlying event effects. The  HERA accelerator facility was operated in the years 1992 to 2007 colliding electrons and positrons with protons.  Jet properties were extensively studied at HERA during data taking~\cite{H1:1998rpm,ZEUS:1998fts,ZEUS:2002sux,ZEUS:2004gcp}, but all of these studies pre-date modern jet substructure~\cite{Altheimer:2012mn,Altheimer:2013yza,Adams:2015hiv,1803.06991,1709.04464}. With new theoretical and methodological advances, novel insight can be extracted from the preserved HERA data \cite{South:2012vh,Britzger:2021xcx}  for precision QCD studies as well as for event generator improvements.

A canonical set of observables used to explore different aspects of the jet radiation pattern are the generalized angularities~\cite{Larkoski:2014pca}:

\begin{equation}
    \lambda^\kappa_\beta = \sum_{i\in\text{jet}}z_i^\kappa\left(\frac{R_i}{R_0}\right)^\beta, 
    \label{eq:jet_angularity}
\end{equation}
with $z_i = p_{\text{T},i}/\pt^{\text{jet}}$ for a particle with momentum $p_{\text{T},i}$ transverse to the incoming beams and clustered inside a jet with distance parameter $R_0$ and transverse momentum $\pt^{\text{jet}}$. The variable $R_i$ is the Euclidean distance between particle $i$ and the jet axis in the pseudorapidity-azimuthal angle plane.
These observables can be further augmented by multiplying the summand in Eq.~\ref{eq:jet_angularity} with the constituent electric charge $q_i$ to form the charged-weighted angularities $\tilde{\lambda}_\beta^\kappa$.

In this work, normalized multi-differential cross sections are measured as a function of six jet angularities. The jets are selected at high transverse momenta in neutral current deep inelastic scattering (DIS) at high photon virtualities \Qs{}, resulting  for the majority of the collisions in single jet events.

The angularities include three infrared and collinear (IRC) safe observables:
the jet broadening ($\lambda^1_1$)
\cite{Catani:1992jc,Rakow:1981qn,Ellis:1986ig}, an intermediate
observable $\lambda^1_{1.5}$, and the jet thrust  ($\lambda^1_2$)
\cite{Farhi:1977sg}. Three IRC unsafe variables are also studied:
the momentum dispersion $\pt\text{D}= \sqrt{\lambda_0^2}$
\cite{CMS:2012rth,Pandolfi:2012ima,CMS:2013wea},
the jet charge $Q_1=\tilde{\lambda}_0^1$
\cite{Field:1977fa,Krohn:2012fg}, and the number of charged constituents $N_c=\tilde{\lambda}_0^0$.
All observables measured in this work are summarized in Table.~\ref{tab:jet_obs}.

\begin{table}[ht]
    \centering
    \small
    \renewcommand{\arraystretch}{1.3}
    \caption{Description of the jet substructure observables measured in this work.}
    \label{tab:jet_obs}
	\begin{tabular}{c|c|c}
        Name/Symbol  & Observable definition & Charge used \\
        \hline
        Logarithm of jet broadening &  $\ln(\lambda_1^1)$ & \multirow{4}{*}{No} \\
        Intermediate observable &  $\ln(\lambda_{1.5}^1)$ & \\ 
        Logarithm of jet thrust &  $\ln(\lambda_2^1)$ & \\
        Momentum dispersion  $\pt\text{D}$ &  $\sqrt{\lambda_0^2}$ &  \\
       \hline 
       Jet charge $Q_1$ &  $\tilde{\lambda}_0^1$ & \multirow{2}{*}{Yes}\\
       Charged particle multiplicity $N_c$ & $\tilde{\lambda}_0^0$  & \\
	\end{tabular}
\end{table}

These observables are chosen because they have been extensively studied in $e^+e^-$ and $pp$ collisions and they cover a diverse set of physics effects.  The momentum dispersion is known to be one of the best probes for separating jets originating from quarks versus jets originating from gluons~\cite{CMS:2013kfa}.  The charged particle  multiplicity is also an excellent quark-versus-gluon jet discriminant~\cite{ATLAS:2016wzt,ATLAS:2017nma} and its evolution with energy scale has been measured at nearly all colliders that have studied jets, including the SPS~\cite{UA1:1986xyj,UA2:1984tht,UA2:1983btx}, PETRA~\cite{JADE:1982ttq,TASSO:1989orr}, PEP~\cite{Derrick:1985du,Petersen:1985hp,SLD:1993mfo,SLD:1996yvs}, TRISTAN~\cite{AMY:1989rdg}, CESR~\cite{CLEO:1997mqo}, LEP ~\cite{OPAL:1991ssr,OPAL:1993uun,OPAL:1995ab,ALEPH:1994hlg,OPAL:1996irm,ALEPH:1995oxo,DELPHI:1995nzf,OPAL:1997dkk,DELPHI:1999gah,OPAL:1999jkz,OPAL:1990vmr,OPAL:2004prv}, Tevatron~\cite{CDF:2001nqb}, and LHC~\cite{ATLAS:2011eid,ATLAS:2011myc,CMS:2012oyn,1602.00988,1906.09254}. Precise theory predictions are provided despite of the quantity not being IRC safe~\cite{Capella:1999ms,Dremin:1999ji}.  The charged constituent multiplicity also is important input to event generator tuning~\cite{ATL-PHYS-PUB-2014-021}.  The jet charge can be used to disentangle quark jets of different flavors~\cite{1509.05190,CMS:2017yer}. Even though it is not collinear-safe, its energy scale dependence can be predicted in perturbation theory~\cite{Krohn:2012fg,Waalewijn:2012sv} and can be used to search for scaling violation~\cite{1509.05190}. The IRC safe observables can be computed in perturbation theory both at a fixed energy scale and as a function of energy scale.  The thrust is formally equivalent to the jet mass, which is the most precisely known observable in $e^+e^-$ and $pp$ collisions~\cite{Frye:2016aiz,Frye:2016okc,Marzani:2017kqd,Marzani:2017mva,Kang:2018jwa,Kang:2018vgn} and may be used to extract the strong coupling constant~\cite{Proceedings:2018jsb}. All IRC safe observables are presented here after a logarithmic transformation, highlighting the region dominated by non-perturbative effects. Results are also reported as a function of the energy scale set by the DIS photon virtuality, thus probing the evolution of jet substructure. 

New machine learning methods are used to simultaneously correct (unfold) all observables for detector effects, where Graph Neural Networks (GNN) are used for the first time to process all of the reconstructed particles inside jets. Making use of the unbinned nature of the data unfolding, both mean and standard deviation of the measured distributions are provided at multiple \Qs{} intervals. They are reported after unfolding and are free of binning effects.

These first measurements, derived from the HERA preserved data, are going to set the foundation for future studies at the future electron-ion collider or possible other electron-proton collider experiments~\cite{Aschenauer:2019uex,PhysRevLett.125.242003,LHeC:2020van,FCC:2018byv}, which all will operate in kinematic regimes complementary to the HERA machine.

\section{Experimental method}\label{sec:h1}
A full description of the H1 detector can be found elsewhere~\cite{Abt:1993wz,Andrieu:1993kh,Abt:1996hi,Abt:1996xv,Appuhn:1996na}. The detector components that are most relevant for this measurement are described below. Results are reported using a right handed coordinate system with positive $z$ direction pointing towards the outgoing proton beam and positive $x$-axis pointing to the center of the HERA ring. The nominal interaction point is located at $z=0$. The polar angle $\theta$ is defined with respect to the $z$ axis, the azimuthal angle $\phi$ is measured in the $xy$ plane. The pseudorapidity is defined as $\eta_{\text{lab}} = -\ln\tan(\theta/2)$. The main sub-detectors used in this analysis are the inner tracking detectors and the Liquid Argon (LAr) calorimeter, which are both immersed in a magnetic field of 1.16 T provided by a superconducting solenoid. The central tracking system, which covers 15$^{\circ}$ $<\theta<$ 165$^{\circ}$ and the full azimuthal angle, consists of drift and proportional chambers that are complemented with a silicon vertex detector in the range $30^{\circ}<\theta<150^{\circ}$~\cite{Pitzl:2000wz}. It yields a transverse momentum resolution for charged particles of $\sigma_{p_\text{T}}/p_\text{T}$ = 0.2$\%$ $p_\text{T}$/GeV$~\oplus~$1.5$\%$. The LAr calorimeter, which covers $4^{\circ}<\theta< 154^{\circ}$ and full azimuthal angle, consists of an electromagnetic section made of lead absorbers and a hadronic section with steel absorbers; both are highly segmented in the transverse and longitudinal directions. Its energy resolution is $\sigma_{E}/E = 11\%/\sqrt{E/\text{GeV}}$~$\oplus$~$1\%$ for leptons~\cite{Andrieu:1994yn} and $\sigma_{E}/E\approx 50\%/\sqrt{E/\text{GeV}}$~$\oplus$~$3\%$ for charged pions~\cite{Andrieu:1993tz}.  In the backward region ($153^\circ < \theta < 177.5^\circ$), energies are measured with a lead-scintillating fiber calorimeter~\cite{Appuhn:1996na}.
Results are reported using data recorded by the H1 detector in the years 2006 and 2007 when protons and electrons/positrons (henceforth referred to as `electrons') were collided at energies of 920 GeV and 27.6 GeV, respectively.  The total integrated luminosity of this data sample corresponds to 228~pb$^{-1}$ \cite{H1:2012wor}.

Events are triggered by requiring a high energy cluster in the electromagnetic part of the LAr calorimeter. The scattered electron is identified as the highest transverse momentum LAr cluster matched to a track passing an isolation criteria~\cite{Adloff:2003uh}. Events containing scattered electrons with energy $E_{e'}>11$ GeV are kept for further analysis, resulting in a trigger efficiency higher than 99.5$\%$~\cite{Aaron:2012qi,Andreev:2014wwa}. Backgrounds from additional processes such as cosmic rays, beam-gas interactions, photoproduction, charged-current DIS and QED Compton processes are rejected after dedicated selection~\cite{Andreev:2014wwa,Andreev:2016tgi}, resulting in negligible background contamination.

The inelasticity and photon virtuality are reconstructed using the $\Sigma$ method~\cite{Bassler:1994uq} defined as

\begin{equation}
    y = \frac{\Sigma_\text{had}}{\Sigma_\text{had} + \Sigma_\text{e'}}, \quad
    Q^2 = \frac{E_{e'}^2\sin^2\theta_{e'}}{1-y} \,,
\end{equation}
where $\Sigma_\text{e'}= E_{e'}(1-\cos\theta_{e'})$  with $\theta_{e^{'}}$  the polar angle of the scattered electron and $\Sigma_\text{had} = \sum(E_{i}-p_{i,z})$ is the difference between the energy and longitudinal momentum sums of the entire hadronic final state (HFS). HFS objects are reconstructed using an energy flow algorithm~\cite{energyflowthesis,energyflowthesis2,energyflowthesis3} after removing energy clusters and tracks associated to the electron. Additionally, events are required to have $45 <\Sigma_\text{had} + \Sigma_\text{e'} < 65$~GeV to suppress initial-state  QED radiation and contributions from photoproduction. 

Jets are defined in the laboratory frame by clustering HFS objects satisfying $-1.5<\eta_{\text{lab}}<2.75$. The \textsc{FastJet}~3.3.2 package~\cite{Cacciari:2011ma,Cacciari:2005hq} is used with longitudinally invariant $k_{\text{T}}$ clustering algorithm~\cite{Catani:1993hr,Ellis:1993tq} using the default $E$-scheme and  distance parameter $R_0 = 1$. All jets in the event with $\pt> 5$~GeV are kept for further analysis.

The unfolding targets (`particle-level') are calculated in the simulation using  final-state particles with proper lifetime $c\tau > 10$~mm and excluding the scattered lepton and photons radiated from one of
the lepton lines. Reconstructed and generator level jets are matched by requiring the distance
$
\Delta R = \sqrt{(\phi_{\rm gen}^{\rm jet}- \phi_{\rm reco}^{\rm jet})^{2} + (\eta_{\rm gen}^{\rm jet}- \eta_{\rm reco}^{\rm jet})^{2}}  <0.9$.

Results are presented after unfolding the data to particle level for events in the kinematic region defined by $Q^2> 150$ GeV, $0.2 < y < 0.7$, $\pt^\text{jet} > 10$ GeV, and $-1.0 < \eta^\text{jet} < 2.5$. 

\section{Monte Carlo simulations}
Monte Carlo (MC) simulations are used to correct the data for detector acceptance and resolution effects as well as to compare theoretical predictions with unfolded results. 

Detector acceptance and resolution effects are estimated using large samples of simulated events, generated with the \textsc{Djangoh}~1.4~\cite{Charchula:1994kf} and \textsc{Rapgap}~3.1~\cite{Jung:1993gf} event generators. Both implement Born level matrix elements and are interfaced with \textsc{Heracles}~\cite{Spiesberger:237380,Kwiatkowski:1990cx,Kwiatkowski:1990es} for QED radiation.  The CTEQ6L Parton distribution function (PDF) set~\cite{Pumplin:2002vw} and the Lund hadronization model~\cite{Andersson:1983ia} with parameters determined by the ALEPH Collaboration~\cite{Schael:2004ux} are used for the non-perturbative components.  \textsc{Djangoh} uses the Colour Dipole Model as implemented in \textsc{Ariadne}~\cite{Lonnblad:1992tz} for higher order emissions, and \textsc{Rapgap} uses parton showers in the leading logarithmic approximation. Each of these generators is combined with a detailed simulation of the H1 detector response based on the \textsc{Geant}3 simulation program~\cite{Brun:1987ma} and reconstructed in the same way as data.  

Additional predictions, yet without a detailed simulation of the H1 detector, are made using a set of state-of-the-art generators developed mostly for $pp$ collisions. Predictions from \textsc{Pythia}~8.3~\cite{Sjostrand:2006za,Sjostrand:2014zea} are used for comparison using the default implementation and two additional parton shower implementations: \textsc{Vincia} \cite{Giele:2007di,Giele:2013ema} and \textsc{Dire} \cite{Hoche:2015sya}. \textsc{Vincia} uses a $\pt$-ordered model for QCD + QED showers based on the antenna formalism while \textsc{Dire} implements a $\pt$ ordered dipole shower similar to \textsc{Ariadne}. The  NNPDF3.1 PDF set \cite{NNPDF:2017mvq} is used for both default and \textsc{Vincia} implementation and the MMHT14nlo68cl PDF set \cite{Harland-Lang:2014zoa} is used for the \textsc{Dire} implementation. Predictions from \textsc{Herwig}~7.2~\cite{Bellm:2015jjp,Bahr:2008pv} are calculated using the cluster hadronisation model~\cite{Bellm:2019zci} with default implementation parameters and alternative \textsc{MatchBox}~\cite{Lonnblad:2012ix} matching and merging~\cite{Bellm:2019zci} schemes. Predictions from a pre-release version of \textsc{Sherpa}~3.0~\cite{Sherpa:2019gpd} are provided by the \textsc{Sherpa} authors  featuring a new cluster hadronisation model~\cite{Chahal:2022rid} and matrix element calculation at next-to-leading order (NLO) obtained from OpenLoops with the \textsc{Sherpa} Dipole Shower~\cite{Schumann:2007mg} based on the truncated shower method~\cite{Hoeche:2009rj,Hoeche:2012yf}. 

\section{Unfolding methodology}
The unfolding procedure is carried out by simultaneously unfolding the six  jet angularities, jet momentum ($\pt, \eta, \phi$), and photon virtuality \Qs{} using the \textsc{OmniFold} method \cite{Andreassen:2019cjw,omnifoldiclr}. \textsc{OmniFold} corrects the data for detector effects using  machine learning in an iterative reweighting process that is unbinned and naturally incorporates high dimensional inputs. This method generalizes the Lucy-Richardson deconvolution technique~\cite{1974AJ79745L,Richardson:72,DAgostini:1994fjx} designed for binned spectra.

At each iteration, \textsc{OmniFold} employs a set of classifiers trained using neural networks to estimate two reweighting functions used during the unfolding procedure. Each iteration has two steps, with one classifier each. In the first step, a classifier is trained to distinguish data from simulated events, resulting in event weights based on the classifier output. The second step updates the simulation by training a classifier to distinguish between the original simulation and the simulation weighted after the first step. This second step only uses the particle-level information of simulated events as inputs, producing a weight map that is a proper function of the particle level inputs. The updated simulation is then used as the starting point for the following \textsc{OmniFold} iteration. The process is repeated a total of 13 times; six iterations are chosen for the final results as that choice yields the lowest overall uncertainty.

The first step uses up to 30 HFS objects clustered inside jets as inputs to a GNN, where HFS objects are represented as nodes of the graph.  This novel hybrid approach has been found to reduce uncertainties by accounting for all possible covariates of the detector response.
For each HFS object with momentum ($\pt$,$\eta$,$\phi$) and electric charge $q$, the kinematic information used for the first step classifier is the set ($z$,$\Delta\eta$,$\Delta\phi$,$q$), with $z=\pt/\pt^{\text{jet}}$, $\Delta\eta = \eta - \eta^\text{jet}$, and $\Delta\phi = \phi - \phi^\text{jet}$. Jet momentum ($\pt^\text{jet}$, $\eta^\text{jet}$,$\phi^\text{jet}$) and \Qs{} information are also included in the aggregation function of the graph implementation. In total, $30 \times 4 + 4 = 124$ features are considered during this first step, requiring the network implementation to learn the differences between data and simulation in a high-dimensional setting. 
Leveraging the success of GNNs for jet classification \cite{Komiske:2018cqr,Qu:2019gqs,Moreno:2019bmu,Moreno:2019neq,Mikuni:2020wpr,Bernreuther:2020vhm,Guo:2020vvt,Dolan:2020qkr,Mikuni:2021pou,Konar:2021zdg,Shimmin:2021pkm,Gong:2022lye} based on particle sets as inputs, the Point Cloud Transformer (\textsc{PCT}) \cite{guo2021pct,Mikuni:2021pou} GNN architecture is chosen. This GNN architecture is optimized to learn the relationship between nearby particles through the use of attention layers, resulting in state-of-the-art classification performance for different types of jets \cite{Mikuni:2021pou}. The local neighborhood in \textsc{PCT} is defined by selecting the five closest neighbors of each particle with distances calculated in the $\eta-\phi$ plane. The impact of adding more than 30 particles was found to be negligible.

In the second step, a simplified classifier architecture is used. The inputs of the classifier are the same jet kinematic information and \Qs{} used in the previous step, but HFS objects are replaced by the jet observables that are reported in the final results. This choice decreases the number of inputs by an order of magnitude and reduces the training time of the classifier by a factor five. Since the first step is tasked to incorporate as much information as possible from the data, the simplified classifier adopted for the second step does not degrade the unfolding results while greatly reducing the computational cost of the method. This simplified classifier is implemented using a fully-connected neural network with three hidden layers with 64, 128, and 64 nodes per layer. 
Different hyperparameter choices of the neural networks were also tested, but no significant difference in the final results was observed. The robustness of the classifiers is improved by training an ensemble of 10 network models for each \textsc{OmniFold} step. The average response of the ensemble is then used to derive the reweighting functions. 

The output of all classifiers is passed through a sigmoid activation function and trained using the binary cross-entropy loss function. The training proceeds until the validation loss, estimated with a statistically independent data set, does not improve for 10 consecutive epochs. All machine learning methods are implemented using the \textsc{Keras}~\cite{keras} and \textsc{TensorFlow}~\cite{tensorflow} libraries, using the \textsc{Adam}~\cite{adam} optimization algorithm. In total around $50\times 10^6$ jets are used during the training of the method split between data and simulation. To derive the unfolded results and uncertainties, 2800 neural networks are trained independently using 128 graphic processing units (GPUs) simultaneously with the Perlmutter supercomputer~\cite{Perlmutter} and \textsc{Horovod}~\cite{sergeev2018horovod} library used for parallel distributed training.

\section{Uncertainties}
Systematic uncertainties on the reconstruction of the data observables are estimated by varying the relevant parameters in the simulation and carrying out the full unfolding procedure for each simulation variant. Uncertainties on HFS objects include the energy scale from two different contributions: HFS objects contained in high $\pt$ jets and other HFS objects. In both cases, the energy-scale uncertainty is $\pm1\%$. Both uncertainty sources are estimated separately \cite{etde_21406988,Andreev:2014wwa} by varying the corresponding HFS energy by $\pm1\%$. An uncertainty of $\pm20$~mrad is assigned to the azimuthal angle determination of HFS objects. Lepton uncertainties are considered in the \Qs{} determination and the uncertainty on the lepton energy scale ranges from $\pm0.5\%$ to $\pm1\%$~\cite{H1:2011unn,Andreev:2014wwa}. Uncertainties on the azimuthal angle of the scattered lepton are estimated to be $\pm1$~mrad~\cite{Aaron:2012qi}. In \ref{app:data_mc}  distributions on reconstructed level and the uncertainties are compared with predictions from \textsc{Djangoh} and \textsc{Rapgap} generators.

Additional uncertainties from the unfolding procedure are estimated to cover a possible bias from the generator choice used to perform the unfolding procedure. A model uncertainty is estimated by unfolding the data with \textsc{Djangoh} instead of the default \textsc{Rapgap}. In addition, a non-closure uncertainty is estimated by unfolding \textsc{Djangoh} as pseudo-data using the \textsc{Rapgap} simulation and comparing to \textsc{Djangoh} at particle level. The combined effects of the unfolding uncertainties typically are below 10$\%$ and increase only in the more extreme regions of the jet angularities.
%I think there is possibly some double counting here, but I also think it is okay.  Imagine data ~ Rapgap.  Then, the first uncertainty is basically just the result of unfolding Rapgap with Djangoh, which is the same as the second uncertainty.
QED corrections  accounting for virtual and real higher-order QED effects as implemented in HERACLES are small and the full effect is taken as an uncertainty, estimated by comparing predictions with and without first order QED radiative effects at the leptonic vertex. The statistical uncertainty is estimated using the bootstrap technique~\cite{10.1214/aos/1176344552}. The unfolding procedure is repeated on 100 pseudo datasets, each defined by resampling the original dataset with replacement. Normalization uncertainties, such as luminosity scale and trigger efficiencies, are not considered since they cancel in the ratio of the normalized differential cross section results.

\section{Results}\label{sec:results}

\begin{figure*}[!htb]
    \centering
        \includegraphics[width=.32\textwidth]{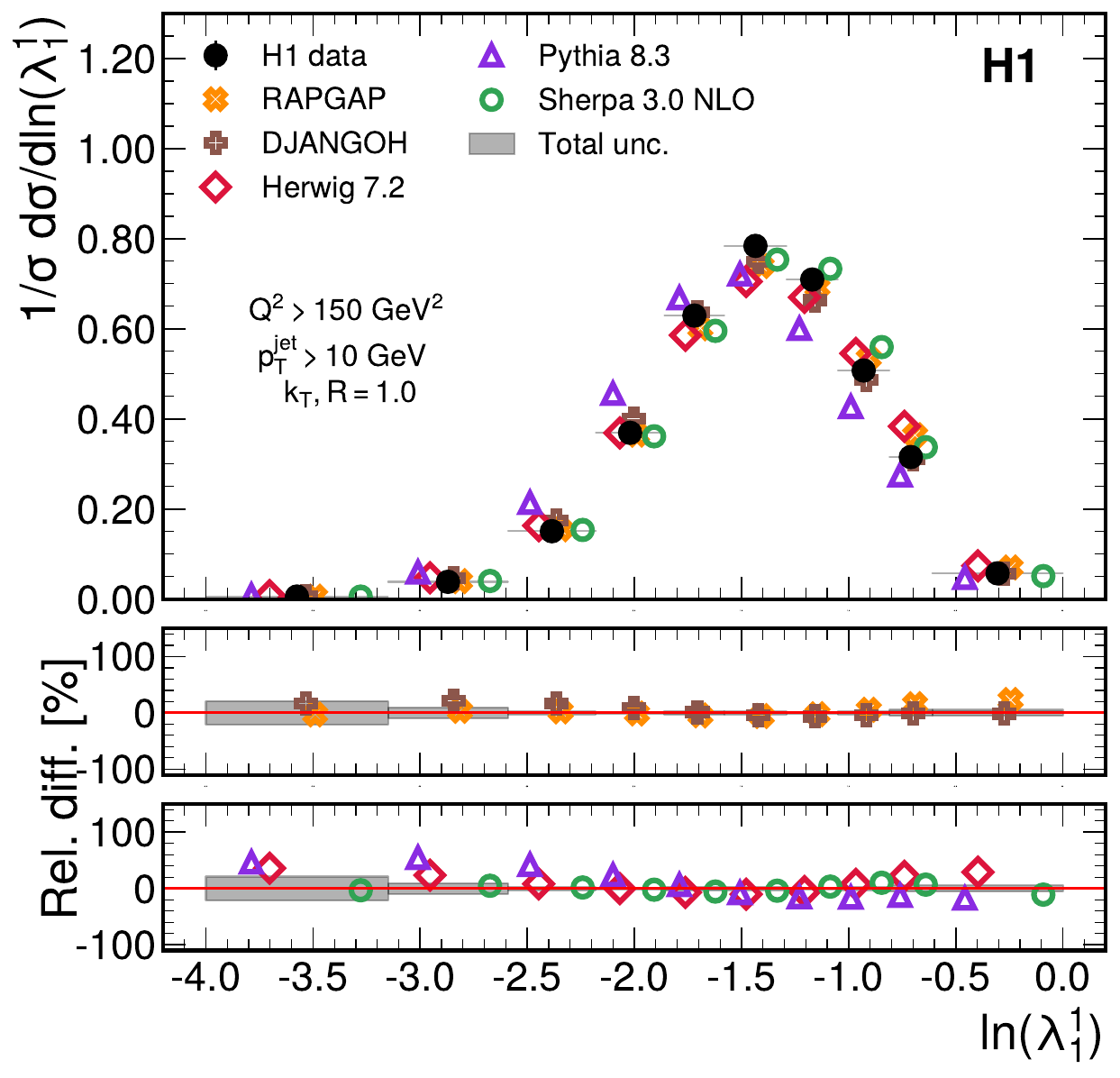}
    \includegraphics[width=.32\textwidth]{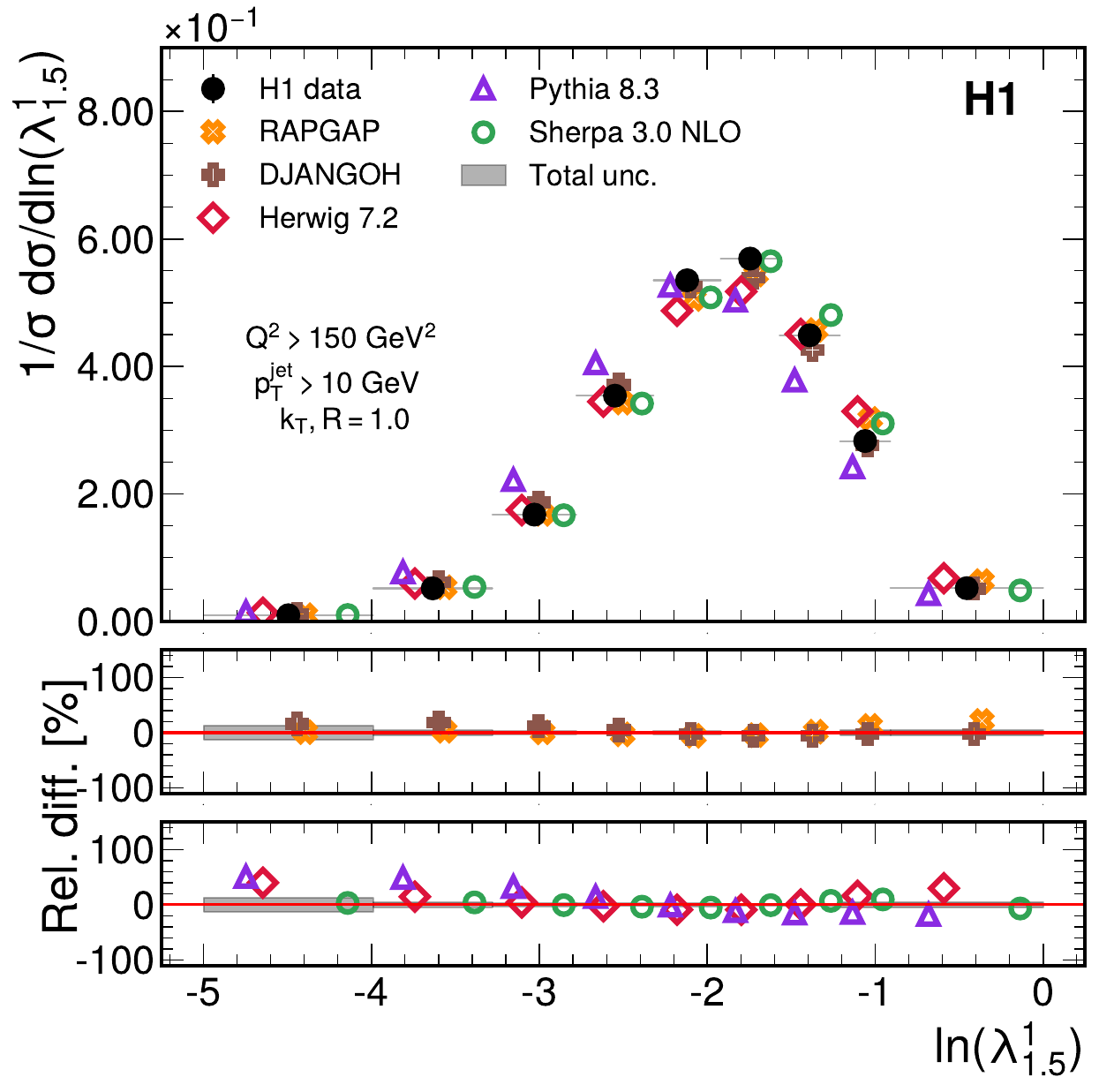}
        \includegraphics[width=.32\textwidth]{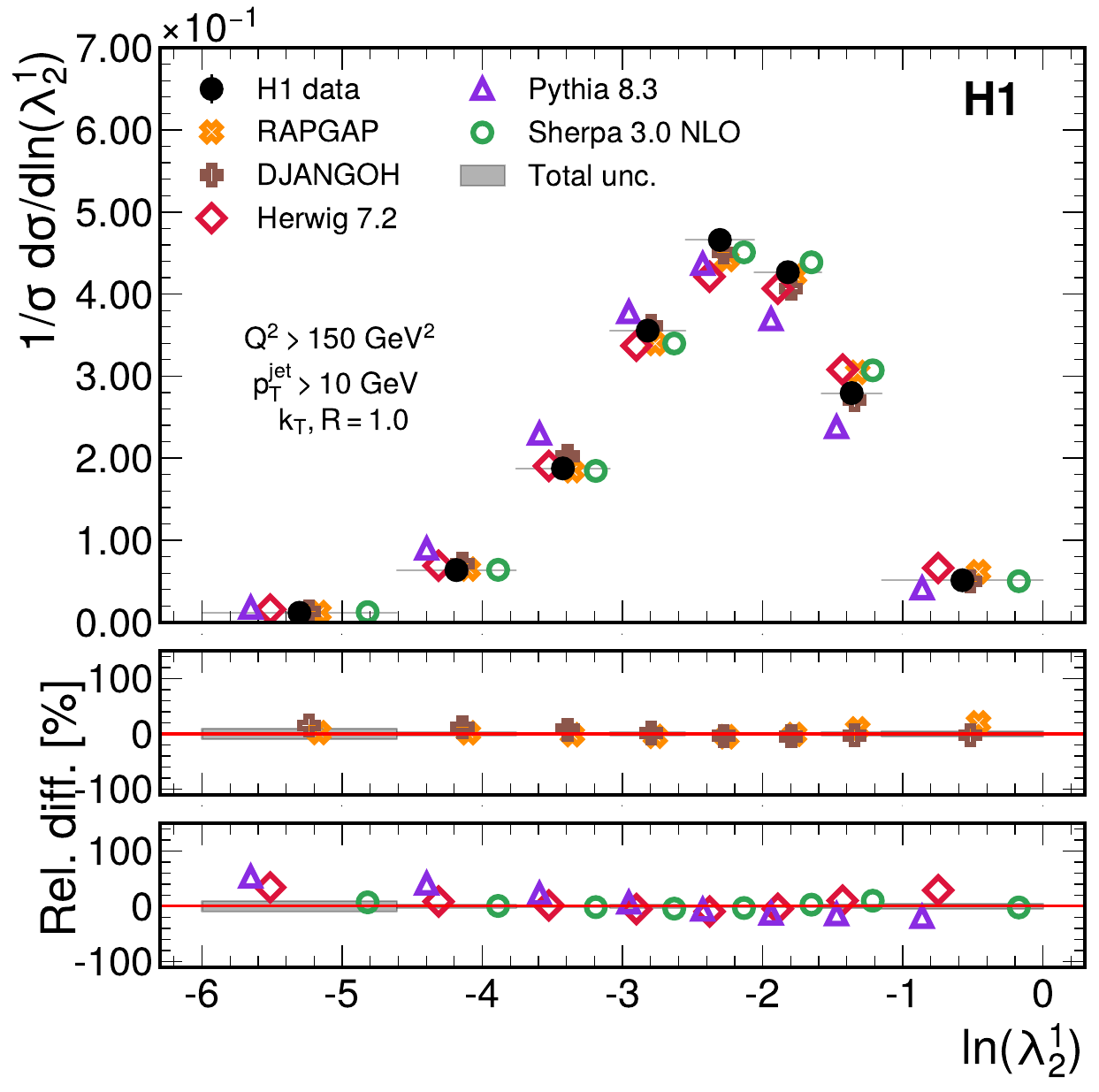}
        \includegraphics[width=.32\textwidth]{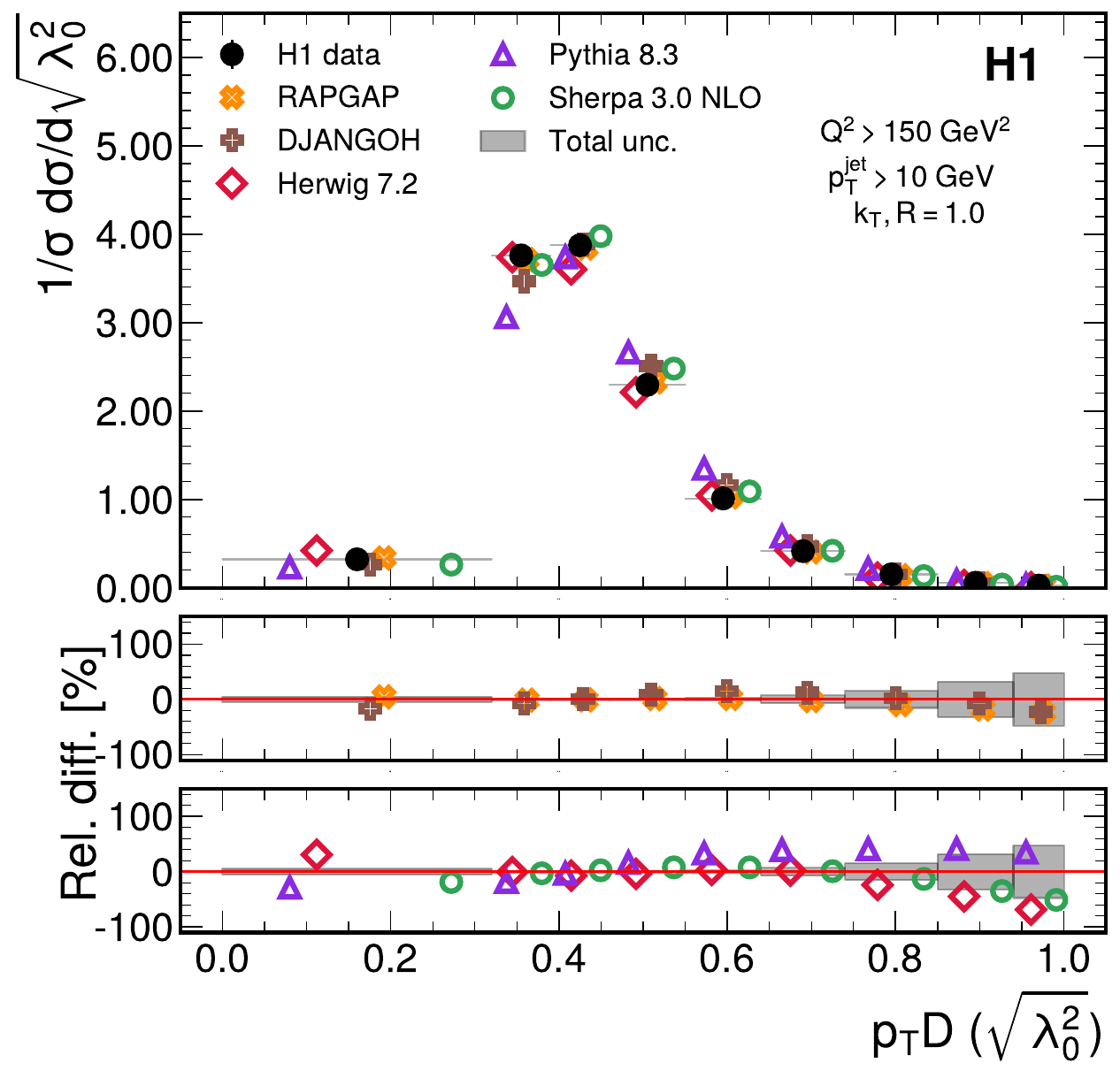}        
        \includegraphics[width=.32\textwidth]{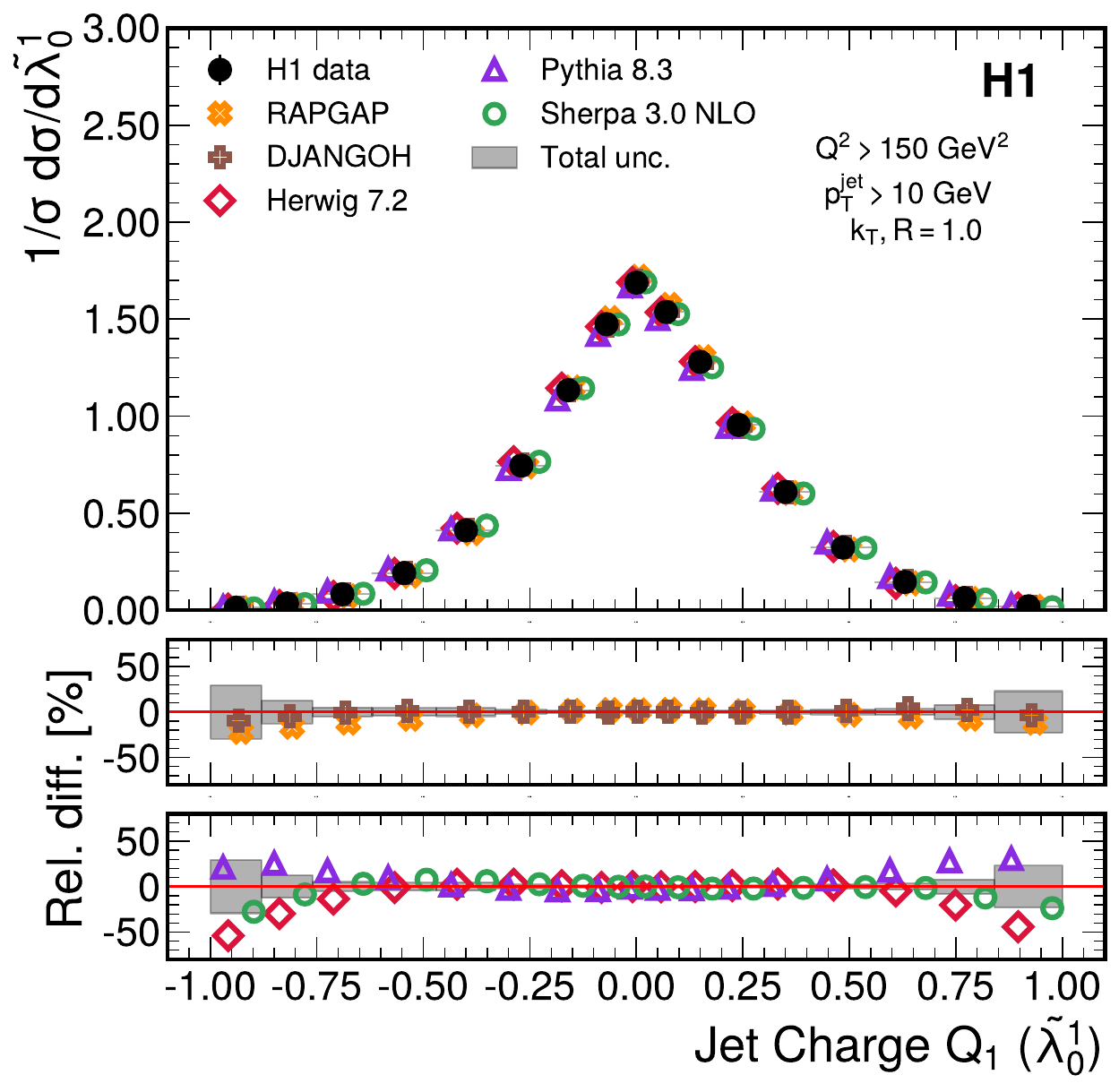}
        \includegraphics[width=.32\textwidth]{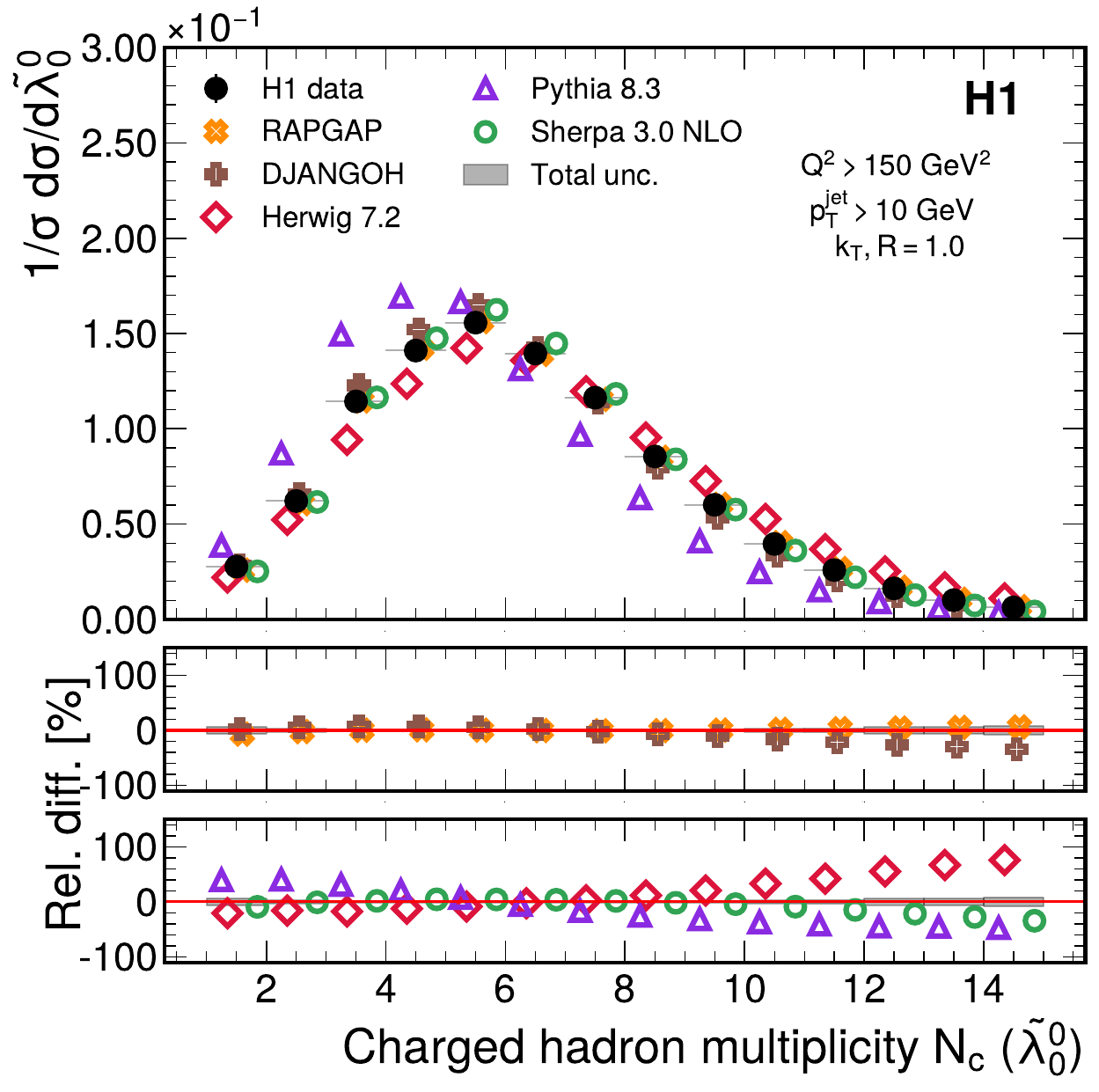}
        
    \caption{Measured cross sections, normalized to the inclusive jet production cross section as a function of the jet angularities measured in this work. Data are shown as solid dots, horizontal bars indicate the bin ranges. Predictions from multiple simulations are shown for comparison, and are offset horizontally for visual clarity. The relative differences between data and predictions are shown in the bottom panels, split between dedicated DIS simulators (middle) and general purpose simulators (bottom). Gray bands represent the total data uncertainties.
        }
    \label{fig:rapgap_data_sys}
\end{figure*}

\begin{figure*}[htb]
    \centering
    \includegraphics[width=.32\textwidth]{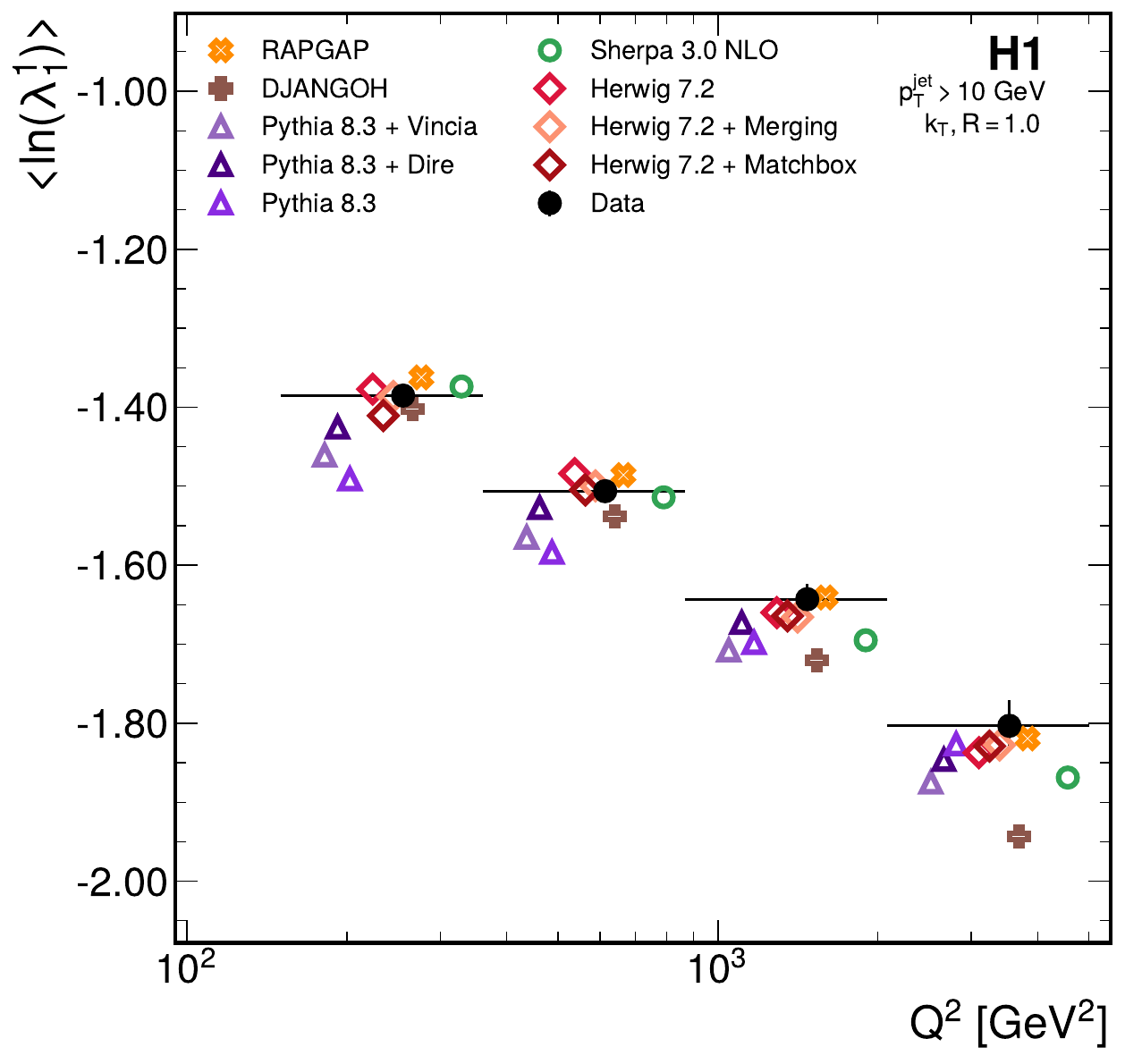}
\includegraphics[width=.32\textwidth]{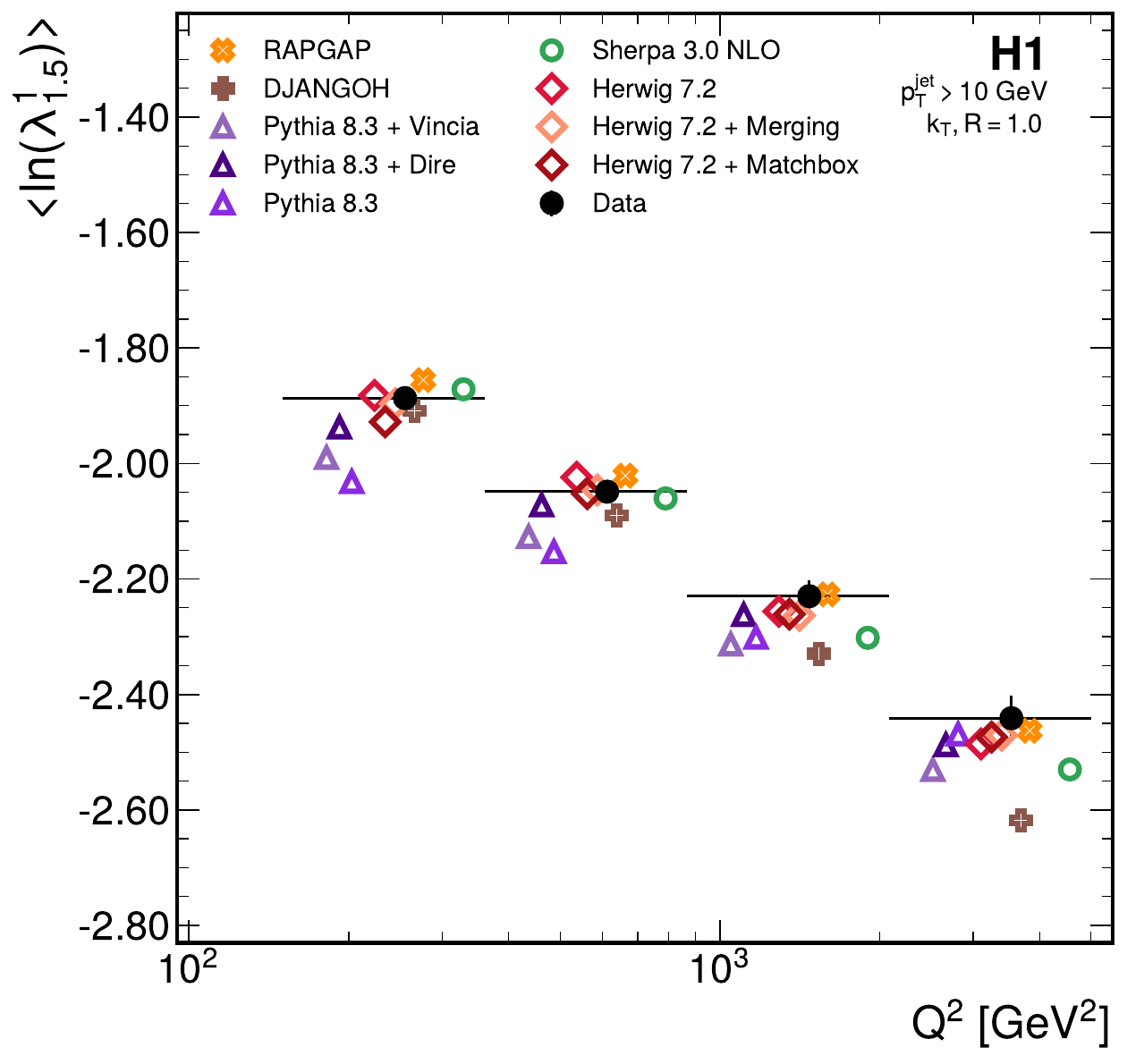}
    \includegraphics[width=.32\textwidth]{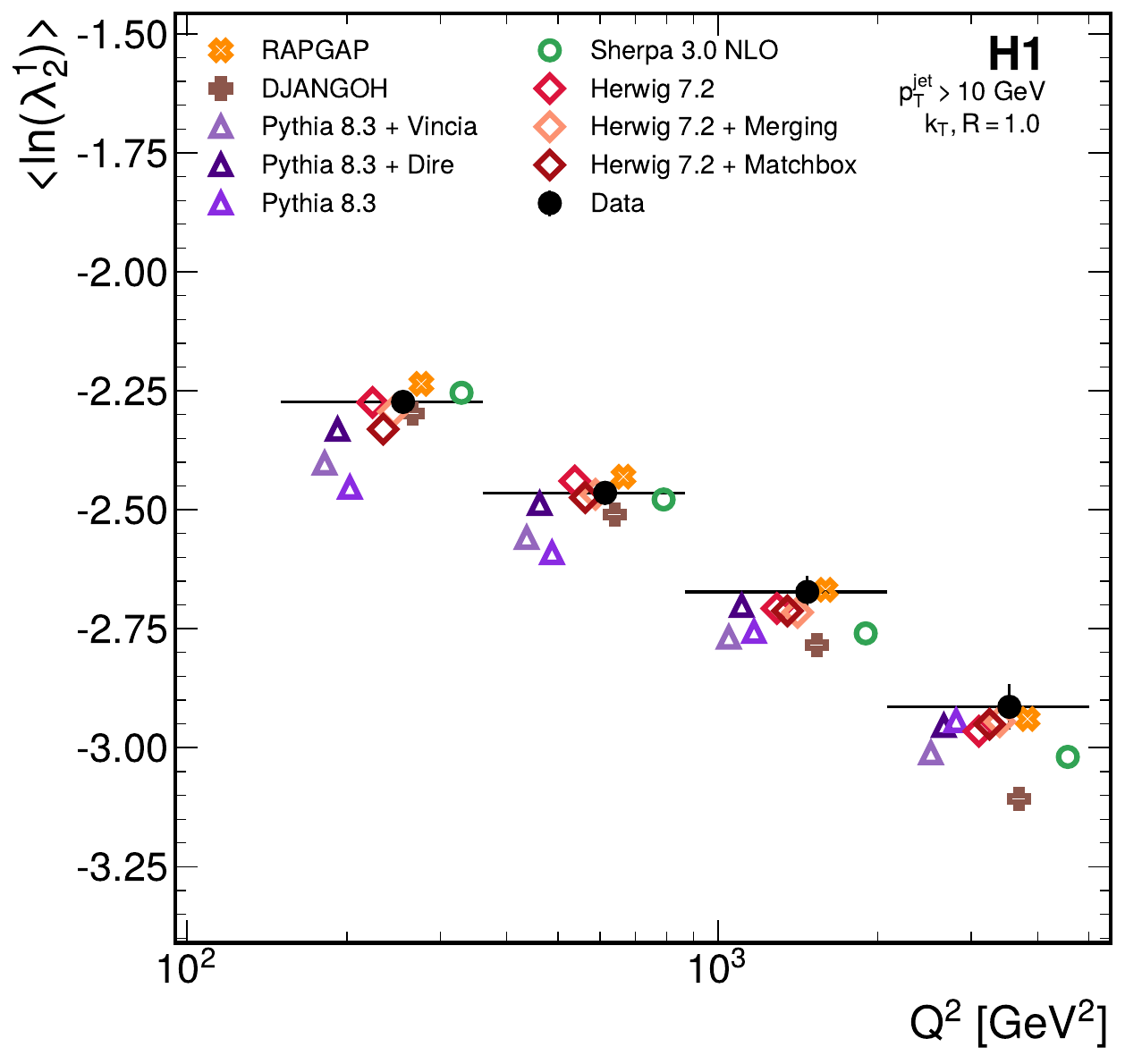}
    \includegraphics[width=.32\textwidth]{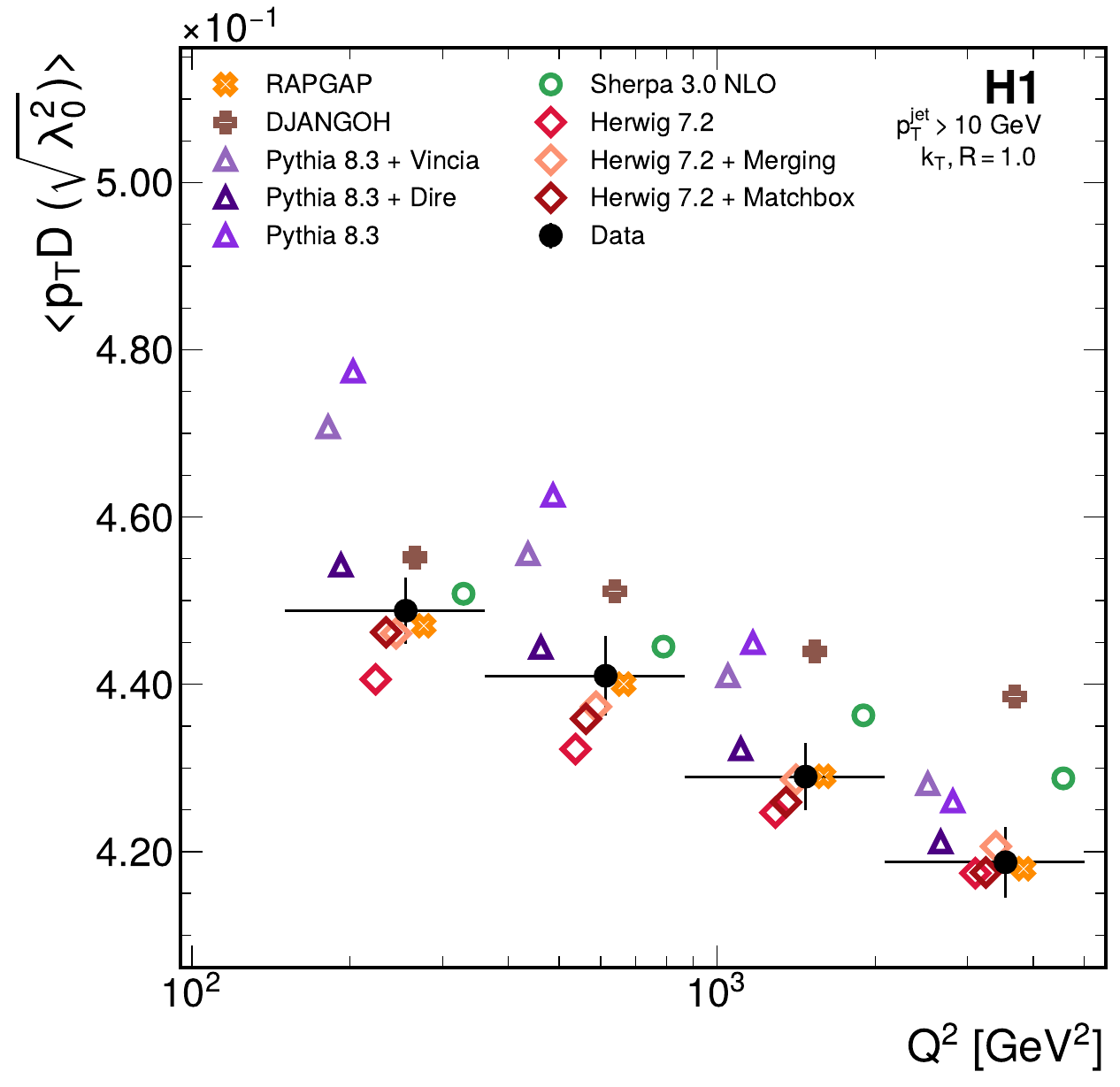}
    \includegraphics[width=.32\textwidth]{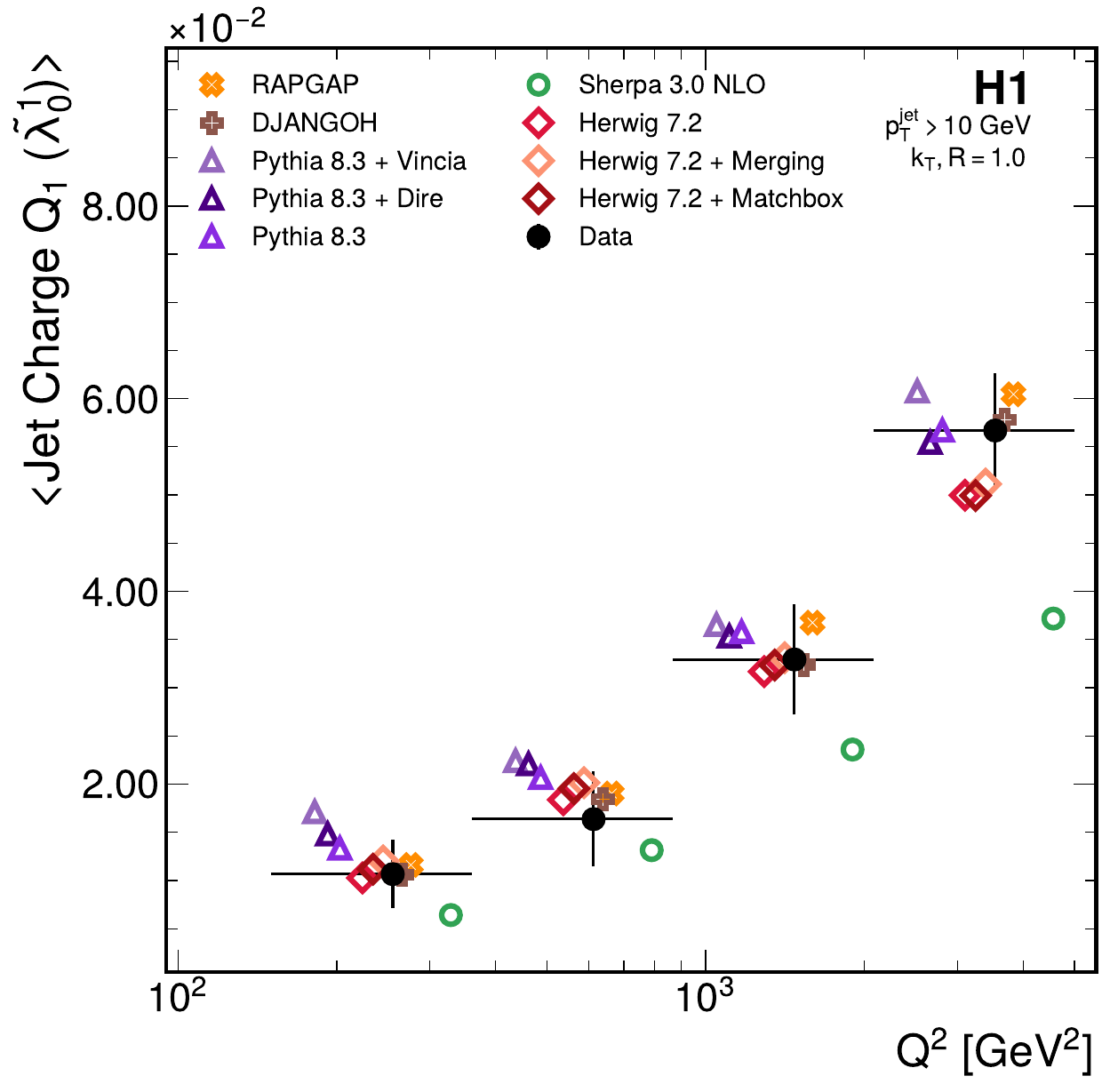}
    \includegraphics[width=.32\textwidth]{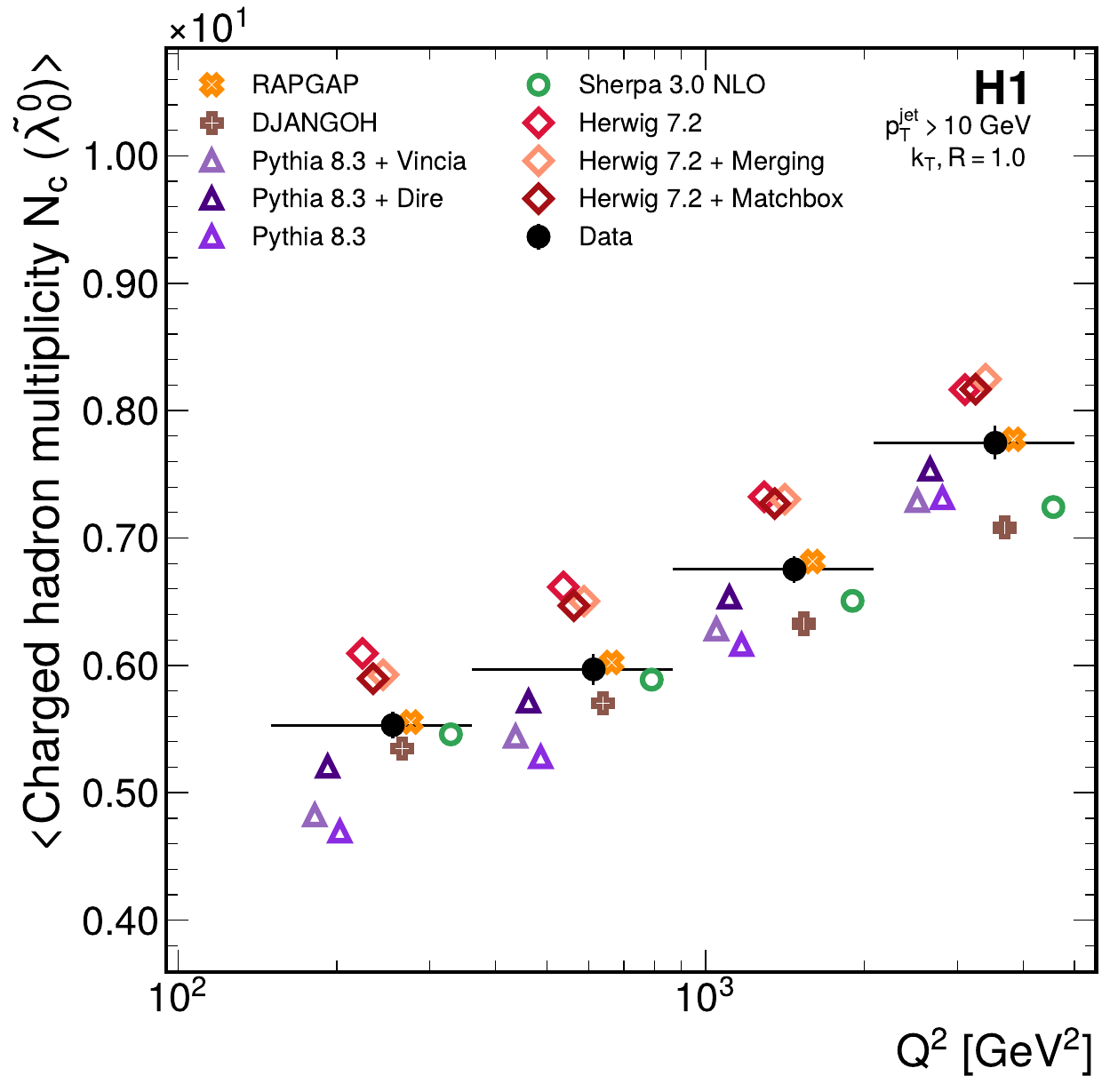}

    \caption{Measured mean of the unfolded jet angularities for multiple \Qs{} intervals. Data are shown as solid dots, horizontal bars indicate the bin ranges. Vertical bars represent the total data uncertainties. Predictions from multiple simulations are shown for comparison, and are offset horizontally for visual clarity.
    }
    \label{fig:rapgap_data_sys_q2_mom1}
\end{figure*}

\begin{figure*}[htb]
    \centering
    \includegraphics[width=.32\textwidth]{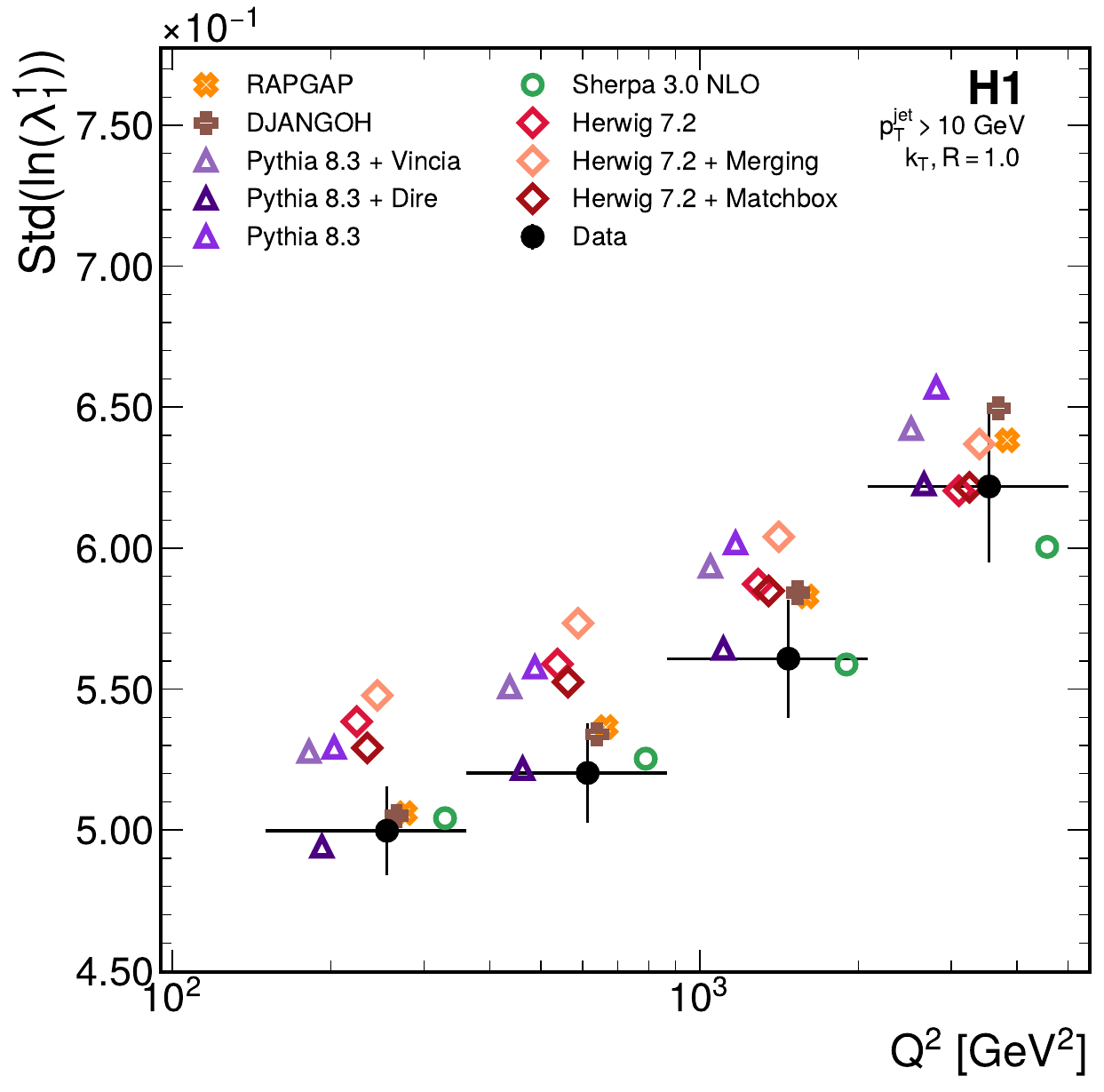}
\includegraphics[width=.32\textwidth]{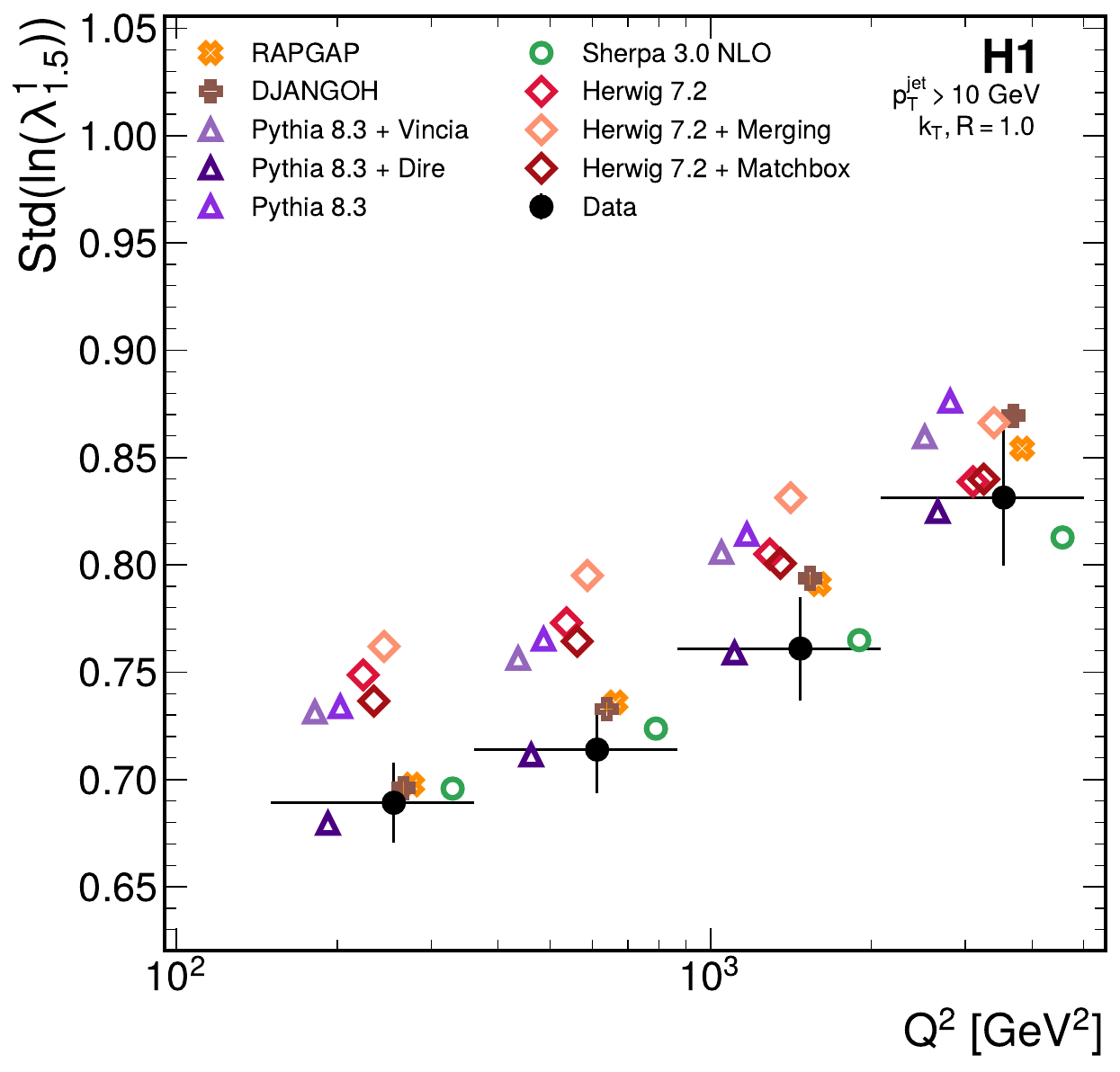}
    \includegraphics[width=.32\textwidth]{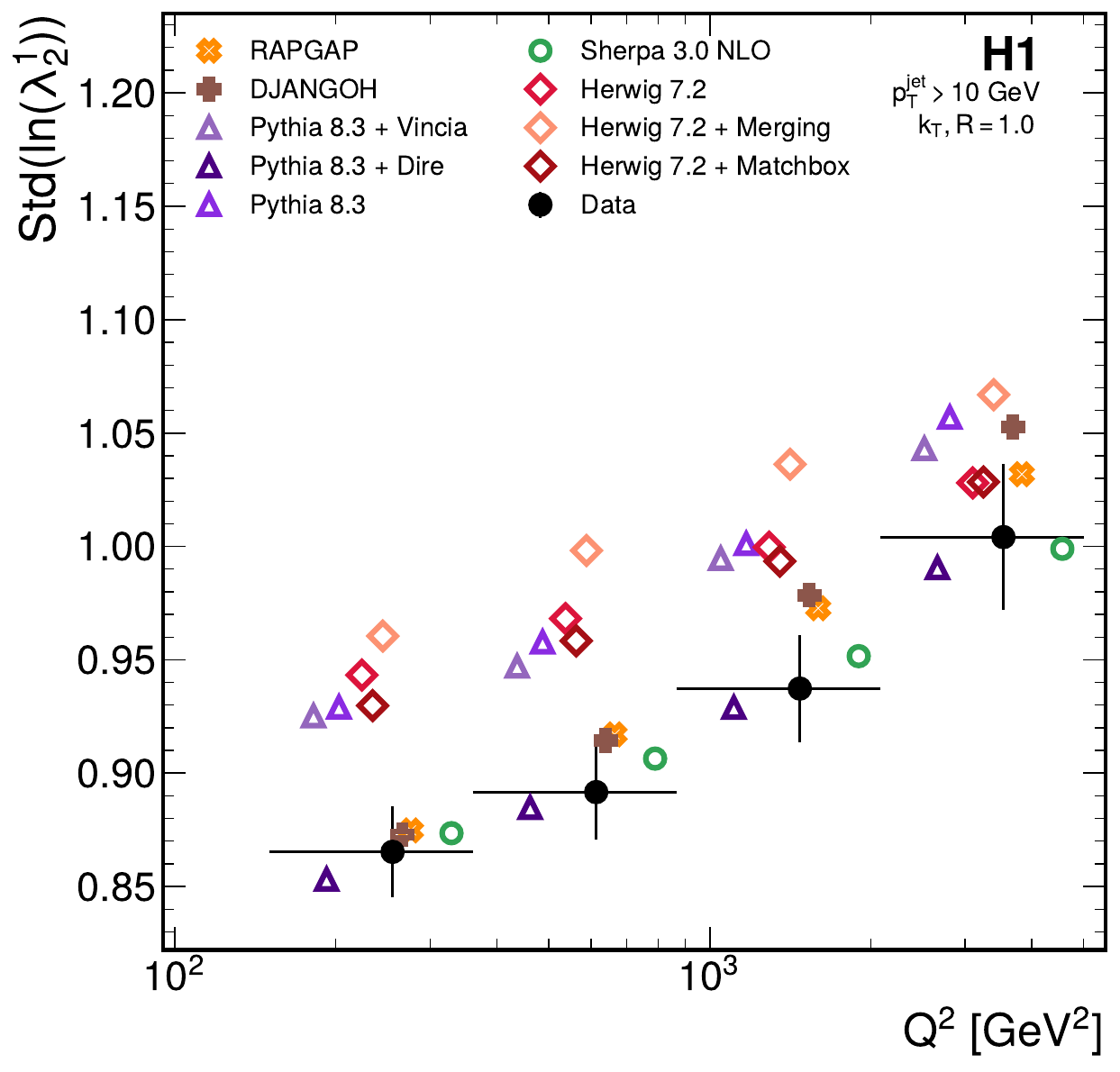}
    \includegraphics[width=.32\textwidth]{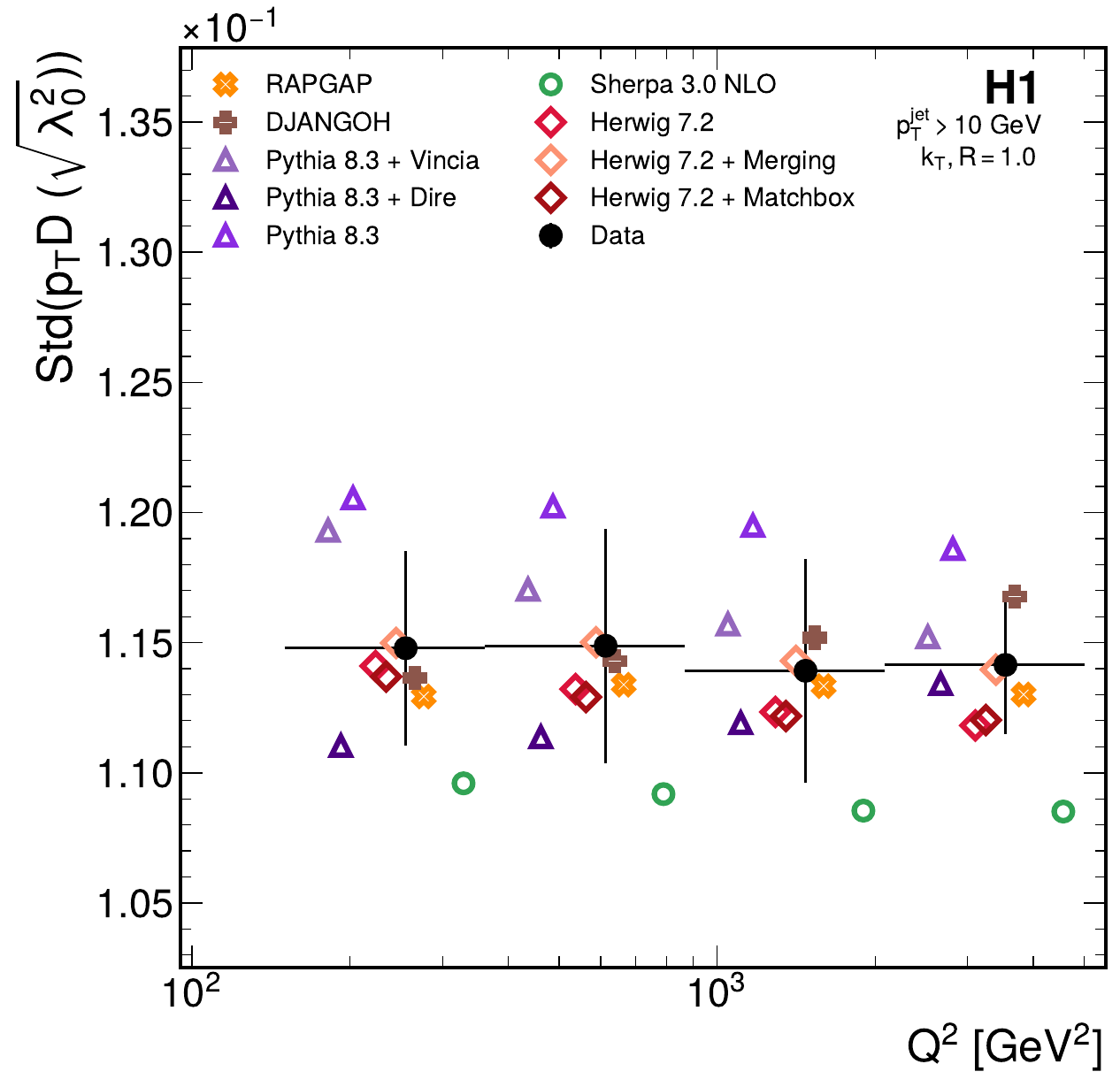}
    \includegraphics[width=.32\textwidth]{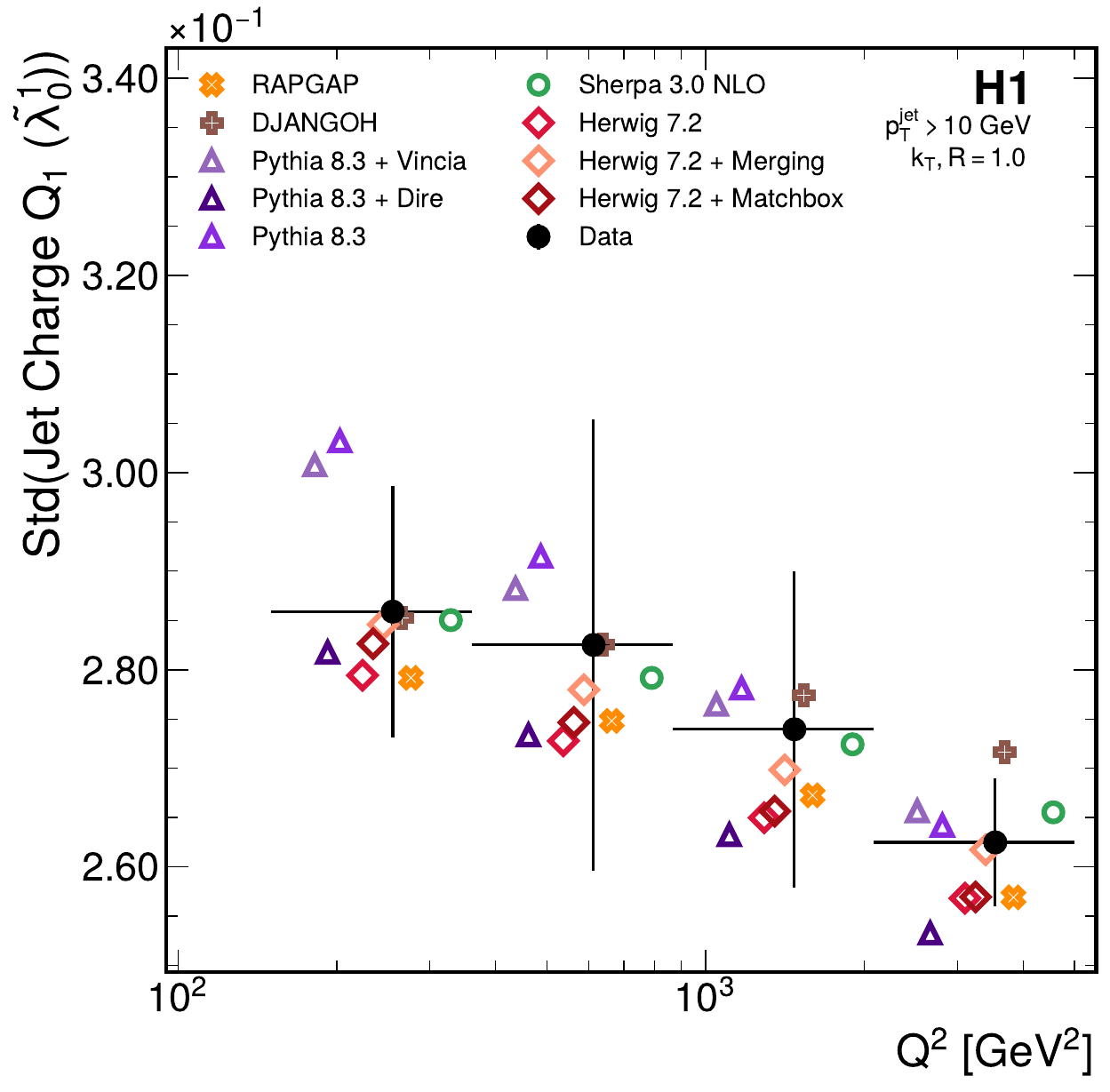}
    \includegraphics[width=.32\textwidth]{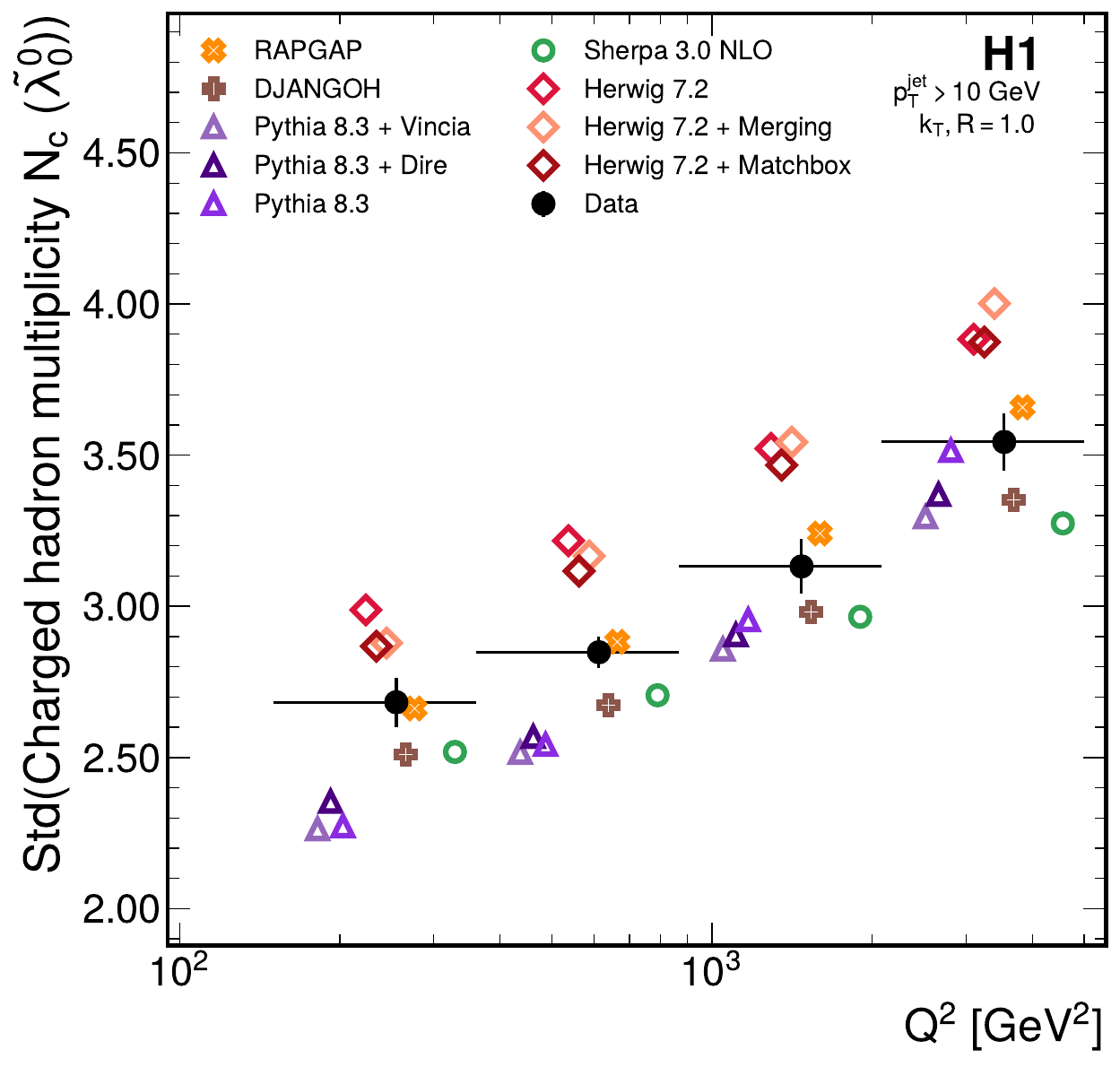}

    \caption{Measured standard deviation of the unfolded jet angularities for multiple \Qs{} intervals. Data are shown as solid dots, horizontal bars indicate the bin ranges. Vertical bars represent the total data uncertainties. Predictions from multiple simulations are shown for comparison, and are offset horizontally for visual clarity.
    }
    \label{fig:rapgap_data_sys_q2_mom2}
\end{figure*}

The results of this analysis are unbinned normalized cross sections as a function of \Qs{} and the generalized jet angularities.  Unfolded results of the normalized differential particle level cross sections of the jet substructure observables are presented in Figs.~\ref{fig:rapgap_data_sys}  in the kinematic region described in Sec.~\ref{sec:h1}. In order to visualize the unfolded distributions bins have been introduced based on the respective detector resolution. All distributions are observed to accumulate around a distinct peak. Mean and standard deviation are calculated from the unbinned, unfolded event sample, and are shown for the six jet angularities in four \Qs{} ranges in Figs.~\ref{fig:rapgap_data_sys_q2_mom1} and \ref{fig:rapgap_data_sys_q2_mom2}, respectively, with numerical results reported in Tables~\ref{tab:mean} and \ref{tab:std}. Distributions of the angularities in the four \Qs{} ranges are shown in \ref{app:q2_int}.

Predictions by the DIS MC generators, \textsc{Djangoh} and \textsc{Rapgap}, and by general purpose simulators, \textsc{Sherpa}, \textsc{Herwig}, \textsc{Pythia} show a good agreement with data for most observables. For \textsc{Herwig}, the largest differences to data are observed in the charged hadron multiplicity, predicting a higher number of particles than observed. In contrast, \textsc{Pythia} underpredicts the charged particle multiplicity.  \textsc{Sherpa} shows a good agreement in all IRC-safe observables and is able to describe the data at low observables values, where contributions from soft-radiation are dominant and challenging to describe, as evidenced by a worse agreement by \textsc{Pythia} and \textsc{Herwig}. 

A larger selection of predictions is used to study the mean and standard deviations as a function of \Qs{} in Figs.~\ref{fig:rapgap_data_sys_q2_mom1} and \ref{fig:rapgap_data_sys_q2_mom2}.
\begin{table*}[h!]
\centering
\caption{Measured means of the jet angularities in multiple $Q^2$ intervals. The statistical (Stat.) and total (Tot.) uncertainties are given, as well as the seven systematic uncertainty sources described in the main text.}
\label{tab:mean}
\renewcommand{\arraystretch}{1.3}
\begin{adjustbox}{width=1\textwidth}
\begin{tabular}{| c | c | c | c || c | c | c | c | c | c | c | c | c |}
\hline
Q$^2$ [GeV$^2$] & $<$$\ln(\lambda_1^1)$$>$ & Stat. & Tot.  & HFS(jet) & HFS(other) & HFS($\phi$) & Lepton(E) & Lepton($\phi$) & Model & Closure\\ 
\hline
$[150.00,360.42]$ & -1.385 & 0.004 & 0.007 & 0.004 & 0.004 & 0.004 & 0.002 & 0.002 & 0.002 & 0.002\\ 
$[360.42,866.03]$ & -1.506 & 0.005 & 0.009 & 0.003 & 0.000 & 0.004 & 0.002 & 0.004 & 0.007 & 0.009\\ 
$[866.03,2080.90]$ & -1.643 & 0.006 & 0.012 & 0.006 & 0.005 & 0.000 & 0.014 & 0.005 & 0.004 & 0.005\\ 
$[2080.90,5000.00]$ & -1.803 & 0.010 & 0.018 & 0.008 & 0.009 & 0.010 & 0.020 & 0.008 & 0.009 & 0.014\\ 
\hline
Q$^2$ [GeV$^2$] & $<$$\ln(\lambda_{1.5}^1)$$>$ & Stat. & Tot.  & HFS(jet) & HFS(other) & HFS($\phi$) & Lepton(E) & Lepton($\phi$) & Model & Closure\\ 
\hline
$[150.00,360.42]$ & -1.887 & 0.006 & 0.007 & 0.006 & 0.006 & 0.004 & 0.003 & 0.004 & 0.003 & 0.002\\ 
$[360.42,866.03]$ & -2.049 & 0.006 & 0.010 & 0.004 & 0.004 & 0.006 & 0.001 & 0.006 & 0.009 & 0.013\\ 
$[866.03,2080.90]$ & -2.230 & 0.008 & 0.012 & 0.009 & 0.006 & 0.004 & 0.020 & 0.007 & 0.007 & 0.006\\ 
$[2080.90,5000.00]$ & -2.441 & 0.013 & 0.016 & 0.010 & 0.012 & 0.013 & 0.026 & 0.011 & 0.008 & 0.008\\ 
\hline
Q$^2$ [GeV$^2$] & $<$$\ln(\lambda_2^1)$$>$ & Stat. & Tot.  & HFS(jet) & HFS(other) & HFS($\phi$) & Lepton(E) & Lepton($\phi$) & Model & Closure\\ 
\hline
$[150.00,360.42]$ & -2.273 & 0.007 & 0.007 & 0.007 & 0.007 & 0.004 & 0.004 & 0.006 & 0.004 & 0.003\\ 
$[360.42,866.03]$ & -2.464 & 0.008 & 0.010 & 0.002 & 0.006 & 0.007 & 0.003 & 0.008 & 0.010 & 0.018\\ 
$[866.03,2080.90]$ & -2.673 & 0.010 & 0.012 & 0.011 & 0.006 & 0.007 & 0.025 & 0.008 & 0.011 & 0.003\\ 
$[2080.90,5000.00]$ & -2.914 & 0.015 & 0.016 & 0.011 & 0.014 & 0.016 & 0.031 & 0.013 & 0.014 & 0.013\\ 
\hline
Q$^2$ [GeV$^2$] & $<$$\sqrt{\lambda_0^2}$$>$ & Stat. & Tot.  & HFS(jet) & HFS(other) & HFS($\phi$) & Lepton(E) & Lepton($\phi$) & Model & Closure\\ 
\hline
$[150.00,360.42]$ & 0.449 & 0.001 & 0.009 & 0.001 & 0.001 & 0.002 & 0.001 & 0.002 & 0.001 & 0.001\\ 
$[360.42,866.03]$ & 0.441 & 0.001 & 0.011 & 0.001 & 0.001 & 0.002 & 0.001 & 0.003 & 0.001 & 0.001\\ 
$[866.03,2080.90]$ & 0.429 & 0.001 & 0.009 & 0.001 & 0.001 & 0.001 & 0.001 & 0.003 & 0.001 & 0.001\\ 
$[2080.90,5000.00]$ & 0.419 & 0.002 & 0.010 & 0.002 & 0.002 & 0.002 & 0.001 & 0.001 & 0.002 & 0.001\\ 
\hline
Q$^2$ [GeV$^2$] & $<$$\tilde{\lambda}_0^1$$>$ & Stat. & Tot.  & HFS(jet) & HFS(other) & HFS($\phi$) & Lepton(E) & Lepton($\phi$) & Model & Closure\\ 
\hline
$[150.00,360.42]$ & 0.011 & 0.001 & 0.332 & 0.001 & 0.001 & 0.001 & 0.001 & 0.001 & 0.001 & 0.001\\ 
$[360.42,866.03]$ & 0.016 & 0.001 & 0.302 & 0.001 & 0.001 & 0.002 & 0.003 & 0.001 & 0.001 & 0.001\\ 
$[866.03,2080.90]$ & 0.033 & 0.002 & 0.174 & 0.001 & 0.001 & 0.003 & 0.004 & 0.001 & 0.001 & 0.001\\ 
$[2080.90,5000.00]$ & 0.057 & 0.003 & 0.104 & 0.002 & 0.003 & 0.002 & 0.001 & 0.001 & 0.002 & 0.002\\ 
\hline
Q$^2$ [GeV$^2$] & $<$$\tilde{\lambda}_0^0$$>$ & Stat. & Tot.  & HFS(jet) & HFS(other) & HFS($\phi$) & Lepton(E) & Lepton($\phi$) & Model & Closure\\ 
\hline
$[150.00,360.42]$ & 5.531 & 0.019 & 0.019 & 0.010 & 0.020 & 0.047 & 0.006 & 0.041 & 0.012 & 0.081\\ 
$[360.42,866.03]$ & 5.969 & 0.024 & 0.021 & 0.021 & 0.023 & 0.020 & 0.020 & 0.036 & 0.070 & 0.081\\ 
$[866.03,2080.90]$ & 6.753 & 0.035 & 0.016 & 0.012 & 0.028 & 0.032 & 0.048 & 0.006 & 0.037 & 0.066\\ 
$[2080.90,5000.00]$ & 7.747 & 0.056 & 0.017 & 0.053 & 0.056 & 0.050 & 0.050 & 0.028 & 0.019 & 0.050\\ 
\hline
\end{tabular}
\end{adjustbox}
\end{table*}
\begin{table*}[h!]
\centering
\caption{Measured standard deviations of the jet angularities in multiple $Q^2$ intervals. The statistical (Stat.) and total (Tot.) uncertainties are given, as well as the seven systematic uncertainty sources described in the main text.}
\label{tab:std}
\renewcommand{\arraystretch}{1.3}
\begin{adjustbox}{width=1\textwidth}
\begin{tabular}{| c | c | c | c || c | c | c | c | c | c | c | c | c |}
\hline
Q$^2$ [GeV$^2$] & std($\ln(\lambda_1^1)$) & Stat. & Tot.  & HFS(jet) & HFS(other) & HFS($\phi$) & Lepton(E) & Lepton($\phi$) & Model & Closure\\ 
\hline
$[150.00,360.42]$ & 0.500 & 0.002 & 0.016 & 0.008 & 0.005 & 0.008 & 0.008 & 0.004 & 0.002 & 0.002\\ 
$[360.42,866.03]$ & 0.520 & 0.003 & 0.018 & 0.009 & 0.005 & 0.009 & 0.009 & 0.006 & 0.003 & 0.002\\ 
$[866.03,2080.90]$ & 0.561 & 0.003 & 0.021 & 0.012 & 0.006 & 0.005 & 0.011 & 0.007 & 0.008 & 0.003\\ 
$[2080.90,5000.00]$ & 0.622 & 0.004 & 0.027 & 0.012 & 0.006 & 0.000 & 0.011 & 0.007 & 0.018 & 0.004\\ 
\hline
Q$^2$ [GeV$^2$] & std($\ln(\lambda_{1.5}^1)$) & Stat. & Tot.  & HFS(jet) & HFS(other) & HFS($\phi$) & Lepton(E) & Lepton($\phi$) & Model & Closure\\ 
\hline
$[150.00,360.42]$ & 0.689 & 0.003 & 0.019 & 0.010 & 0.005 & 0.010 & 0.009 & 0.004 & 0.002 & 0.002\\ 
$[360.42,866.03]$ & 0.714 & 0.004 & 0.020 & 0.010 & 0.005 & 0.010 & 0.010 & 0.006 & 0.004 & 0.003\\ 
$[866.03,2080.90]$ & 0.761 & 0.004 & 0.024 & 0.014 & 0.006 & 0.005 & 0.012 & 0.008 & 0.010 & 0.004\\ 
$[2080.90,5000.00]$ & 0.831 & 0.005 & 0.032 & 0.015 & 0.009 & 0.002 & 0.013 & 0.008 & 0.021 & 0.002\\ 
\hline
Q$^2$ [GeV$^2$] & std($\ln(\lambda_2^1)$) & Stat. & Tot.  & HFS(jet) & HFS(other) & HFS($\phi$) & Lepton(E) & Lepton($\phi$) & Model & Closure\\ 
\hline
$[150.00,360.42]$ & 0.865 & 0.004 & 0.020 & 0.011 & 0.005 & 0.012 & 0.010 & 0.003 & 0.002 & 0.002\\ 
$[360.42,866.03]$ & 0.892 & 0.004 & 0.021 & 0.010 & 0.004 & 0.011 & 0.010 & 0.005 & 0.004 & 0.005\\ 
$[866.03,2080.90]$ & 0.937 & 0.005 & 0.024 & 0.014 & 0.003 & 0.004 & 0.012 & 0.008 & 0.010 & 0.003\\ 
$[2080.90,5000.00]$ & 1.004 & 0.006 & 0.032 & 0.016 & 0.010 & 0.003 & 0.013 & 0.009 & 0.019 & 0.005\\ 
\hline
Q$^2$ [GeV$^2$] & std($\sqrt{\lambda_0^2}$) & Stat. & Tot.  & HFS(jet) & HFS(other) & HFS($\phi$) & Lepton(E) & Lepton($\phi$) & Model & Closure\\ 
\hline
$[150.00,360.42]$ & 0.115 & 0.001 & 0.004 & 0.002 & 0.001 & 0.002 & 0.002 & 0.002 & 0.001 & 0.001\\ 
$[360.42,866.03]$ & 0.115 & 0.001 & 0.005 & 0.002 & 0.001 & 0.002 & 0.001 & 0.002 & 0.001 & 0.001\\ 
$[866.03,2080.90]$ & 0.114 & 0.001 & 0.004 & 0.002 & 0.000 & 0.002 & 0.000 & 0.002 & 0.002 & 0.001\\ 
$[2080.90,5000.00]$ & 0.114 & 0.001 & 0.003 & 0.000 & 0.001 & 0.001 & 0.001 & 0.001 & 0.002 & 0.001\\ 
\hline
Q$^2$ [GeV$^2$] & std($\tilde{\lambda}_0^1$) & Stat. & Tot.  & HFS(jet) & HFS(other) & HFS($\phi$) & Lepton(E) & Lepton($\phi$) & Model & Closure\\ 
\hline
$[150.00,360.42]$ & 0.286 & 0.001 & 0.013 & 0.001 & 0.001 & 0.001 & 0.001 & 0.001 & 0.001 & 0.012\\ 
$[360.42,866.03]$ & 0.283 & 0.001 & 0.023 & 0.001 & 0.001 & 0.001 & 0.001 & 0.002 & 0.000 & 0.023\\ 
$[866.03,2080.90]$ & 0.274 & 0.001 & 0.016 & 0.000 & 0.001 & 0.001 & 0.001 & 0.002 & 0.001 & 0.016\\ 
$[2080.90,5000.00]$ & 0.262 & 0.002 & 0.006 & 0.002 & 0.001 & 0.001 & 0.000 & 0.002 & 0.000 & 0.005\\ 
\hline
Q$^2$ [GeV$^2$] & std($\tilde{\lambda}_0^0$) & Stat. & Tot.  & HFS(jet) & HFS(other) & HFS($\phi$) & Lepton(E) & Lepton($\phi$) & Model & Closure\\ 
\hline
$[150.00,360.42]$ & 2.683 & 0.013 & 0.082 & 0.012 & 0.034 & 0.023 & 0.006 & 0.036 & 0.043 & 0.038\\ 
$[360.42,866.03]$ & 2.848 & 0.012 & 0.052 & 0.012 & 0.003 & 0.009 & 0.011 & 0.026 & 0.011 & 0.038\\ 
$[866.03,2080.90]$ & 3.132 & 0.020 & 0.090 & 0.030 & 0.015 & 0.019 & 0.021 & 0.018 & 0.068 & 0.029\\ 
$[2080.90,5000.00]$ & 3.544 & 0.031 & 0.096 & 0.028 & 0.007 & 0.029 & 0.033 & 0.030 & 0.066 & 0.015\\ 
\hline
\end{tabular}
\end{adjustbox}
\end{table*}
\textsc{Rapgap} shows a good agreement with data in all distributions and \Qs{} intervals, while \textsc{Djangoh} shows worse agreement with data at high \Qs{} values. Of the three \textsc{Pythia} prediction variants, where different parton shower models are studied, the \textsc{Dire} parton shower provides the best overall description of the data. The \textsc{Herwig} variants, where NLO matrix elements and details of the matching procedure are studied, all are in overall good agreement with data. Large deviations from data are observed in the charged hadron multiplicity, which has a high sensitivity to the modeling of the hadronization process.  Predictions from \textsc{Sherpa} show a better agreement with data at lower \Qs{} values and  larger discrepancies at the highest \Qs{} interval. \textsc{Sherpa} is also able to provide a good description of the standard deviation of the IRC jet angularities together with DIS decidated generators and the \textsc{Pythia} implementation with \textsc{Dire} parton shower. In contrast, all other implementations predict a larger standard deviation than observed.

\section{Conclusion}\label{se:conclusion}
A first measurement of jet angularities in neutral current DIS events with $Q^{2} > 150$~GeV$^{2}$ and $0.2 < y < 0.7$, as well as divided into four \Qs{} intervals is presented.  The distributions as well as mean and standard deviation of these jet substructure observables are sensitive to perturbative QCD effects and to hadronization models used to describe jet formation. 

The process of unfolding detector effects makes use of novel machine learning methods. All  measured distributions are simultaneously unfolded using the \textsc{OmniFold} approach. Kinematic information from each reconstructed particle clustered inside a jet is input to a dedicated Graph Neural Network implementation that learns the correlation between particles clustered inside a jet. On particle level, however, the jet substructure is represented by the generalized angularities alone. The success of this approach demonstrates that different sets of unbinned observables at detector level and particle level can be used to determine the unfolded distributions. The training of over two thousand neural networks is carried out on the Perlmutter supercomputer using 128 GPUs simultaneously during the unfolding procedure.

While the unfolding procedure is unbinned, results are provided as histograms for both single- and multi-differential cross sections to ease the comparison with different theory predictions. Mean and standard deviation of all observables are calculated after unfolding in multiple $Q^2$ ranges. Theory predictions from dedicated DIS simulators provide an overall good description of all measured quantities. In particular, a good agreement between data and~\textsc{Rapgap} simulation is observed for all distributions and \Qs{} intervals. General purpose simulators are also able to describe the data well. \textsc{Pythia} interfaced with \textsc{Dire} parton shower, \textsc{Herwig}, and \textsc{Sherpa} predictions all show a good agreement with most of the jet angularities studied. The largest differences to data are observed for the charged hadron multiplicity, which is expected to be most sensitive to non-perturbative effects.
The H1 data can be used to further improve the precision of event generators and analytic calculations. The $Q^2$ dependence is measured and can be compared to scale evolution predictions.

The measurements can serve as a complementary guide for jet substructure studies at the future Electron-Ion Collider, which is going to record much larger data samples in a similar kinematic region.  Results presented in this work make use of all detector objects during the unfolding procedure, learning the relationship between reconstructed objects using neural networks. This points to a possible more general algorithm for data unfolding. In the generalization, all particles or jets in a collisions could be made available after unfolding. This would greatly simplify future comparisons to  predictions, in particular for specialized theories valid only in small phase-space regions, or for new observables not known to date.

\section*{Acknowledgements}

We are grateful to the HERA machine group whose outstanding efforts have made this experiment possible. We thank the engineers and technicians for their work in constructing and maintaining the H1 detector, our funding agencies for financial support, the DESY technical staff for continual assistance and the DESY directorate for support and for the hospitality which they extend to the non-DESY members of the collaboration.

We express our thanks to all those involved in securing not only the H1 data but also the software and working environment for long term use, allowing the unique H1 data set to continue to be explored. The transfer from experiment specific to central resources with long term support, including both storage and batch systems, has also been crucial to this enterprise. We therefore also acknowledge the role played by DESY-IT and all people involved during this transition and their future role in the years to come.

\par\noindent$^{a}$ supported by the U.S. DOE Office of Science
\par\noindent$^{b}$ supported by FNRS-FWO-Vlaanderen, IISN-IIKW and IWT and by Interuniversity Attraction Poles Programme, Belgian Science Policy
\par\noindent$^{c}$ supported by the UK Science and Technology Facilities Council, and formerly by the UK Particle Physics and Astronomy Research Council
\par\noindent$^{d}$ supported by the Romanian National Authority for Scientific Research under the contract PN 09370101
\par\noindent$^{e}$ supported by the Bundesministerium für Bildung und Forschung, FRG, under contract numbers 05H09GUF, 05H09VHC, 05H09VHF, 05H16PEA
\par\noindent$^{f}$ partially supported by Polish Ministry of Science and Higher Education, grant DPN/N168/DESY/2009
\par\noindent$^{g}$ partially supported by Ministry of Science of Montenegro, no. 05-1/3-3352
\par\noindent$^{h}$ supported by the Ministry of Education of the Czech Republic under the project INGO-LG14033
\par\noindent$^{i}$ supported by CONACYT, México, grant 48778-F
\par\noindent$^{j}$ supported by the Swiss National Science Foundation

\begin{flushleft}
\bibliographystyle{elsarticle-num-names} 
\bibliography{desy23-034}

\begin{thebibliography}{144}
\expandafter\ifx\csname natexlab\endcsname\relax\def\natexlab#1{#1}\fi
\providecommand{\url}[1]{\texttt{#1}}
\providecommand{\href}[2]{#2}
\providecommand{\path}[1]{#1}
\providecommand{\DOIprefix}{doi:}
\providecommand{\ArXivprefix}{arXiv:}
\providecommand{\URLprefix}{URL: }
\providecommand{\Pubmedprefix}{pmid:}
\providecommand{\doi}[1]{\href{http://dx.doi.org/#1}{\path{#1}}}
\providecommand{\Pubmed}[1]{\href{pmid:#1}{\path{#1}}}
\providecommand{\bibinfo}[2]{#2}
\ifx\xfnm\relax \def\xfnm[#1]{\unskip,\space#1}\fi
%Type = Article
\bibitem[{Gross et~al.(2022)}]{Gross:2022hyw}
\bibinfo{author}{F.~Gross}, et~al.,
\newblock \bibinfo{title}{{50 Years of Quantum Chromodynamics}}
  (\bibinfo{year}{2022}). \href{http://arxiv.org/abs/2212.11107}{{\tt
  arXiv:2212.11107}}.
%Type = Article
\bibitem[{Adloff et~al.(1999)}]{H1:1998rpm}
\bibinfo{author}{C.~Adloff}, et~al. (\bibinfo{collaboration}{H1}),
\newblock \bibinfo{title}{{Measurement of internal jet structure in dijet
  production in deep inelastic scattering at HERA}},
\newblock \bibinfo{journal}{Nucl. Phys. B} \bibinfo{volume}{545}
  (\bibinfo{year}{1999}) \bibinfo{pages}{3--20}.
  \DOIprefix\doi{10.1016/S0550-3213(99)00118-2}.
  \href{http://arxiv.org/abs/hep-ex/9901010}{{\tt arXiv:hep-ex/9901010}}.
%Type = Article
\bibitem[{Breitweg et~al.(1999)}]{ZEUS:1998fts}
\bibinfo{author}{J.~Breitweg}, et~al. (\bibinfo{collaboration}{ZEUS}),
\newblock \bibinfo{title}{{Measurement of jet shapes in high $Q^{2}$ deep
  inelastic scattering at HERA}},
\newblock \bibinfo{journal}{Eur. Phys. J. C} \bibinfo{volume}{8}
  (\bibinfo{year}{1999}) \bibinfo{pages}{367--380}.
  \DOIprefix\doi{10.1007/s100520050471}.
  \href{http://arxiv.org/abs/hep-ex/9804001}{{\tt arXiv:hep-ex/9804001}}.
%Type = Article
\bibitem[{Chekanov et~al.(2003)}]{ZEUS:2002sux}
\bibinfo{author}{S.~Chekanov}, et~al. (\bibinfo{collaboration}{ZEUS}),
\newblock \bibinfo{title}{{Measurement of subjet multiplicities in neutral
  current deep inelastic scattering at HERA and determination of alpha(s)}},
\newblock \bibinfo{journal}{Phys. Lett. B} \bibinfo{volume}{558}
  (\bibinfo{year}{2003}) \bibinfo{pages}{41--58}.
  \DOIprefix\doi{10.1016/S0370-2693(03)00216-8}.
  \href{http://arxiv.org/abs/hep-ex/0212030}{{\tt arXiv:hep-ex/0212030}}.
%Type = Article
\bibitem[{Chekanov et~al.(2004)}]{ZEUS:2004gcp}
\bibinfo{author}{S.~Chekanov}, et~al. (\bibinfo{collaboration}{ZEUS}),
\newblock \bibinfo{title}{{Substructure dependence of jet cross sections at
  HERA and determination of alpha(s)}},
\newblock \bibinfo{journal}{Nucl. Phys. B} \bibinfo{volume}{700}
  (\bibinfo{year}{2004}) \bibinfo{pages}{3--50}.
  \DOIprefix\doi{10.1016/j.nuclphysb.2004.08.049}.
  \href{http://arxiv.org/abs/hep-ex/0405065}{{\tt arXiv:hep-ex/0405065}}.
%Type = Article
\bibitem[{Altheimer et~al.(2012)}]{Altheimer:2012mn}
\bibinfo{author}{A.~Altheimer}, et~al.,
\newblock \bibinfo{title}{{Jet Substructure at the Tevatron and LHC: New
  results, new tools, new benchmarks}},
\newblock \bibinfo{journal}{J. Phys. G} \bibinfo{volume}{39}
  (\bibinfo{year}{2012}) \bibinfo{pages}{063001}.
  \DOIprefix\doi{10.1088/0954-3899/39/6/063001}.
  \href{http://arxiv.org/abs/1201.0008}{{\tt arXiv:1201.0008}}.
%Type = Article
\bibitem[{Altheimer et~al.(2014)}]{Altheimer:2013yza}
\bibinfo{author}{A.~Altheimer}, et~al.,
\newblock \bibinfo{title}{{Boosted Objects and Jet Substructure at the LHC.
  Report of BOOST2012, held at IFIC Valencia, 23rd-27th of July 2012}},
\newblock \bibinfo{journal}{Eur. Phys. J. C} \bibinfo{volume}{74}
  (\bibinfo{year}{2014}) \bibinfo{pages}{2792}.
  \DOIprefix\doi{10.1140/epjc/s10052-014-2792-8}.
  \href{http://arxiv.org/abs/1311.2708}{{\tt arXiv:1311.2708}}.
%Type = Article
\bibitem[{Adams et~al.(2015)}]{Adams:2015hiv}
\bibinfo{author}{D.~Adams}, et~al.,
\newblock \bibinfo{title}{{Towards an Understanding of the Correlations in Jet
  Substructure}},
\newblock \bibinfo{journal}{Eur. Phys. J. C} \bibinfo{volume}{75}
  (\bibinfo{year}{2015}) \bibinfo{pages}{409}.
  \DOIprefix\doi{10.1140/epjc/s10052-015-3587-2}.
  \href{http://arxiv.org/abs/1504.00679}{{\tt arXiv:1504.00679}}.
%Type = Article
\bibitem[{Kogler et~al.(2019)}]{1803.06991}
\bibinfo{author}{R.~Kogler}, et~al.,
\newblock \bibinfo{title}{{Jet Substructure at the Large Hadron Collider:
  Experimental Review}},
\newblock \bibinfo{journal}{Rev. Mod. Phys.} \bibinfo{volume}{91}
  (\bibinfo{year}{2019}) \bibinfo{pages}{045003}.
  \DOIprefix\doi{10.1103/RevModPhys.91.045003}.
  \href{http://arxiv.org/abs/1803.06991}{{\tt arXiv:1803.06991}}.
%Type = Article
\bibitem[{Larkoski et~al.(2020)Larkoski, Moult, and Nachman}]{1709.04464}
\bibinfo{author}{A.~J. Larkoski}, \bibinfo{author}{I.~Moult},
  \bibinfo{author}{B.~Nachman},
\newblock \bibinfo{title}{Jet substructure at the large hadron collider: A
  review of recent advances in theory and machine learning},
\newblock \bibinfo{journal}{Physics Reports} \bibinfo{volume}{841}
  (\bibinfo{year}{2020}) \bibinfo{pages}{1--63}. \URLprefix
  \url{https://www.sciencedirect.com/science/article/pii/S0370157319303643}.
  \DOIprefix\doi{https://doi.org/10.1016/j.physrep.2019.11.001},
  \bibinfo{note}{jet substructure at the Large Hadron Collider: A review of
  recent advances in theory and machine learning}.
%Type = Article
\bibitem[{South and Steder(2012)}]{South:2012vh}
\bibinfo{author}{D.~M. South}, \bibinfo{author}{M.~Steder}
  (\bibinfo{collaboration}{H1}),
\newblock \bibinfo{title}{{The H1 Data Preservation Project}},
\newblock \bibinfo{journal}{J. Phys. Conf. Ser.} \bibinfo{volume}{396}
  (\bibinfo{year}{2012}) \bibinfo{pages}{062019}.
  \DOIprefix\doi{10.1088/1742-6596/396/6/062019}.
  \href{http://arxiv.org/abs/1206.5200}{{\tt arXiv:1206.5200}}.
%Type = Article
\bibitem[{Britzger et~al.(2021)Britzger, Levonian, Schmitt, and
  South}]{Britzger:2021xcx}
\bibinfo{author}{D.~Britzger}, \bibinfo{author}{S.~Levonian},
  \bibinfo{author}{S.~Schmitt}, \bibinfo{author}{D.~South}
  (\bibinfo{collaboration}{H1}),
\newblock \bibinfo{title}{{Preservation through modernisation: The software of
  the H1 experiment at HERA}},
\newblock \bibinfo{journal}{EPJ Web Conf.} \bibinfo{volume}{251}
  (\bibinfo{year}{2021}) \bibinfo{pages}{03004}.
  \DOIprefix\doi{10.1051/epjconf/202125103004}.
  \href{http://arxiv.org/abs/2106.11058}{{\tt arXiv:2106.11058}}.
%Type = Article
\bibitem[{Larkoski et~al.(2014)Larkoski, Thaler, and
  Waalewijn}]{Larkoski:2014pca}
\bibinfo{author}{A.~J. Larkoski}, \bibinfo{author}{J.~Thaler},
  \bibinfo{author}{W.~J. Waalewijn},
\newblock \bibinfo{title}{{Gaining (Mutual) Information about Quark/Gluon
  Discrimination}},
\newblock \bibinfo{journal}{JHEP} \bibinfo{volume}{11} (\bibinfo{year}{2014})
  \bibinfo{pages}{129}. \DOIprefix\doi{10.1007/JHEP11(2014)129}.
  \href{http://arxiv.org/abs/1408.3122}{{\tt arXiv:1408.3122}}.
%Type = Article
\bibitem[{Catani et~al.(1992)Catani, Turnock, and Webber}]{Catani:1992jc}
\bibinfo{author}{S.~Catani}, \bibinfo{author}{G.~Turnock},
  \bibinfo{author}{B.~R. Webber},
\newblock \bibinfo{title}{{Jet broadening measures in $e^{+} e^{-}$
  annihilation}},
\newblock \bibinfo{journal}{Phys. Lett. B} \bibinfo{volume}{295}
  (\bibinfo{year}{1992}) \bibinfo{pages}{269--276}.
  \DOIprefix\doi{10.1016/0370-2693(92)91565-Q}.
%Type = Article
\bibitem[{Rakow and Webber(1981)}]{Rakow:1981qn}
\bibinfo{author}{P.~E.~L. Rakow}, \bibinfo{author}{B.~R. Webber},
\newblock \bibinfo{title}{{Transverse Momentum Moments of Hadron Distributions
  in {QCD} Jets}},
\newblock \bibinfo{journal}{Nucl. Phys. B} \bibinfo{volume}{191}
  (\bibinfo{year}{1981}) \bibinfo{pages}{63--74}.
  \DOIprefix\doi{10.1016/0550-3213(81)90286-8}.
%Type = Article
\bibitem[{Ellis and Webber(1986)}]{Ellis:1986ig}
\bibinfo{author}{R.~K. Ellis}, \bibinfo{author}{B.~R. Webber},
\newblock \bibinfo{title}{{QCD Jet Broadening in Hadron Hadron Collisions}},
\newblock \bibinfo{journal}{Conf. Proc. C} \bibinfo{volume}{860623}
  (\bibinfo{year}{1986}) \bibinfo{pages}{74}.
%Type = Article
\bibitem[{Farhi(1977)}]{Farhi:1977sg}
\bibinfo{author}{E.~Farhi},
\newblock \bibinfo{title}{{A QCD Test for Jets}},
\newblock \bibinfo{journal}{Phys. Rev. Lett.} \bibinfo{volume}{39}
  (\bibinfo{year}{1977}) \bibinfo{pages}{1587--1588}.
  \DOIprefix\doi{10.1103/PhysRevLett.39.1587}.
%Type = Article
\bibitem[{Chatrchyan et~al.(2012)}]{CMS:2012rth}
\bibinfo{author}{S.~Chatrchyan}, et~al. (\bibinfo{collaboration}{CMS}),
\newblock \bibinfo{title}{{Search for a Higgs boson in the decay channel $H$ to
  ZZ(*) to $q$ qbar $\ell^-$ l+ in $pp$ collisions at $\sqrt{s}=7$ TeV}},
\newblock \bibinfo{journal}{JHEP} \bibinfo{volume}{04} (\bibinfo{year}{2012})
  \bibinfo{pages}{036}. \DOIprefix\doi{10.1007/JHEP04(2012)036}.
  \href{http://arxiv.org/abs/1202.1416}{{\tt arXiv:1202.1416}}.
%Type = Phdthesis
\bibitem[{Pandolfi(2012)}]{Pandolfi:2012ima}
\bibinfo{author}{F.~Pandolfi}, \bibinfo{title}{{Search for the Standard Model
  Higgs Boson in the $H \to ZZ \to l^{+}l^{-}q\overline{q}$ Decay Channel at
  CMS}}, Ph.D. thesis, Zurich, ETH, \bibinfo{address}{New York},
  \bibinfo{year}{2012}. \DOIprefix\doi{10.1007/978-3-319-00903-2}.
%Type = Techreport
\bibitem[{CMS(2013)}]{CMS:2013wea}
\bibinfo{title}{{Pileup Jet Identification}}, \bibinfo{type}{Technical Report},
  CERN, \bibinfo{address}{Geneva}, \bibinfo{year}{2013}. \URLprefix
  \url{https://cds.cern.ch/record/1581583}.
%Type = Article
\bibitem[{Field and Feynman(1978)}]{Field:1977fa}
\bibinfo{author}{R.~D. Field}, \bibinfo{author}{R.~P. Feynman},
\newblock \bibinfo{title}{{A Parametrization of the Properties of Quark Jets}},
\newblock \bibinfo{journal}{Nucl. Phys. B} \bibinfo{volume}{136}
  (\bibinfo{year}{1978}) \bibinfo{pages}{1}.
  \DOIprefix\doi{10.1016/0550-3213(78)90015-9}.
%Type = Article
\bibitem[{Krohn et~al.(2013)Krohn, Schwartz, Lin, and Waalewijn}]{Krohn:2012fg}
\bibinfo{author}{D.~Krohn}, \bibinfo{author}{M.~D. Schwartz},
  \bibinfo{author}{T.~Lin}, \bibinfo{author}{W.~J. Waalewijn},
\newblock \bibinfo{title}{{Jet Charge at the LHC}},
\newblock \bibinfo{journal}{Phys. Rev. Lett.} \bibinfo{volume}{110}
  (\bibinfo{year}{2013}) \bibinfo{pages}{212001}.
  \DOIprefix\doi{10.1103/PhysRevLett.110.212001}.
  \href{http://arxiv.org/abs/1209.2421}{{\tt arXiv:1209.2421}}.
%Type = Misc
\bibitem[{CMS(2013)}]{CMS:2013kfa}
\bibinfo{title}{{Performance of quark/gluon discrimination in 8 TeV pp data}},
  \bibinfo{howpublished}{\url{https://cds.cern.ch/record/1599732}},
  \bibinfo{year}{2013}.
%Type = Misc
\bibitem[{ATL(2016)}]{ATLAS:2016wzt}
\bibinfo{title}{{Discrimination of Light Quark and Gluon Jets in $pp$
  collisions at $\sqrt{s} = 8$ TeV with the ATLAS Detector}},
  \bibinfo{howpublished}{\url{https://cds.cern.ch/record/2200202}},
  \bibinfo{year}{2016}.
%Type = Misc
\bibitem[{ATL(2017)}]{ATLAS:2017nma}
\bibinfo{title}{{Quark versus Gluon Jet Tagging Using Charged Particle
  Multiplicity with the ATLAS Detector}},
  \bibinfo{howpublished}{\url{https://cds.cern.ch/record/2263679}},
  \bibinfo{year}{2017}.
%Type = Article
\bibitem[{Arnison et~al.(1986)}]{UA1:1986xyj}
\bibinfo{author}{G.~Arnison}, et~al. (\bibinfo{collaboration}{UA1}),
\newblock \bibinfo{title}{{Analysis of the Fragmentation Properties of Quark
  and Gluon Jets at the CERN SPS $p \bar{p}$ Collider}},
\newblock \bibinfo{journal}{Nucl. Phys. B} \bibinfo{volume}{276}
  (\bibinfo{year}{1986}) \bibinfo{pages}{253--271}.
  \DOIprefix\doi{10.1016/0550-3213(86)90296-8}.
%Type = Article
\bibitem[{Bagnaia et~al.(1984)}]{UA2:1984tht}
\bibinfo{author}{P.~Bagnaia}, et~al. (\bibinfo{collaboration}{UA2,
  Bern-CERN-Copenhagen-Orsay-Pavia-Saclay}),
\newblock \bibinfo{title}{{Measurement of Jet Fragmentation Properties at the
  {CERN} $\bar{p} p$ Collider}},
\newblock \bibinfo{journal}{Phys. Lett. B} \bibinfo{volume}{144}
  (\bibinfo{year}{1984}) \bibinfo{pages}{291--296}.
  \DOIprefix\doi{10.1016/0370-2693(84)91823-9}.
%Type = Article
\bibitem[{Bagnaia et~al.(1983)}]{UA2:1983btx}
\bibinfo{author}{P.~Bagnaia}, et~al. (\bibinfo{collaboration}{UA2}),
\newblock \bibinfo{title}{{Measurement of Production and Properties of Jets at
  the CERN anti-p p Collider}},
\newblock \bibinfo{journal}{Z. Phys. C} \bibinfo{volume}{20}
  (\bibinfo{year}{1983}) \bibinfo{pages}{117--134}.
  \DOIprefix\doi{10.1007/BF01573214}.
%Type = Article
\bibitem[{Bartel et~al.(1983)}]{JADE:1982ttq}
\bibinfo{author}{W.~Bartel}, et~al. (\bibinfo{collaboration}{JADE}),
\newblock \bibinfo{title}{{Experimental Evidence for Differences in $p_T$
  Between Quark Jets and Gluon Jets}},
\newblock \bibinfo{journal}{Phys. Lett. B} \bibinfo{volume}{123}
  (\bibinfo{year}{1983}) \bibinfo{pages}{460--466}.
  \DOIprefix\doi{10.1016/0370-2693(83)90994-2}.
%Type = Article
\bibitem[{Braunschweig et~al.(1989)}]{TASSO:1989orr}
\bibinfo{author}{W.~Braunschweig}, et~al. (\bibinfo{collaboration}{TASSO}),
\newblock \bibinfo{title}{{Charged Multiplicity Distributions and Correlations
  in e+ e- Annihilation at PETRA Energies}},
\newblock \bibinfo{journal}{Z. Phys. C} \bibinfo{volume}{45}
  (\bibinfo{year}{1989}) \bibinfo{pages}{193}.
  \DOIprefix\doi{10.1007/BF01674450}.
%Type = Article
\bibitem[{Derrick et~al.(1985)}]{Derrick:1985du}
\bibinfo{author}{M.~Derrick}, et~al.,
\newblock \bibinfo{title}{{Comparison of Charged Particle Multiplicities in
  Quark and Gluon Jets Produced in $e^+ e^-$ Annihilation at 29-{GeV}}},
\newblock \bibinfo{journal}{Phys. Lett. B} \bibinfo{volume}{165}
  (\bibinfo{year}{1985}) \bibinfo{pages}{449--453}.
  \DOIprefix\doi{10.1016/0370-2693(85)91263-8}.
%Type = Article
\bibitem[{Petersen et~al.(1985)}]{Petersen:1985hp}
\bibinfo{author}{A.~Petersen}, et~al.,
\newblock \bibinfo{title}{{Inclusive Charged Particle Distribution in Nearly
  Threefold Symmetric Three Jet Events at $e$({CM}) = 29-{GeV}}},
\newblock \bibinfo{journal}{Phys. Rev. Lett.} \bibinfo{volume}{55}
  (\bibinfo{year}{1985}) \bibinfo{pages}{1954}.
  \DOIprefix\doi{10.1103/PhysRevLett.55.1954}.
%Type = Article
\bibitem[{Abe et~al.(1994)}]{SLD:1993mfo}
\bibinfo{author}{K.~Abe}, et~al. (\bibinfo{collaboration}{SLD}),
\newblock \bibinfo{title}{{Measurement of the charged multiplicity of Z0
  ---\ensuremath{>} b anti-b events}},
\newblock \bibinfo{journal}{Phys. Rev. Lett.} \bibinfo{volume}{72}
  (\bibinfo{year}{1994}) \bibinfo{pages}{3145--3149}.
  \DOIprefix\doi{10.1103/PhysRevLett.72.3145}.
  \href{http://arxiv.org/abs/hep-ex/9405004}{{\tt arXiv:hep-ex/9405004}}.
%Type = Article
\bibitem[{Abe et~al.(1996)}]{SLD:1996yvs}
\bibinfo{author}{K.~Abe}, et~al. (\bibinfo{collaboration}{SLD}),
\newblock \bibinfo{title}{{Measurement of the charged multiplicities in b, c
  and light quark events from Z0 decays}},
\newblock \bibinfo{journal}{Phys. Lett. B} \bibinfo{volume}{386}
  (\bibinfo{year}{1996}) \bibinfo{pages}{475--485}.
  \DOIprefix\doi{10.1016/0370-2693(96)01025-8}.
  \href{http://arxiv.org/abs/hep-ex/9608008}{{\tt arXiv:hep-ex/9608008}}.
%Type = Article
\bibitem[{Kim et~al.(1989)}]{AMY:1989rdg}
\bibinfo{author}{Y.-K. Kim}, et~al. (\bibinfo{collaboration}{AMY}),
\newblock \bibinfo{title}{{A comparison of quark and gluon jets produced in
  high-energy $e^+e^-$ annihilations}},
\newblock \bibinfo{journal}{Phys. Rev. Lett.} \bibinfo{volume}{63}
  (\bibinfo{year}{1989}) \bibinfo{pages}{1772}.
  \DOIprefix\doi{10.1103/PhysRevLett.63.1772}.
%Type = Article
\bibitem[{Alam et~al.(1997)}]{CLEO:1997mqo}
\bibinfo{author}{M.~S. Alam}, et~al. (\bibinfo{collaboration}{CLEO}),
\newblock \bibinfo{title}{{Study of gluon versus quark fragmentation in Upsilon
  --\ensuremath{>} g g gamma and e+ e- --\ensuremath{>} q anti-q gamma events
  at s**(1/2) = 10-GeV}},
\newblock \bibinfo{journal}{Phys. Rev. D} \bibinfo{volume}{56}
  (\bibinfo{year}{1997}) \bibinfo{pages}{17--22}.
  \DOIprefix\doi{10.1103/PhysRevD.56.17}.
  \href{http://arxiv.org/abs/hep-ex/9701006}{{\tt arXiv:hep-ex/9701006}}.
%Type = Article
\bibitem[{Alexander et~al.(1991)}]{OPAL:1991ssr}
\bibinfo{author}{G.~Alexander}, et~al. (\bibinfo{collaboration}{OPAL}),
\newblock \bibinfo{title}{{A Direct observation of quark - gluon jet
  differences at LEP}},
\newblock \bibinfo{journal}{Phys. Lett. B} \bibinfo{volume}{265}
  (\bibinfo{year}{1991}) \bibinfo{pages}{462--474}.
  \DOIprefix\doi{10.1016/0370-2693(91)90082-2}.
%Type = Article
\bibitem[{Acton et~al.(1993)}]{OPAL:1993uun}
\bibinfo{author}{P.~D. Acton}, et~al. (\bibinfo{collaboration}{OPAL}),
\newblock \bibinfo{title}{{A Study of differences between quark and gluon jets
  using vertex tagging of quark jets}},
\newblock \bibinfo{journal}{Z. Phys. C} \bibinfo{volume}{58}
  (\bibinfo{year}{1993}) \bibinfo{pages}{387--404}.
  \DOIprefix\doi{10.1007/BF01557696}.
%Type = Article
\bibitem[{Akers et~al.(1995)}]{OPAL:1995ab}
\bibinfo{author}{R.~Akers}, et~al. (\bibinfo{collaboration}{OPAL}),
\newblock \bibinfo{title}{{A Model independent measurement of quark and gluon
  jet properties and differences}},
\newblock \bibinfo{journal}{Z. Phys. C} \bibinfo{volume}{68}
  (\bibinfo{year}{1995}) \bibinfo{pages}{179--202}.
  \DOIprefix\doi{10.1007/BF01566667}.
%Type = Article
\bibitem[{Buskulic et~al.(1995)}]{ALEPH:1994hlg}
\bibinfo{author}{D.~Buskulic}, et~al. (\bibinfo{collaboration}{ALEPH}),
\newblock \bibinfo{title}{{Study of the subjet structure of quark and gluon
  jets}},
\newblock \bibinfo{journal}{Phys. Lett. B} \bibinfo{volume}{346}
  (\bibinfo{year}{1995}) \bibinfo{pages}{389--398}.
  \DOIprefix\doi{10.1016/0370-2693(95)00037-L}.
%Type = Article
\bibitem[{Alexander et~al.(1996)}]{OPAL:1996irm}
\bibinfo{author}{G.~Alexander}, et~al. (\bibinfo{collaboration}{OPAL}),
\newblock \bibinfo{title}{{Test of QCD analytic predictions for the
  multiplicity ratio between gluon and quark jets}},
\newblock \bibinfo{journal}{Phys. Lett. B} \bibinfo{volume}{388}
  (\bibinfo{year}{1996}) \bibinfo{pages}{659--672}.
  \DOIprefix\doi{10.1016/S0370-2693(96)01319-6}.
%Type = Article
\bibitem[{Buskulic et~al.(1996)}]{ALEPH:1995oxo}
\bibinfo{author}{D.~Buskulic}, et~al. (\bibinfo{collaboration}{ALEPH}),
\newblock \bibinfo{title}{{Quark and gluon jet properties in symmetric three
  jet events}},
\newblock \bibinfo{journal}{Phys. Lett. B} \bibinfo{volume}{384}
  (\bibinfo{year}{1996}) \bibinfo{pages}{353--364}.
  \DOIprefix\doi{10.1016/0370-2693(96)00849-0}.
%Type = Article
\bibitem[{Abreu et~al.(1996)}]{DELPHI:1995nzf}
\bibinfo{author}{P.~Abreu}, et~al. (\bibinfo{collaboration}{DELPHI}),
\newblock \bibinfo{title}{{Energy dependence of the differences between the
  quark and gluon jet fragmentation}},
\newblock \bibinfo{journal}{Z. Phys. C} \bibinfo{volume}{70}
  (\bibinfo{year}{1996}) \bibinfo{pages}{179--196}.
  \DOIprefix\doi{10.1007/s002880050095}.
%Type = Article
\bibitem[{Ackerstaff et~al.(1998)}]{OPAL:1997dkk}
\bibinfo{author}{K.~Ackerstaff}, et~al. (\bibinfo{collaboration}{OPAL}),
\newblock \bibinfo{title}{{Multiplicity distributions of gluon and quark jets
  and tests of QCD analytic predictions}},
\newblock \bibinfo{journal}{Eur. Phys. J. C} \bibinfo{volume}{1}
  (\bibinfo{year}{1998}) \bibinfo{pages}{479--494}.
  \DOIprefix\doi{10.1007/s100520050097}.
  \href{http://arxiv.org/abs/hep-ex/9708029}{{\tt arXiv:hep-ex/9708029}}.
%Type = Article
\bibitem[{Abreu et~al.(1999)}]{DELPHI:1999gah}
\bibinfo{author}{P.~Abreu}, et~al. (\bibinfo{collaboration}{DELPHI}),
\newblock \bibinfo{title}{{The Scale dependence of the hadron multiplicity in
  quark and gluon jets and a precise determination of C(A) / C(F)}},
\newblock \bibinfo{journal}{Phys. Lett. B} \bibinfo{volume}{449}
  (\bibinfo{year}{1999}) \bibinfo{pages}{383--400}.
  \DOIprefix\doi{10.1016/S0370-2693(99)00112-4}.
  \href{http://arxiv.org/abs/hep-ex/9903073}{{\tt arXiv:hep-ex/9903073}}.
%Type = Article
\bibitem[{Abbiendi et~al.(1999)}]{OPAL:1999jkz}
\bibinfo{author}{G.~Abbiendi}, et~al. (\bibinfo{collaboration}{OPAL}),
\newblock \bibinfo{title}{{Experimental properties of gluon and quark jets from
  a point source}},
\newblock \bibinfo{journal}{Eur. Phys. J. C} \bibinfo{volume}{11}
  (\bibinfo{year}{1999}) \bibinfo{pages}{217--238}.
  \DOIprefix\doi{10.1007/s100520050628}.
  \href{http://arxiv.org/abs/hep-ex/9903027}{{\tt arXiv:hep-ex/9903027}}.
%Type = Article
\bibitem[{Akrawy et~al.(1990)}]{OPAL:1990vmr}
\bibinfo{author}{M.~Z. Akrawy}, et~al. (\bibinfo{collaboration}{OPAL}),
\newblock \bibinfo{title}{{A Study of coherence of soft gluons in hadron
  jets}},
\newblock \bibinfo{journal}{Phys. Lett. B} \bibinfo{volume}{247}
  (\bibinfo{year}{1990}) \bibinfo{pages}{617--628}.
  \DOIprefix\doi{10.1016/0370-2693(90)91911-T}.
%Type = Article
\bibitem[{Abbiendi et~al.(2004)}]{OPAL:2004prv}
\bibinfo{author}{G.~Abbiendi}, et~al. (\bibinfo{collaboration}{OPAL}),
\newblock \bibinfo{title}{{Scaling violations of quark and gluon jet
  fragmentation functions in e+ e- annihilations at s**(1/2) = 91.2-GeV and
  183-GeV to 209-GeV}},
\newblock \bibinfo{journal}{Eur. Phys. J. C} \bibinfo{volume}{37}
  (\bibinfo{year}{2004}) \bibinfo{pages}{25--47}.
  \DOIprefix\doi{10.1140/epjc/s2004-01964-4}.
  \href{http://arxiv.org/abs/hep-ex/0404026}{{\tt arXiv:hep-ex/0404026}}.
%Type = Article
\bibitem[{Affolder et~al.(2001)}]{CDF:2001nqb}
\bibinfo{author}{T.~Affolder}, et~al. (\bibinfo{collaboration}{CDF}),
\newblock \bibinfo{title}{{Charged Particle Multiplicity in Jets in $p\bar{p}$
  Collisions at $\sqrt{s} = 1.8$ TeV}},
\newblock \bibinfo{journal}{Phys. Rev. Lett.} \bibinfo{volume}{87}
  (\bibinfo{year}{2001}) \bibinfo{pages}{211804}.
  \DOIprefix\doi{10.1103/PhysRevLett.87.211804}.
%Type = Article
\bibitem[{Aad et~al.(2011{\natexlab{a}})}]{ATLAS:2011eid}
\bibinfo{author}{G.~Aad}, et~al. (\bibinfo{collaboration}{ATLAS}),
\newblock \bibinfo{title}{{Properties of jets measured from tracks in
  proton-proton collisions at center-of-mass energy $\sqrt{s}=7$ TeV with the
  ATLAS detector}},
\newblock \bibinfo{journal}{Phys. Rev. D} \bibinfo{volume}{84}
  (\bibinfo{year}{2011}{\natexlab{a}}) \bibinfo{pages}{054001}.
  \DOIprefix\doi{10.1103/PhysRevD.84.054001}.
  \href{http://arxiv.org/abs/1107.3311}{{\tt arXiv:1107.3311}}.
%Type = Article
\bibitem[{Aad et~al.(2011{\natexlab{b}})}]{ATLAS:2011myc}
\bibinfo{author}{G.~Aad}, et~al. (\bibinfo{collaboration}{ATLAS}),
\newblock \bibinfo{title}{{Measurement of the jet fragmentation function and
  transverse profile in proton-proton collisions at a center-of-mass energy of
  7 TeV with the ATLAS detector}},
\newblock \bibinfo{journal}{Eur. Phys. J. C} \bibinfo{volume}{71}
  (\bibinfo{year}{2011}{\natexlab{b}}) \bibinfo{pages}{1795}.
  \DOIprefix\doi{10.1140/epjc/s10052-011-1795-y}.
  \href{http://arxiv.org/abs/1109.5816}{{\tt arXiv:1109.5816}}.
%Type = Article
\bibitem[{Chatrchyan et~al.(2012)}]{CMS:2012oyn}
\bibinfo{author}{S.~Chatrchyan}, et~al. (\bibinfo{collaboration}{CMS}),
\newblock \bibinfo{title}{{Shape, Transverse Size, and Charged Hadron
  Multiplicity of Jets in pp Collisions at 7 TeV}},
\newblock \bibinfo{journal}{JHEP} \bibinfo{volume}{06} (\bibinfo{year}{2012})
  \bibinfo{pages}{160}. \DOIprefix\doi{10.1007/JHEP06(2012)160}.
  \href{http://arxiv.org/abs/1204.3170}{{\tt arXiv:1204.3170}}.
%Type = Article
\bibitem[{{ATLAS Collaboration}(2016)}]{1602.00988}
\bibinfo{author}{{ATLAS Collaboration}},
\newblock \bibinfo{title}{{Measurement of the charged particle multiplicity
  inside jets from $\sqrt{s}=8$ TeV $pp$ collisions with the ATLAS detector}},
\newblock \bibinfo{journal}{Eur. Phys. J. C} \bibinfo{volume}{76}
  (\bibinfo{year}{2016}) \bibinfo{pages}{1}.
  \DOIprefix\doi{10.1140/epjc/s10052-016-4126-5}.
  \href{http://arxiv.org/abs/1602.00988}{{\tt arXiv:1602.00988}}.
%Type = Article
\bibitem[{{ATLAS Collaboration}(2019)}]{1906.09254}
\bibinfo{author}{{ATLAS Collaboration}},
\newblock \bibinfo{title}{{Properties of jet fragmentation using charged
  particles measured with the ATLAS detector in $pp$ collisions at
  $\sqrt{s}=13$ TeV}},
\newblock \bibinfo{journal}{Phys. Rev. D} \bibinfo{volume}{100}
  (\bibinfo{year}{2019}) \bibinfo{pages}{052011}.
  \DOIprefix\doi{10.1103/PhysRevD.100.052011}.
  \href{http://arxiv.org/abs/1906.09254}{{\tt arXiv:1906.09254}}.
%Type = Article
\bibitem[{Capella et~al.(2000)Capella, Dremin, Gary, Nechitailo, and Tran
  Thanh~Van}]{Capella:1999ms}
\bibinfo{author}{A.~Capella}, \bibinfo{author}{I.~M. Dremin},
  \bibinfo{author}{J.~W. Gary}, \bibinfo{author}{V.~A. Nechitailo},
  \bibinfo{author}{J.~Tran Thanh~Van},
\newblock \bibinfo{title}{{Evolution of average multiplicities of quark and
  gluon jets}},
\newblock \bibinfo{journal}{Phys. Rev. D} \bibinfo{volume}{61}
  (\bibinfo{year}{2000}) \bibinfo{pages}{074009}.
  \DOIprefix\doi{10.1103/PhysRevD.61.074009}.
  \href{http://arxiv.org/abs/hep-ph/9910226}{{\tt arXiv:hep-ph/9910226}}.
%Type = Article
\bibitem[{Dremin and Gary(1999)}]{Dremin:1999ji}
\bibinfo{author}{I.~M. Dremin}, \bibinfo{author}{J.~W. Gary},
\newblock \bibinfo{title}{{Energy dependence of mean multiplicities in gluon
  and quark jets at the next-to-next-to-next-to leading order}},
\newblock \bibinfo{journal}{Phys. Lett. B} \bibinfo{volume}{459}
  (\bibinfo{year}{1999}) \bibinfo{pages}{341--346}.
  \DOIprefix\doi{10.1016/S0370-2693(99)00713-3}.
  \href{http://arxiv.org/abs/hep-ph/9905477}{{\tt arXiv:hep-ph/9905477}},
  \bibinfo{note}{[Erratum: Phys.Lett.B 463, 346--346 (1999)]}.
%Type = Techreport
\bibitem[{ATL(2014)}]{ATL-PHYS-PUB-2014-021}
\bibinfo{title}{{ATLAS Pythia 8 tunes to 7 TeV data}}, \bibinfo{type}{Technical
  Report}, CERN, \bibinfo{address}{Geneva}, \bibinfo{year}{2014}. \URLprefix
  \url{https://cds.cern.ch/record/1966419}.
%Type = Article
\bibitem[{{ATLAS Collaboration}(2016)}]{1509.05190}
\bibinfo{author}{{ATLAS Collaboration}},
\newblock \bibinfo{title}{{Measurement of jet charge in dijet events from
  $sqrt{s}=8$ TeV $pp$ collisions with the ATLAS detector}},
\newblock \bibinfo{journal}{Phys. Rev. D} \bibinfo{volume}{93}
  (\bibinfo{year}{2016}) \bibinfo{pages}{052003}.
  \DOIprefix\doi{10.1103/PhysRevD.93.052003}.
  \href{http://arxiv.org/abs/1509.05190}{{\tt arXiv:1509.05190}}.
%Type = Article
\bibitem[{Sirunyan et~al.(2017)}]{CMS:2017yer}
\bibinfo{author}{A.~M. Sirunyan}, et~al. (\bibinfo{collaboration}{CMS}),
\newblock \bibinfo{title}{{Measurements of jet charge with dijet events in pp
  collisions at $\sqrt{s}=8$ TeV}},
\newblock \bibinfo{journal}{JHEP} \bibinfo{volume}{10} (\bibinfo{year}{2017})
  \bibinfo{pages}{131}. \DOIprefix\doi{10.1007/JHEP10(2017)131}.
  \href{http://arxiv.org/abs/1706.05868}{{\tt arXiv:1706.05868}}.
%Type = Article
\bibitem[{Waalewijn(2012)}]{Waalewijn:2012sv}
\bibinfo{author}{W.~J. Waalewijn},
\newblock \bibinfo{title}{{Calculating the Charge of a Jet}},
\newblock \bibinfo{journal}{Phys. Rev. D} \bibinfo{volume}{86}
  (\bibinfo{year}{2012}) \bibinfo{pages}{094030}.
  \DOIprefix\doi{10.1103/PhysRevD.86.094030}.
  \href{http://arxiv.org/abs/1209.3019}{{\tt arXiv:1209.3019}}.
%Type = Article
\bibitem[{Frye et~al.(2016{\natexlab{a}})Frye, Larkoski, Schwartz, and
  Yan}]{Frye:2016aiz}
\bibinfo{author}{C.~Frye}, \bibinfo{author}{A.~J. Larkoski},
  \bibinfo{author}{M.~D. Schwartz}, \bibinfo{author}{K.~Yan},
\newblock \bibinfo{title}{{Factorization for groomed jet substructure beyond
  the next-to-leading logarithm}},
\newblock \bibinfo{journal}{JHEP} \bibinfo{volume}{07}
  (\bibinfo{year}{2016}{\natexlab{a}}) \bibinfo{pages}{064}.
  \DOIprefix\doi{10.1007/JHEP07(2016)064}.
  \href{http://arxiv.org/abs/1603.09338}{{\tt arXiv:1603.09338}}.
%Type = Article
\bibitem[{Frye et~al.(2016{\natexlab{b}})Frye, Larkoski, Schwartz, and
  Yan}]{Frye:2016okc}
\bibinfo{author}{C.~Frye}, \bibinfo{author}{A.~J. Larkoski},
  \bibinfo{author}{M.~D. Schwartz}, \bibinfo{author}{K.~Yan},
\newblock \bibinfo{title}{{Precision physics with pile-up insensitive
  observables}}  (\bibinfo{year}{2016}{\natexlab{b}}).
  \href{http://arxiv.org/abs/1603.06375}{{\tt arXiv:1603.06375}}.
%Type = Article
\bibitem[{Marzani et~al.(2018)Marzani, Schunk, and Soyez}]{Marzani:2017kqd}
\bibinfo{author}{S.~Marzani}, \bibinfo{author}{L.~Schunk},
  \bibinfo{author}{G.~Soyez},
\newblock \bibinfo{title}{{The jet mass distribution after Soft Drop}},
\newblock \bibinfo{journal}{Eur. Phys. J.} \bibinfo{volume}{C78}
  (\bibinfo{year}{2018}) \bibinfo{pages}{96}.
  \DOIprefix\doi{10.1140/epjc/s10052-018-5579-5}.
  \href{http://arxiv.org/abs/1712.05105}{{\tt arXiv:1712.05105}}.
%Type = Article
\bibitem[{Marzani et~al.(2017)Marzani, Schunk, and Soyez}]{Marzani:2017mva}
\bibinfo{author}{S.~Marzani}, \bibinfo{author}{L.~Schunk},
  \bibinfo{author}{G.~Soyez},
\newblock \bibinfo{title}{{A study of jet mass distributions with grooming}},
\newblock \bibinfo{journal}{JHEP} \bibinfo{volume}{07} (\bibinfo{year}{2017})
  \bibinfo{pages}{132}. \DOIprefix\doi{10.1007/JHEP07(2017)132}.
  \href{http://arxiv.org/abs/1704.02210}{{\tt arXiv:1704.02210}}.
%Type = Article
\bibitem[{Kang et~al.(2018)Kang, Lee, Liu, and Ringer}]{Kang:2018jwa}
\bibinfo{author}{Z.-B. Kang}, \bibinfo{author}{K.~Lee},
  \bibinfo{author}{X.~Liu}, \bibinfo{author}{F.~Ringer},
\newblock \bibinfo{title}{{The groomed and ungroomed jet mass distribution for
  inclusive jet production at the LHC}},
\newblock \bibinfo{journal}{JHEP} \bibinfo{volume}{10} (\bibinfo{year}{2018})
  \bibinfo{pages}{137}. \DOIprefix\doi{10.1007/JHEP10(2018)137}.
  \href{http://arxiv.org/abs/1803.03645}{{\tt arXiv:1803.03645}}.
%Type = Article
\bibitem[{Kang et~al.(2019)Kang, Lee, Liu, and Ringer}]{Kang:2018vgn}
\bibinfo{author}{Z.-B. Kang}, \bibinfo{author}{K.~Lee},
  \bibinfo{author}{X.~Liu}, \bibinfo{author}{F.~Ringer},
\newblock \bibinfo{title}{{Soft drop groomed jet angularities at the LHC}},
\newblock \bibinfo{journal}{Phys. Lett. B} \bibinfo{volume}{793}
  (\bibinfo{year}{2019}) \bibinfo{pages}{41--47}.
  \DOIprefix\doi{10.1016/j.physletb.2019.04.018}.
  \href{http://arxiv.org/abs/1811.06983}{{\tt arXiv:1811.06983}}.
%Type = Article
\bibitem[{Andersen et~al.(2018)}]{Proceedings:2018jsb}
\bibinfo{author}{J.~R. Andersen}, et~al.,
\newblock \bibinfo{title}{{Les Houches 2017: Physics at TeV Colliders Standard
  Model Working Group Report}}  (\bibinfo{year}{2018}).
  \href{http://arxiv.org/abs/1803.07977}{{\tt arXiv:1803.07977}}.
%Type = Article
\bibitem[{Aschenauer et~al.(2020)Aschenauer, Lee, Page, and
  Ringer}]{Aschenauer:2019uex}
\bibinfo{author}{E.-C. Aschenauer}, \bibinfo{author}{K.~Lee},
  \bibinfo{author}{B.~S. Page}, \bibinfo{author}{F.~Ringer},
\newblock \bibinfo{title}{{Jet angularities in photoproduction at the
  Electron-Ion Collider}},
\newblock \bibinfo{journal}{Phys. Rev. D} \bibinfo{volume}{101}
  (\bibinfo{year}{2020}) \bibinfo{pages}{054028}.
  \DOIprefix\doi{10.1103/PhysRevD.101.054028}.
  \href{http://arxiv.org/abs/1910.11460}{{\tt arXiv:1910.11460}}.
%Type = Article
\bibitem[{Kang et~al.(2020)Kang, Liu, Mantry, and
  Shao}]{PhysRevLett.125.242003}
\bibinfo{author}{Z.-B. Kang}, \bibinfo{author}{X.~Liu},
  \bibinfo{author}{S.~Mantry}, \bibinfo{author}{D.~Y. Shao},
\newblock \bibinfo{title}{Jet charge: A flavor prism for spin asymmetries at
  the electron-ion collider},
\newblock \bibinfo{journal}{Phys. Rev. Lett.} \bibinfo{volume}{125}
  (\bibinfo{year}{2020}) \bibinfo{pages}{242003}. \URLprefix
  \url{https://link.aps.org/doi/10.1103/PhysRevLett.125.242003}.
  \DOIprefix\doi{10.1103/PhysRevLett.125.242003}.
%Type = Article
\bibitem[{Agostini et~al.(2021)}]{LHeC:2020van}
\bibinfo{author}{P.~Agostini}, et~al. (\bibinfo{collaboration}{LHeC, FCC-he
  Study Group}),
\newblock \bibinfo{title}{{The Large Hadron\textendash{}Electron Collider at
  the HL-LHC}},
\newblock \bibinfo{journal}{J. Phys. G} \bibinfo{volume}{48}
  (\bibinfo{year}{2021}) \bibinfo{pages}{110501}.
  \DOIprefix\doi{10.1088/1361-6471/abf3ba}.
  \href{http://arxiv.org/abs/2007.14491}{{\tt arXiv:2007.14491}}.
%Type = Article
\bibitem[{Abada et~al.(2019)}]{FCC:2018byv}
\bibinfo{author}{A.~Abada}, et~al. (\bibinfo{collaboration}{FCC}),
\newblock \bibinfo{title}{{FCC Physics Opportunities}: {Future Circular
  Collider Conceptual Design Report Volume 1}},
\newblock \bibinfo{journal}{Eur. Phys. J. C} \bibinfo{volume}{79}
  (\bibinfo{year}{2019}) \bibinfo{pages}{474}.
  \DOIprefix\doi{10.1140/epjc/s10052-019-6904-3}.
%Type = Article
\bibitem[{Abt et~al.(1993)}]{Abt:1993wz}
\bibinfo{author}{I.~Abt}, et~al. (\bibinfo{collaboration}{H1}),
\newblock \bibinfo{title}{{The H1 detector at HERA}},
\newblock \bibinfo{journal}{DESY-93-103}  (\bibinfo{year}{1993}).
%Type = Article
\bibitem[{Andrieu et~al.(1993)}]{Andrieu:1993kh}
\bibinfo{author}{B.~Andrieu}, et~al. (\bibinfo{collaboration}{H1 Calorimeter
  Group}),
\newblock \bibinfo{title}{{The H1 liquid argon calorimeter system}},
\newblock \bibinfo{journal}{Nucl. Instrum. Meth. A} \bibinfo{volume}{336}
  (\bibinfo{year}{1993}) \bibinfo{pages}{460--498}.
  \DOIprefix\doi{10.1016/0168-9002(93)91257-N}.
%Type = Article
\bibitem[{Abt et~al.(1997{\natexlab{a}})}]{Abt:1996hi}
\bibinfo{author}{I.~Abt}, et~al. (\bibinfo{collaboration}{H1}),
\newblock \bibinfo{title}{{The H1 detector at HERA}},
\newblock \bibinfo{journal}{Nucl. Instrum. Meth. A} \bibinfo{volume}{386}
  (\bibinfo{year}{1997}{\natexlab{a}}) \bibinfo{pages}{310--347}.
  \DOIprefix\doi{10.1016/S0168-9002(96)00893-5}.
%Type = Article
\bibitem[{Abt et~al.(1997{\natexlab{b}})}]{Abt:1996xv}
\bibinfo{author}{I.~Abt}, et~al. (\bibinfo{collaboration}{H1}),
\newblock \bibinfo{title}{{The Tracking, calorimeter and muon detectors of the
  H1 experiment at HERA}},
\newblock \bibinfo{journal}{Nucl. Instrum. Meth. A} \bibinfo{volume}{386}
  (\bibinfo{year}{1997}{\natexlab{b}}) \bibinfo{pages}{348--396}.
  \DOIprefix\doi{10.1016/S0168-9002(96)00894-7}.
%Type = Article
\bibitem[{Appuhn et~al.(1997)}]{Appuhn:1996na}
\bibinfo{author}{R.~D. Appuhn}, et~al. (\bibinfo{collaboration}{H1 SPACAL
  Group}),
\newblock \bibinfo{title}{{The H1 lead / scintillating fiber calorimeter}},
\newblock \bibinfo{journal}{Nucl. Instrum. Meth. A} \bibinfo{volume}{386}
  (\bibinfo{year}{1997}) \bibinfo{pages}{397--408}.
  \DOIprefix\doi{10.1016/S0168-9002(96)01171-0}.
%Type = Article
\bibitem[{Pitzl et~al.(2000)}]{Pitzl:2000wz}
\bibinfo{author}{D.~Pitzl}, et~al.,
\newblock \bibinfo{title}{{The H1 silicon vertex detector}},
\newblock \bibinfo{journal}{Nucl. Instrum. Meth. A} \bibinfo{volume}{454}
  (\bibinfo{year}{2000}) \bibinfo{pages}{334--349}.
  \DOIprefix\doi{10.1016/S0168-9002(00)00488-5}.
  \href{http://arxiv.org/abs/hep-ex/0002044}{{\tt arXiv:hep-ex/0002044}}.
%Type = Article
\bibitem[{Andrieu et~al.(1994)}]{Andrieu:1994yn}
\bibinfo{author}{B.~Andrieu}, et~al. (\bibinfo{collaboration}{H1 Calorimeter
  Group}),
\newblock \bibinfo{title}{{Beam tests and calibration of the H1 liquid argon
  calorimeter with electrons}},
\newblock \bibinfo{journal}{Nucl. Instrum. Meth. A} \bibinfo{volume}{350}
  (\bibinfo{year}{1994}) \bibinfo{pages}{57--72}.
  \DOIprefix\doi{10.1016/0168-9002(94)91155-X}.
%Type = Article
\bibitem[{Andrieu et~al.(1993)}]{Andrieu:1993tz}
\bibinfo{author}{B.~Andrieu}, et~al. (\bibinfo{collaboration}{H1 Calorimeter
  Group}),
\newblock \bibinfo{title}{{Results from pion calibration runs for the H1 liquid
  argon calorimeter and comparisons with simulations}},
\newblock \bibinfo{journal}{Nucl. Instrum. Meth. A} \bibinfo{volume}{336}
  (\bibinfo{year}{1993}) \bibinfo{pages}{499--509}.
  \DOIprefix\doi{10.1016/0168-9002(93)91258-O}.
%Type = Article
\bibitem[{Aaron et~al.(2012)}]{H1:2012wor}
\bibinfo{author}{F.~D. Aaron}, et~al. (\bibinfo{collaboration}{H1}),
\newblock \bibinfo{title}{{Determination of the Integrated Luminosity at HERA
  using Elastic QED Compton Events}},
\newblock \bibinfo{journal}{Eur. Phys. J. C} \bibinfo{volume}{72}
  (\bibinfo{year}{2012}) \bibinfo{pages}{2163}.
  \DOIprefix\doi{10.1140/epjc/s10052-012-2163-2}.
  \href{http://arxiv.org/abs/1205.2448}{{\tt arXiv:1205.2448}},
  \bibinfo{note}{[Erratum: Eur.Phys.J.C 74, 2733 (2014)]}.
%Type = Article
\bibitem[{Adloff et~al.(2003)}]{Adloff:2003uh}
\bibinfo{author}{C.~Adloff}, et~al. (\bibinfo{collaboration}{H1}),
\newblock \bibinfo{title}{{Measurement and QCD analysis of neutral and charged
  current cross-sections at HERA}},
\newblock \bibinfo{journal}{Eur. Phys. J. C} \bibinfo{volume}{30}
  (\bibinfo{year}{2003}) \bibinfo{pages}{1--32}.
  \DOIprefix\doi{10.1140/epjc/s2003-01257-6}.
  \href{http://arxiv.org/abs/hep-ex/0304003}{{\tt arXiv:hep-ex/0304003}}.
%Type = Article
\bibitem[{Aaron et~al.(2012)}]{Aaron:2012qi}
\bibinfo{author}{F.~D. Aaron}, et~al. (\bibinfo{collaboration}{H1}),
\newblock \bibinfo{title}{{Inclusive Deep Inelastic Scattering at High $Q^2$
  with Longitudinally Polarised Lepton Beams at HERA}},
\newblock \bibinfo{journal}{JHEP} \bibinfo{volume}{09} (\bibinfo{year}{2012})
  \bibinfo{pages}{061}. \DOIprefix\doi{10.1007/JHEP09(2012)061}.
  \href{http://arxiv.org/abs/1206.7007}{{\tt arXiv:1206.7007}}.
%Type = Article
\bibitem[{Andreev et~al.(2015)}]{Andreev:2014wwa}
\bibinfo{author}{V.~Andreev}, et~al. (\bibinfo{collaboration}{H1}),
\newblock \bibinfo{title}{{Measurement of multijet production in $ep$
  collisions at high $Q^2$ and determination of the strong coupling $\alpha
  _s$}},
\newblock \bibinfo{journal}{Eur. Phys. J. C} \bibinfo{volume}{75}
  (\bibinfo{year}{2015}) \bibinfo{pages}{65}.
  \DOIprefix\doi{10.1140/epjc/s10052-014-3223-6}.
  \href{http://arxiv.org/abs/1406.4709}{{\tt arXiv:1406.4709}}.
%Type = Article
\bibitem[{Andreev et~al.(2017)}]{Andreev:2016tgi}
\bibinfo{author}{V.~Andreev}, et~al. (\bibinfo{collaboration}{H1}),
\newblock \bibinfo{title}{{Measurement of Jet Production Cross Sections in
  Deep-inelastic $ep$ Scattering at HERA}},
\newblock \bibinfo{journal}{Eur. Phys. J. C} \bibinfo{volume}{77}
  (\bibinfo{year}{2017}) \bibinfo{pages}{215}.
  \DOIprefix\doi{10.1140/epjc/s10052-017-4717-9}.
  \href{http://arxiv.org/abs/1611.03421}{{\tt arXiv:1611.03421}},
  \bibinfo{note}{[Erratum: Eur.Phys.J.C 81, 739 (2021)]}.
%Type = Article
\bibitem[{Bassler and Bernardi(1995)}]{Bassler:1994uq}
\bibinfo{author}{U.~Bassler}, \bibinfo{author}{G.~Bernardi},
\newblock \bibinfo{title}{{On the kinematic reconstruction of deep inelastic
  scattering at HERA: The Sigma method}},
\newblock \bibinfo{journal}{Nucl. Instrum. Meth. A} \bibinfo{volume}{361}
  (\bibinfo{year}{1995}) \bibinfo{pages}{197--208}.
  \DOIprefix\doi{10.1016/0168-9002(95)00173-5}.
  \href{http://arxiv.org/abs/hep-ex/9412004}{{\tt arXiv:hep-ex/9412004}}.
%Type = Masterthesis
\bibitem[{Peez(2003)}]{energyflowthesis}
\bibinfo{author}{M.~Peez}, \bibinfo{title}{{Search for deviations from the
  standard model in high transverse energy processes at the electron proton
  collider HERA}}, \bibinfo{type}{Phd thesis}, Lyon 1 U., \bibinfo{year}{2003}.
  \DOIprefix\doi{10.3204/DESY-THESIS-2003-023}.
%Type = Masterthesis
\bibitem[{Hellwig(2004)}]{energyflowthesis2}
\bibinfo{author}{S.~Hellwig}, \bibinfo{title}{{Untersuchung der $D^* -
  \pi_{slow}$ Double Tagging Methode in Charmanalysen}}, Master's thesis,
  Hamburg U., \bibinfo{year}{2004}.
%Type = Masterthesis
\bibitem[{Portheault(2005)}]{energyflowthesis3}
\bibinfo{author}{B.~Portheault}, \bibinfo{title}{{First measurement of charged
  and neutral current cross sections with the polarized positron beam at HERA
  II and QCD-electroweak analyses}}, \bibinfo{type}{Phd thesis}, Paris XI Orsay
  U., \bibinfo{year}{2005}.
%Type = Article
\bibitem[{Cacciari et~al.(2012)Cacciari, Salam, and Soyez}]{Cacciari:2011ma}
\bibinfo{author}{M.~Cacciari}, \bibinfo{author}{G.~P. Salam},
  \bibinfo{author}{G.~Soyez},
\newblock \bibinfo{title}{{FastJet User Manual}},
\newblock \bibinfo{journal}{Eur. Phys. J. C} \bibinfo{volume}{72}
  (\bibinfo{year}{2012}) \bibinfo{pages}{1896}.
  \DOIprefix\doi{10.1140/epjc/s10052-012-1896-2}.
  \href{http://arxiv.org/abs/1111.6097}{{\tt arXiv:1111.6097}}.
%Type = Article
\bibitem[{Cacciari and Salam(2006)}]{Cacciari:2005hq}
\bibinfo{author}{M.~Cacciari}, \bibinfo{author}{G.~P. Salam},
\newblock \bibinfo{title}{{Dispelling the $N^{3}$ myth for the $k_t$
  jet-finder}},
\newblock \bibinfo{journal}{Phys. Lett. B} \bibinfo{volume}{641}
  (\bibinfo{year}{2006}) \bibinfo{pages}{57--61}.
  \DOIprefix\doi{10.1016/j.physletb.2006.08.037}.
  \href{http://arxiv.org/abs/hep-ph/0512210}{{\tt arXiv:hep-ph/0512210}}.
%Type = Article
\bibitem[{Catani et~al.(1993)Catani, Dokshitzer, Seymour, and
  Webber}]{Catani:1993hr}
\bibinfo{author}{S.~Catani}, \bibinfo{author}{Y.~L. Dokshitzer},
  \bibinfo{author}{M.~H. Seymour}, \bibinfo{author}{B.~R. Webber},
\newblock \bibinfo{title}{{Longitudinally invariant $K_t$ clustering algorithms
  for hadron hadron collisions}},
\newblock \bibinfo{journal}{Nucl. Phys. B} \bibinfo{volume}{406}
  (\bibinfo{year}{1993}) \bibinfo{pages}{187--224}.
  \DOIprefix\doi{10.1016/0550-3213(93)90166-M}.
%Type = Article
\bibitem[{Ellis and Soper(1993)}]{Ellis:1993tq}
\bibinfo{author}{S.~D. Ellis}, \bibinfo{author}{D.~E. Soper},
\newblock \bibinfo{title}{{Successive combination jet algorithm for hadron
  collisions}},
\newblock \bibinfo{journal}{Phys. Rev. D} \bibinfo{volume}{48}
  (\bibinfo{year}{1993}) \bibinfo{pages}{3160--3166}.
  \DOIprefix\doi{10.1103/PhysRevD.48.3160}.
  \href{http://arxiv.org/abs/hep-ph/9305266}{{\tt arXiv:hep-ph/9305266}}.
%Type = Article
\bibitem[{Charchula et~al.(1994)Charchula, Schuler, and
  Spiesberger}]{Charchula:1994kf}
\bibinfo{author}{K.~Charchula}, \bibinfo{author}{G.~A. Schuler},
  \bibinfo{author}{H.~Spiesberger},
\newblock \bibinfo{title}{{Combined QED and QCD radiative effects in deep
  inelastic lepton - proton scattering: The Monte Carlo generator DJANGO6}},
\newblock \bibinfo{journal}{Comput. Phys. Commun.} \bibinfo{volume}{81}
  (\bibinfo{year}{1994}) \bibinfo{pages}{381--402}.
  \DOIprefix\doi{10.1016/0010-4655(94)90086-8}.
%Type = Article
\bibitem[{Jung(1995)}]{Jung:1993gf}
\bibinfo{author}{H.~Jung},
\newblock \bibinfo{title}{{Hard diffractive scattering in high-energy e p
  collisions and the Monte Carlo generator RAPGAP}},
\newblock \bibinfo{journal}{Comput. Phys. Commun.} \bibinfo{volume}{86}
  (\bibinfo{year}{1995}) \bibinfo{pages}{147--161}.
  \DOIprefix\doi{10.1016/0010-4655(94)00150-Z}.
%Type = Inproceedings
\bibitem[{Spiesberger et~al.(1992)}]{Spiesberger:237380}
\bibinfo{author}{H.~Spiesberger}, et~al.,
\newblock \bibinfo{title}{{Radiative corrections at HERA}},
\newblock in: \bibinfo{booktitle}{Workshop on Physics at HERA},
  \bibinfo{year}{1992}.
%Type = Article
\bibitem[{Kwiatkowski et~al.(1991)Kwiatkowski, Spiesberger, and
  Mohring}]{Kwiatkowski:1990cx}
\bibinfo{author}{A.~Kwiatkowski}, \bibinfo{author}{H.~Spiesberger},
  \bibinfo{author}{H.~J. Mohring},
\newblock \bibinfo{title}{{Characteristics of radiative events in deep
  inelastic e p scattering at HERA}},
\newblock \bibinfo{journal}{Z. Phys. C} \bibinfo{volume}{50}
  (\bibinfo{year}{1991}) \bibinfo{pages}{165--178}.
  \DOIprefix\doi{10.1007/BF01558572}.
%Type = Article
\bibitem[{Kwiatkowski et~al.(1992)Kwiatkowski, Spiesberger, and
  Mohring}]{Kwiatkowski:1990es}
\bibinfo{author}{A.~Kwiatkowski}, \bibinfo{author}{H.~Spiesberger},
  \bibinfo{author}{H.~J. Mohring},
\newblock \bibinfo{title}{{Heracles: An Event Generator for $e p$ Interactions
  at {HERA} Energies Including Radiative Processes: Version 1.0}},
\newblock \bibinfo{journal}{Comput. Phys. Commun.} \bibinfo{volume}{69}
  (\bibinfo{year}{1992}) \bibinfo{pages}{155--172}.
  \DOIprefix\doi{10.1016/0010-4655(92)90136-M}.
%Type = Article
\bibitem[{Pumplin et~al.(2002)Pumplin, Stump, Huston, Lai, Nadolsky, and
  Tung}]{Pumplin:2002vw}
\bibinfo{author}{J.~Pumplin}, \bibinfo{author}{D.~R. Stump},
  \bibinfo{author}{J.~Huston}, \bibinfo{author}{H.~L. Lai},
  \bibinfo{author}{P.~M. Nadolsky}, \bibinfo{author}{W.~K. Tung},
\newblock \bibinfo{title}{{New generation of parton distributions with
  uncertainties from global QCD analysis}},
\newblock \bibinfo{journal}{JHEP} \bibinfo{volume}{07} (\bibinfo{year}{2002})
  \bibinfo{pages}{012}. \DOIprefix\doi{10.1088/1126-6708/2002/07/012}.
  \href{http://arxiv.org/abs/hep-ph/0201195}{{\tt arXiv:hep-ph/0201195}}.
%Type = Article
\bibitem[{"Andersson et~al.(1983)"Andersson, Gustafson, Ingelman, and
  Sj\"ostrand}]{Andersson:1983ia}
\bibinfo{author}{B.~"Andersson}, \bibinfo{author}{G.~Gustafson},
  \bibinfo{author}{G.~Ingelman}, \bibinfo{author}{T.~Sj\"ostrand},
\newblock \bibinfo{title}{{Parton Fragmentation and String Dynamics}},
\newblock \bibinfo{journal}{Phys. Rept.} \bibinfo{volume}{97}
  (\bibinfo{year}{1983}) \bibinfo{pages}{31--145}.
  \DOIprefix\doi{10.1016/0370-1573(83)90080-7}.
%Type = Article
\bibitem[{Schael et~al.(2005)}]{Schael:2004ux}
\bibinfo{author}{S.~Schael}, et~al. (\bibinfo{collaboration}{ALEPH}),
\newblock \bibinfo{title}{{Bose-Einstein correlations in W-pair decays with an
  event-mixing technique}},
\newblock \bibinfo{journal}{Phys. Lett. B} \bibinfo{volume}{606}
  (\bibinfo{year}{2005}) \bibinfo{pages}{265--275}.
  \DOIprefix\doi{10.1016/j.physletb.2004.12.018}.
%Type = Article
\bibitem[{L\"onnblad(1992)}]{Lonnblad:1992tz}
\bibinfo{author}{L.~L\"onnblad},
\newblock \bibinfo{title}{{ARIADNE version 4: A Program for simulation of QCD
  cascades implementing the color dipole model}},
\newblock \bibinfo{journal}{Comput. Phys. Commun.} \bibinfo{volume}{71}
  (\bibinfo{year}{1992}) \bibinfo{pages}{15--31}.
  \DOIprefix\doi{10.1016/0010-4655(92)90068-A}.
%Type = Article
\bibitem[{Brun et~al.(1987)Brun, Bruyant, Maire, McPherson, and
  Zanarini}]{Brun:1987ma}
\bibinfo{author}{R.~Brun}, \bibinfo{author}{F.~Bruyant},
  \bibinfo{author}{M.~Maire}, \bibinfo{author}{A.~C. McPherson},
  \bibinfo{author}{P.~Zanarini},
\newblock \bibinfo{title}{{GEANT3}}  (\bibinfo{year}{1987}).
%Type = Article
\bibitem[{Sj\"ostrand et~al.(2006)Sj\"ostrand, Mrenna, and
  Skands}]{Sjostrand:2006za}
\bibinfo{author}{T.~Sj\"ostrand}, \bibinfo{author}{S.~Mrenna},
  \bibinfo{author}{P.~Z. Skands},
\newblock \bibinfo{title}{{PYTHIA 6.4 Physics and Manual}},
\newblock \bibinfo{journal}{JHEP} \bibinfo{volume}{05} (\bibinfo{year}{2006})
  \bibinfo{pages}{026}. \DOIprefix\doi{10.1088/1126-6708/2006/05/026}.
  \href{http://arxiv.org/abs/hep-ph/0603175}{{\tt arXiv:hep-ph/0603175}}.
%Type = Article
\bibitem[{Sj\"ostrand et~al.(2015)Sj\"ostrand, Ask, Christiansen, Corke, Desai,
  Ilten, Mrenna, Prestel, Rasmussen, and Skands}]{Sjostrand:2014zea}
\bibinfo{author}{T.~Sj\"ostrand}, \bibinfo{author}{S.~Ask},
  \bibinfo{author}{J.~R. Christiansen}, \bibinfo{author}{R.~Corke},
  \bibinfo{author}{N.~Desai}, \bibinfo{author}{P.~Ilten},
  \bibinfo{author}{S.~Mrenna}, \bibinfo{author}{S.~Prestel},
  \bibinfo{author}{C.~O. Rasmussen}, \bibinfo{author}{P.~Z. Skands},
\newblock \bibinfo{title}{{An introduction to PYTHIA 8.2}},
\newblock \bibinfo{journal}{Comput. Phys. Commun.} \bibinfo{volume}{191}
  (\bibinfo{year}{2015}) \bibinfo{pages}{159--177}.
  \DOIprefix\doi{10.1016/j.cpc.2015.01.024}.
  \href{http://arxiv.org/abs/1410.3012}{{\tt arXiv:1410.3012}}.
%Type = Article
\bibitem[{Giele et~al.(2008)Giele, Kosower, and Skands}]{Giele:2007di}
\bibinfo{author}{W.~T. Giele}, \bibinfo{author}{D.~A. Kosower},
  \bibinfo{author}{P.~Z. Skands},
\newblock \bibinfo{title}{{A simple shower and matching algorithm}},
\newblock \bibinfo{journal}{Phys. Rev. D} \bibinfo{volume}{78}
  (\bibinfo{year}{2008}) \bibinfo{pages}{014026}.
  \DOIprefix\doi{10.1103/PhysRevD.78.014026}.
  \href{http://arxiv.org/abs/0707.3652}{{\tt arXiv:0707.3652}}.
%Type = Article
\bibitem[{Giele et~al.(2013)Giele, Hartgring, Kosower, Laenen, Larkoski,
  Lopez-Villarejo, Ritzmann, and Skands}]{Giele:2013ema}
\bibinfo{author}{W.~T. Giele}, \bibinfo{author}{L.~Hartgring},
  \bibinfo{author}{D.~A. Kosower}, \bibinfo{author}{E.~Laenen},
  \bibinfo{author}{A.~J. Larkoski}, \bibinfo{author}{J.~J. Lopez-Villarejo},
  \bibinfo{author}{M.~Ritzmann}, \bibinfo{author}{P.~Skands},
\newblock \bibinfo{title}{{The VINCIA Parton Shower}},
\newblock \bibinfo{journal}{PoS} \bibinfo{volume}{DIS2013}
  (\bibinfo{year}{2013}) \bibinfo{pages}{165}.
  \DOIprefix\doi{10.22323/1.191.0165}.
  \href{http://arxiv.org/abs/1307.1060}{{\tt arXiv:1307.1060}}.
%Type = Article
\bibitem[{H\"oche and Prestel(2015)}]{Hoche:2015sya}
\bibinfo{author}{S.~H\"oche}, \bibinfo{author}{S.~Prestel},
\newblock \bibinfo{title}{{The midpoint between dipole and parton showers}},
\newblock \bibinfo{journal}{Eur. Phys. J. C} \bibinfo{volume}{75}
  (\bibinfo{year}{2015}) \bibinfo{pages}{461}.
  \DOIprefix\doi{10.1140/epjc/s10052-015-3684-2}.
  \href{http://arxiv.org/abs/1506.05057}{{\tt arXiv:1506.05057}}.
%Type = Article
\bibitem[{Ball et~al.(2017)}]{NNPDF:2017mvq}
\bibinfo{author}{R.~D. Ball}, et~al. (\bibinfo{collaboration}{NNPDF}),
\newblock \bibinfo{title}{{Parton distributions from high-precision collider
  data}},
\newblock \bibinfo{journal}{Eur. Phys. J. C} \bibinfo{volume}{77}
  (\bibinfo{year}{2017}) \bibinfo{pages}{663}.
  \DOIprefix\doi{10.1140/epjc/s10052-017-5199-5}.
  \href{http://arxiv.org/abs/1706.00428}{{\tt arXiv:1706.00428}}.
%Type = Article
\bibitem[{Harland-Lang et~al.(2015)Harland-Lang, Martin, Motylinski, and
  Thorne}]{Harland-Lang:2014zoa}
\bibinfo{author}{L.~A. Harland-Lang}, \bibinfo{author}{A.~D. Martin},
  \bibinfo{author}{P.~Motylinski}, \bibinfo{author}{R.~S. Thorne},
\newblock \bibinfo{title}{{Parton distributions in the LHC era: MMHT 2014
  PDFs}},
\newblock \bibinfo{journal}{Eur. Phys. J. C} \bibinfo{volume}{75}
  (\bibinfo{year}{2015}) \bibinfo{pages}{204}.
  \DOIprefix\doi{10.1140/epjc/s10052-015-3397-6}.
  \href{http://arxiv.org/abs/1412.3989}{{\tt arXiv:1412.3989}}.
%Type = Article
\bibitem[{Bellm et~al.(2016)}]{Bellm:2015jjp}
\bibinfo{author}{J.~Bellm}, et~al.,
\newblock \bibinfo{title}{{Herwig 7.0/Herwig++ 3.0 release note}},
\newblock \bibinfo{journal}{Eur. Phys. J. C} \bibinfo{volume}{76}
  (\bibinfo{year}{2016}) \bibinfo{pages}{196}.
  \DOIprefix\doi{10.1140/epjc/s10052-016-4018-8}.
  \href{http://arxiv.org/abs/1512.01178}{{\tt arXiv:1512.01178}}.
%Type = Article
\bibitem[{Bahr et~al.(2008)}]{Bahr:2008pv}
\bibinfo{author}{M.~Bahr}, et~al.,
\newblock \bibinfo{title}{{Herwig++ Physics and Manual}},
\newblock \bibinfo{journal}{Eur. Phys. J. C} \bibinfo{volume}{58}
  (\bibinfo{year}{2008}) \bibinfo{pages}{639--707}.
  \DOIprefix\doi{10.1140/epjc/s10052-008-0798-9}.
  \href{http://arxiv.org/abs/0803.0883}{{\tt arXiv:0803.0883}}.
%Type = Article
\bibitem[{Bellm et~al.(2020)}]{Bellm:2019zci}
\bibinfo{author}{J.~Bellm}, et~al.,
\newblock \bibinfo{title}{{Herwig 7.2 release note}},
\newblock \bibinfo{journal}{Eur. Phys. J. C} \bibinfo{volume}{80}
  (\bibinfo{year}{2020}) \bibinfo{pages}{452}.
  \DOIprefix\doi{10.1140/epjc/s10052-020-8011-x}.
  \href{http://arxiv.org/abs/1912.06509}{{\tt arXiv:1912.06509}}.
%Type = Article
\bibitem[{L\"onnblad and Prestel(2013)}]{Lonnblad:2012ix}
\bibinfo{author}{L.~L\"onnblad}, \bibinfo{author}{S.~Prestel},
\newblock \bibinfo{title}{{Merging Multi-leg NLO Matrix Elements with Parton
  Showers}},
\newblock \bibinfo{journal}{JHEP} \bibinfo{volume}{03} (\bibinfo{year}{2013})
  \bibinfo{pages}{166}. \DOIprefix\doi{10.1007/JHEP03(2013)166}.
  \href{http://arxiv.org/abs/1211.7278}{{\tt arXiv:1211.7278}}.
%Type = Article
\bibitem[{Bothmann et~al.(2019)}]{Sherpa:2019gpd}
\bibinfo{author}{E.~Bothmann}, et~al. (\bibinfo{collaboration}{Sherpa}),
\newblock \bibinfo{title}{{Event Generation with Sherpa 2.2}},
\newblock \bibinfo{journal}{SciPost Phys.} \bibinfo{volume}{7}
  (\bibinfo{year}{2019}) \bibinfo{pages}{034}.
  \DOIprefix\doi{10.21468/SciPostPhys.7.3.034}.
  \href{http://arxiv.org/abs/1905.09127}{{\tt arXiv:1905.09127}}.
%Type = Article
\bibitem[{Chahal and Krauss(2022)}]{Chahal:2022rid}
\bibinfo{author}{G.~S. Chahal}, \bibinfo{author}{F.~Krauss},
\newblock \bibinfo{title}{{Cluster Hadronisation in Sherpa}},
\newblock \bibinfo{journal}{SciPost Phys.} \bibinfo{volume}{13}
  (\bibinfo{year}{2022}) \bibinfo{pages}{019}.
  \DOIprefix\doi{10.21468/SciPostPhys.13.2.019}.
  \href{http://arxiv.org/abs/2203.11385}{{\tt arXiv:2203.11385}}.
%Type = Article
\bibitem[{Schumann and Krauss(2008)}]{Schumann:2007mg}
\bibinfo{author}{S.~Schumann}, \bibinfo{author}{F.~Krauss},
\newblock \bibinfo{title}{{A Parton shower algorithm based on Catani-Seymour
  dipole factorisation}},
\newblock \bibinfo{journal}{JHEP} \bibinfo{volume}{03} (\bibinfo{year}{2008})
  \bibinfo{pages}{038}. \DOIprefix\doi{10.1088/1126-6708/2008/03/038}.
  \href{http://arxiv.org/abs/0709.1027}{{\tt arXiv:0709.1027}}.
%Type = Article
\bibitem[{"H\"oche et~al.(2009)"H\"oche, Krauss, Schumann, and
  Siegert}]{Hoeche:2009rj}
\bibinfo{author}{S.~"H\"oche}, \bibinfo{author}{F.~Krauss},
  \bibinfo{author}{S.~Schumann}, \bibinfo{author}{F.~Siegert},
\newblock \bibinfo{title}{{QCD matrix elements and truncated showers}},
\newblock \bibinfo{journal}{JHEP} \bibinfo{volume}{05} (\bibinfo{year}{2009})
  \bibinfo{pages}{053}. \DOIprefix\doi{10.1088/1126-6708/2009/05/053}.
  \href{http://arxiv.org/abs/0903.1219}{{\tt arXiv:0903.1219}}.
%Type = Article
\bibitem[{"H\"oche et~al.(2013)"H\"oche, Krauss, Sch\"onherr, and
  Siegert}]{Hoeche:2012yf}
\bibinfo{author}{S.~"H\"oche}, \bibinfo{author}{F.~Krauss},
  \bibinfo{author}{M.~Sch\"onherr}, \bibinfo{author}{F.~Siegert},
\newblock \bibinfo{title}{{QCD matrix elements + parton showers: The NLO
  case}},
\newblock \bibinfo{journal}{JHEP} \bibinfo{volume}{04} (\bibinfo{year}{2013})
  \bibinfo{pages}{027}. \DOIprefix\doi{10.1007/JHEP04(2013)027}.
  \href{http://arxiv.org/abs/1207.5030}{{\tt arXiv:1207.5030}}.
%Type = Article
\bibitem[{Andreassen et~al.(2020)Andreassen, Komiske, Metodiev, Nachman, and
  Thaler}]{Andreassen:2019cjw}
\bibinfo{author}{A.~Andreassen}, \bibinfo{author}{P.~T. Komiske},
  \bibinfo{author}{E.~M. Metodiev}, \bibinfo{author}{B.~Nachman},
  \bibinfo{author}{J.~Thaler},
\newblock \bibinfo{title}{{OmniFold: A Method to Simultaneously Unfold All
  Observables}},
\newblock \bibinfo{journal}{Phys. Rev. Lett.} \bibinfo{volume}{124}
  (\bibinfo{year}{2020}) \bibinfo{pages}{182001}.
  \DOIprefix\doi{10.1103/PhysRevLett.124.182001}.
  \href{http://arxiv.org/abs/1911.09107}{{\tt arXiv:1911.09107}}.
%Type = Inproceedings
\bibitem[{Andreassen et~al.(2021)Andreassen, Komiske, Metodiev, Nachman,
  Suresh, and Thaler}]{omnifoldiclr}
\bibinfo{author}{A.~Andreassen}, \bibinfo{author}{P.~T. Komiske},
  \bibinfo{author}{E.~M. Metodiev}, \bibinfo{author}{B.~Nachman},
  \bibinfo{author}{A.~Suresh}, \bibinfo{author}{J.~Thaler},
\newblock \bibinfo{title}{{Scaffolding Simulations with Deep Learning for
  High-dimensional Deconvolution}},
\newblock in: \bibinfo{booktitle}{9th International Conference on Learning
  Representations}, \bibinfo{year}{2021}.
  \href{http://arxiv.org/abs/2105.04448}{{\tt arXiv:2105.04448}}.
%Type = Article
\bibitem[{Lucy(1974)}]{1974AJ79745L}
\bibinfo{author}{L.~B. Lucy},
\newblock \bibinfo{title}{{An iterative technique for the rectification of
  observed distributions}},
\newblock \bibinfo{journal}{Astron. J.} \bibinfo{volume}{79}
  (\bibinfo{year}{1974}) \bibinfo{pages}{745--754}.
  \DOIprefix\doi{10.1086/111605}.
%Type = Article
\bibitem[{{Richardson}(1972)}]{Richardson:72}
\bibinfo{author}{W.~H. {Richardson}},
\newblock \bibinfo{title}{{Bayesian-Based Iterative Method of Image
  Restoration}},
\newblock \bibinfo{journal}{Journal of the Optical Society of America
  (1917-1983)} \bibinfo{volume}{62} (\bibinfo{year}{1972}) \bibinfo{pages}{55}.
%Type = Article
\bibitem[{D'Agostini(1995)}]{DAgostini:1994fjx}
\bibinfo{author}{G.~D'Agostini},
\newblock \bibinfo{title}{{A Multidimensional unfolding method based on Bayes'
  theorem}},
\newblock \bibinfo{journal}{Nucl. Instrum. Meth. A} \bibinfo{volume}{362}
  (\bibinfo{year}{1995}) \bibinfo{pages}{487--498}.
  \DOIprefix\doi{10.1016/0168-9002(95)00274-X}.
%Type = Article
\bibitem[{Komiske et~al.(2019)Komiske, Metodiev, and Thaler}]{Komiske:2018cqr}
\bibinfo{author}{P.~T. Komiske}, \bibinfo{author}{E.~M. Metodiev},
  \bibinfo{author}{J.~Thaler},
\newblock \bibinfo{title}{{Energy Flow Networks: Deep Sets for Particle Jets}},
\newblock \bibinfo{journal}{JHEP} \bibinfo{volume}{01} (\bibinfo{year}{2019})
  \bibinfo{pages}{121}. \DOIprefix\doi{10.1007/JHEP01(2019)121}.
  \href{http://arxiv.org/abs/1810.05165}{{\tt arXiv:1810.05165}}.
%Type = Article
\bibitem[{Qu and Gouskos(2020)}]{Qu:2019gqs}
\bibinfo{author}{H.~Qu}, \bibinfo{author}{L.~Gouskos},
\newblock \bibinfo{title}{{ParticleNet: Jet Tagging via Particle Clouds}},
\newblock \bibinfo{journal}{Phys. Rev. D} \bibinfo{volume}{101}
  (\bibinfo{year}{2020}) \bibinfo{pages}{056019}.
  \DOIprefix\doi{10.1103/PhysRevD.101.056019}.
  \href{http://arxiv.org/abs/1902.08570}{{\tt arXiv:1902.08570}}.
%Type = Article
\bibitem[{Moreno et~al.(2020{\natexlab{a}})Moreno, Cerri, Duarte, Newman,
  Nguyen, Periwal, Pierini, Serikova, Spiropulu, and Vlimant}]{Moreno:2019bmu}
\bibinfo{author}{E.~A. Moreno}, \bibinfo{author}{O.~Cerri},
  \bibinfo{author}{J.~M. Duarte}, \bibinfo{author}{H.~B. Newman},
  \bibinfo{author}{T.~Q. Nguyen}, \bibinfo{author}{A.~Periwal},
  \bibinfo{author}{M.~Pierini}, \bibinfo{author}{A.~Serikova},
  \bibinfo{author}{M.~Spiropulu}, \bibinfo{author}{J.-R. Vlimant},
\newblock \bibinfo{title}{{JEDI-net: a jet identification algorithm based on
  interaction networks}},
\newblock \bibinfo{journal}{Eur. Phys. J. C} \bibinfo{volume}{80}
  (\bibinfo{year}{2020}{\natexlab{a}}) \bibinfo{pages}{58}.
  \DOIprefix\doi{10.1140/epjc/s10052-020-7608-4}.
  \href{http://arxiv.org/abs/1908.05318}{{\tt arXiv:1908.05318}}.
%Type = Article
\bibitem[{Moreno et~al.(2020{\natexlab{b}})Moreno, Nguyen, Vlimant, Cerri,
  Newman, Periwal, Spiropulu, Duarte, and Pierini}]{Moreno:2019neq}
\bibinfo{author}{E.~A. Moreno}, \bibinfo{author}{T.~Q. Nguyen},
  \bibinfo{author}{J.-R. Vlimant}, \bibinfo{author}{O.~Cerri},
  \bibinfo{author}{H.~B. Newman}, \bibinfo{author}{A.~Periwal},
  \bibinfo{author}{M.~Spiropulu}, \bibinfo{author}{J.~M. Duarte},
  \bibinfo{author}{M.~Pierini},
\newblock \bibinfo{title}{{Interaction networks for the identification of
  boosted $H \rightarrow b\overline{b}$ decays}},
\newblock \bibinfo{journal}{Phys. Rev. D} \bibinfo{volume}{102}
  (\bibinfo{year}{2020}{\natexlab{b}}) \bibinfo{pages}{012010}.
  \DOIprefix\doi{10.1103/PhysRevD.102.012010}.
  \href{http://arxiv.org/abs/1909.12285}{{\tt arXiv:1909.12285}}.
%Type = Article
\bibitem[{Mikuni and Canelli(2020)}]{Mikuni:2020wpr}
\bibinfo{author}{V.~Mikuni}, \bibinfo{author}{F.~Canelli},
\newblock \bibinfo{title}{{ABCNet: An attention-based method for particle
  tagging}},
\newblock \bibinfo{journal}{Eur. Phys. J. Plus} \bibinfo{volume}{135}
  (\bibinfo{year}{2020}) \bibinfo{pages}{463}.
  \DOIprefix\doi{10.1140/epjp/s13360-020-00497-3}.
  \href{http://arxiv.org/abs/2001.05311}{{\tt arXiv:2001.05311}}.
%Type = Article
\bibitem[{Bernreuther et~al.(2021)Bernreuther, Finke, Kahlhoefer, Kr\"amer, and
  M\"uck}]{Bernreuther:2020vhm}
\bibinfo{author}{E.~Bernreuther}, \bibinfo{author}{T.~Finke},
  \bibinfo{author}{F.~Kahlhoefer}, \bibinfo{author}{M.~Kr\"amer},
  \bibinfo{author}{A.~M\"uck},
\newblock \bibinfo{title}{{Casting a graph net to catch dark showers}},
\newblock \bibinfo{journal}{SciPost Phys.} \bibinfo{volume}{10}
  (\bibinfo{year}{2021}) \bibinfo{pages}{046}.
  \DOIprefix\doi{10.21468/SciPostPhys.10.2.046}.
  \href{http://arxiv.org/abs/2006.08639}{{\tt arXiv:2006.08639}}.
%Type = Article
\bibitem[{Guo et~al.(2021)Guo, Li, Li, and Zhang}]{Guo:2020vvt}
\bibinfo{author}{J.~Guo}, \bibinfo{author}{J.~Li}, \bibinfo{author}{T.~Li},
  \bibinfo{author}{R.~Zhang},
\newblock \bibinfo{title}{{Boosted Higgs boson jet reconstruction via a graph
  neural network}},
\newblock \bibinfo{journal}{Phys. Rev. D} \bibinfo{volume}{103}
  (\bibinfo{year}{2021}) \bibinfo{pages}{116025}.
  \DOIprefix\doi{10.1103/PhysRevD.103.116025}.
  \href{http://arxiv.org/abs/2010.05464}{{\tt arXiv:2010.05464}}.
%Type = Article
\bibitem[{Dolan and Ore(2021)}]{Dolan:2020qkr}
\bibinfo{author}{M.~J. Dolan}, \bibinfo{author}{A.~Ore},
\newblock \bibinfo{title}{{Equivariant Energy Flow Networks for Jet Tagging}},
\newblock \bibinfo{journal}{Phys. Rev. D} \bibinfo{volume}{103}
  (\bibinfo{year}{2021}) \bibinfo{pages}{074022}.
  \DOIprefix\doi{10.1103/PhysRevD.103.074022}.
  \href{http://arxiv.org/abs/2012.00964}{{\tt arXiv:2012.00964}}.
%Type = Article
\bibitem[{Mikuni and Canelli(2021)}]{Mikuni:2021pou}
\bibinfo{author}{V.~Mikuni}, \bibinfo{author}{F.~Canelli},
\newblock \bibinfo{title}{{Point cloud transformers applied to collider
  physics}},
\newblock \bibinfo{journal}{Mach. Learn. Sci. Tech.} \bibinfo{volume}{2}
  (\bibinfo{year}{2021}) \bibinfo{pages}{035027}.
  \DOIprefix\doi{10.1088/2632-2153/ac07f6}.
  \href{http://arxiv.org/abs/2102.05073}{{\tt arXiv:2102.05073}}.
%Type = Article
\bibitem[{Konar et~al.(2021)Konar, Ngairangbam, and Spannowsky}]{Konar:2021zdg}
\bibinfo{author}{P.~Konar}, \bibinfo{author}{V.~S. Ngairangbam},
  \bibinfo{author}{M.~Spannowsky},
\newblock \bibinfo{title}{{Energy-weighted Message Passing: an infra-red and
  collinear safe graph neural network algorithm}}  (\bibinfo{year}{2021}).
  \href{http://arxiv.org/abs/2109.14636}{{\tt arXiv:2109.14636}}.
%Type = Article
\bibitem[{Shimmin(2021)}]{Shimmin:2021pkm}
\bibinfo{author}{C.~Shimmin},
\newblock \bibinfo{title}{{Particle Convolution for High Energy Physics}}
  (\bibinfo{year}{2021}). \href{http://arxiv.org/abs/2107.02908}{{\tt
  arXiv:2107.02908}}.
%Type = Article
\bibitem[{Gong et~al.(2022)Gong, Meng, Zhang, Qu, Li, Qian, Du, Ma, and
  Liu}]{Gong:2022lye}
\bibinfo{author}{S.~Gong}, \bibinfo{author}{Q.~Meng},
  \bibinfo{author}{J.~Zhang}, \bibinfo{author}{H.~Qu}, \bibinfo{author}{C.~Li},
  \bibinfo{author}{S.~Qian}, \bibinfo{author}{W.~Du}, \bibinfo{author}{Z.-M.
  Ma}, \bibinfo{author}{T.-Y. Liu},
\newblock \bibinfo{title}{{An Efficient Lorentz Equivariant Graph Neural
  Network for Jet Tagging}}  (\bibinfo{year}{2022}).
  \href{http://arxiv.org/abs/2201.08187}{{\tt arXiv:2201.08187}}.
%Type = Article
\bibitem[{Guo et~al.(2021)Guo, Cai, Liu, Mu, Martin, and Hu}]{guo2021pct}
\bibinfo{author}{M.-H. Guo}, \bibinfo{author}{J.-X. Cai},
  \bibinfo{author}{Z.-N. Liu}, \bibinfo{author}{T.-J. Mu},
  \bibinfo{author}{R.~R. Martin}, \bibinfo{author}{S.-M. Hu},
\newblock \bibinfo{title}{Pct: Point cloud transformer},
\newblock \bibinfo{journal}{Computational Visual Media} \bibinfo{volume}{7}
  (\bibinfo{year}{2021}) \bibinfo{pages}{187--199}.
%Type = Misc
\bibitem[{Chollet(2017)}]{keras}
\bibinfo{author}{F.~Chollet}, \bibinfo{title}{Keras},
  \bibinfo{howpublished}{\url{https://github.com/fchollet/keras}},
  \bibinfo{year}{2017}.
%Type = Inproceedings
\bibitem[{Abadi et~al.(2016)Abadi, Barham, Chen, Chen, Davis, Dean, Devin,
  Ghemawat, Irving, Isard et~al.}]{tensorflow}
\bibinfo{author}{M.~Abadi}, \bibinfo{author}{P.~Barham},
  \bibinfo{author}{J.~Chen}, \bibinfo{author}{Z.~Chen},
  \bibinfo{author}{A.~Davis}, \bibinfo{author}{J.~Dean},
  \bibinfo{author}{M.~Devin}, \bibinfo{author}{S.~Ghemawat},
  \bibinfo{author}{G.~Irving}, \bibinfo{author}{M.~Isard}, et~al.,
\newblock \bibinfo{title}{Tensorflow: A system for large-scale machine
  learning.},
\newblock in: \bibinfo{booktitle}{OSDI}, volume~\bibinfo{volume}{16},
  \bibinfo{year}{2016}, pp. \bibinfo{pages}{265--283}.
%Type = Article
\bibitem[{Kingma and Ba(2014)}]{adam}
\bibinfo{author}{D.~Kingma}, \bibinfo{author}{J.~Ba},
\newblock \bibinfo{title}{Adam: A method for stochastic optimization}
  (\bibinfo{year}{2014}). \href{http://arxiv.org/abs/1412.6980}{{\tt
  arXiv:1412.6980}}.
%Type = Misc
\bibitem[{Per(2022)}]{Perlmutter}
\bibinfo{title}{{Perlmutter} system},
  \bibinfo{howpublished}{\url{https://docs.nersc.gov/systems/perlmutter}},
  \bibinfo{year}{2022}. \bibinfo{note}{Accessed: 2022-05-04}.
%Type = Article
\bibitem[{Sergeev and Balso(2018)}]{sergeev2018horovod}
\bibinfo{author}{A.~Sergeev}, \bibinfo{author}{M.~D. Balso},
\newblock \bibinfo{title}{Horovod: fast and easy distributed deep learning in
  {TensorFlow}},
\newblock \bibinfo{journal}{arXiv preprint arXiv:1802.05799}
  (\bibinfo{year}{2018}).
%Type = Misc
\bibitem[{Kogler(2011)}]{etde_21406988}
\bibinfo{author}{R.~Kogler}, \bibinfo{title}{Measurement of jet production in
  deep-inelastic ep scattering at hera},
  \bibinfo{howpublished}{\url{https://www-h1.desy.de/psfiles/theses/h1th-590.pdf}},
  \bibinfo{year}{2011}.
%Type = Article
\bibitem[{Aaron et~al.(2011)}]{H1:2011unn}
\bibinfo{author}{F.~D. Aaron}, et~al. (\bibinfo{collaboration}{H1}),
\newblock \bibinfo{title}{{Measurement of $D^{*\pm}$ Meson Production and
  Determination of $F_2^{c\bar c}$ at low $Q^2$ in Deep-Inelastic Scattering at
  HERA}},
\newblock \bibinfo{journal}{Eur. Phys. J. C} \bibinfo{volume}{71}
  (\bibinfo{year}{2011}) \bibinfo{pages}{1769}.
  \DOIprefix\doi{10.1140/epjc/s10052-011-1769-0}.
  \href{http://arxiv.org/abs/1106.1028}{{\tt arXiv:1106.1028}},
  \bibinfo{note}{[Erratum: Eur.Phys.J.C 72, 2252 (2012)]}.
%Type = Article
\bibitem[{Efron(1979)}]{10.1214/aos/1176344552}
\bibinfo{author}{B.~Efron},
\newblock \bibinfo{title}{{Bootstrap Methods: Another Look at the Jackknife}},
\newblock \bibinfo{journal}{Annals Statist.} \bibinfo{volume}{7}
  (\bibinfo{year}{1979}) \bibinfo{pages}{1--26}.
  \DOIprefix\doi{10.1214/aos/1176344552}.

\end{thebibliography}
\end{flushleft}

\appendix
\section{Data and simulation comparison at reconstruction level}\label{app:data_mc}
\begin{figure*}[htb]
    \centering
    \includegraphics[width=0.32\textwidth]{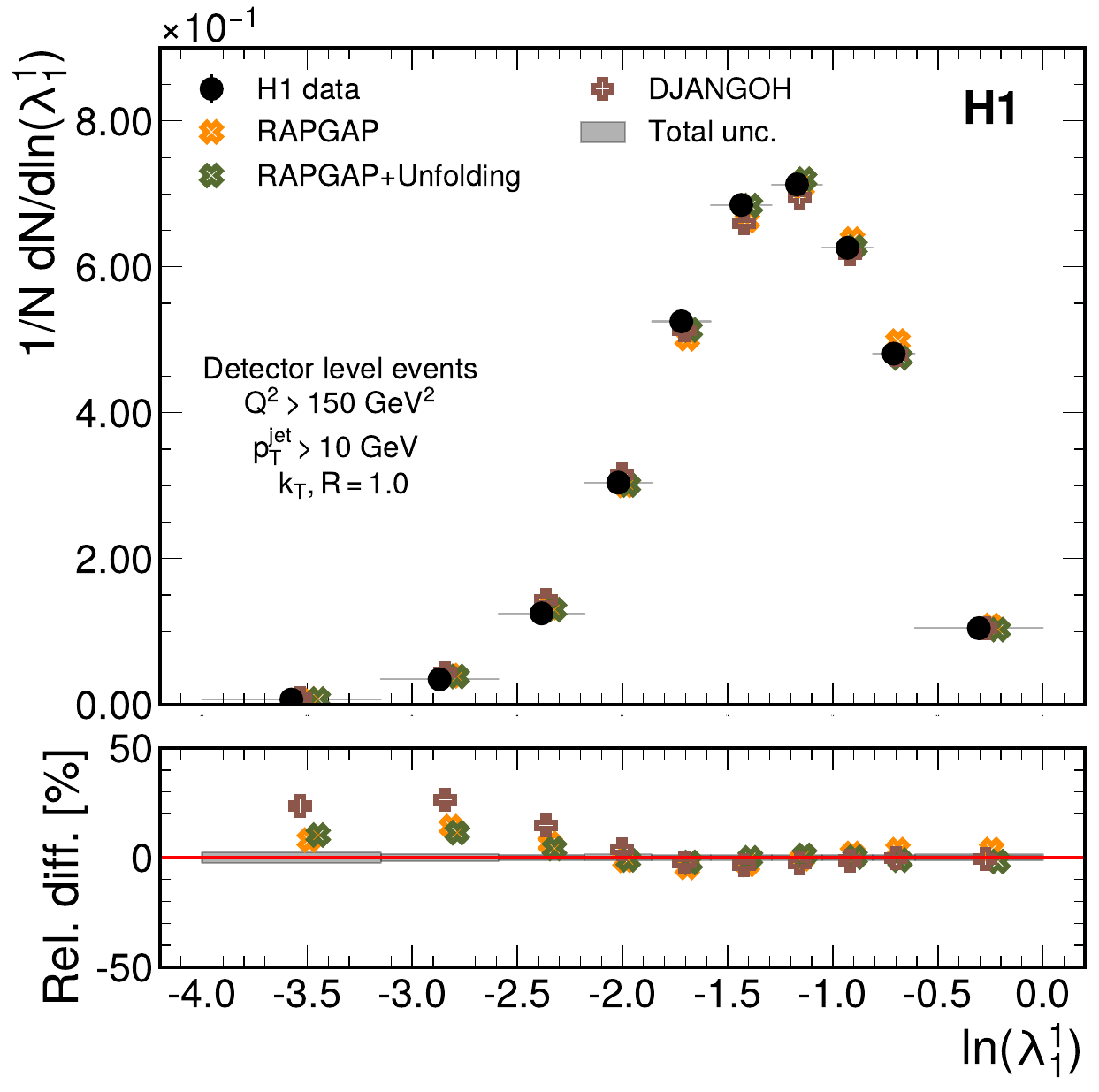}
  \includegraphics[width=0.32\textwidth]{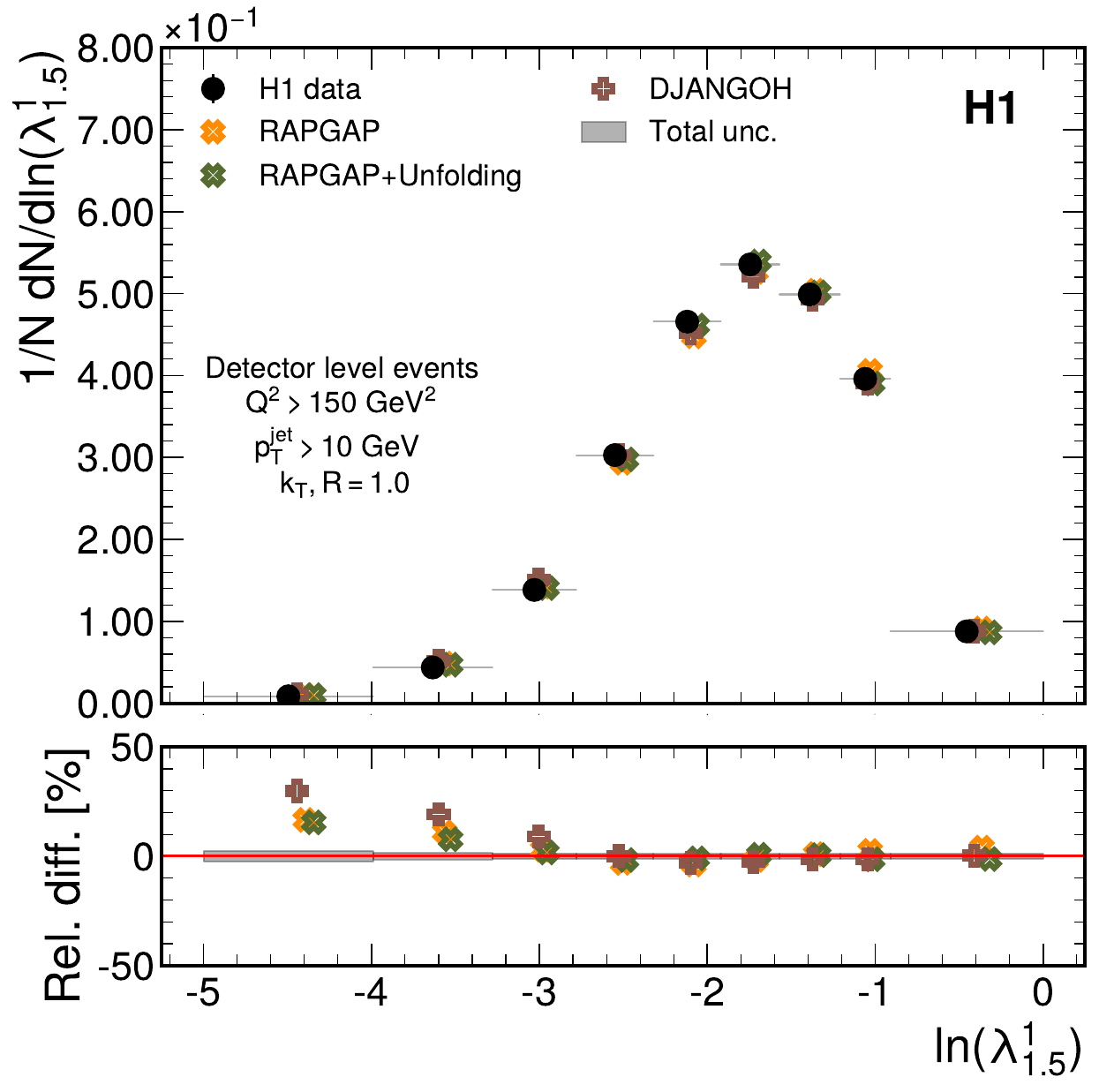}    
    \includegraphics[width=0.32\textwidth]{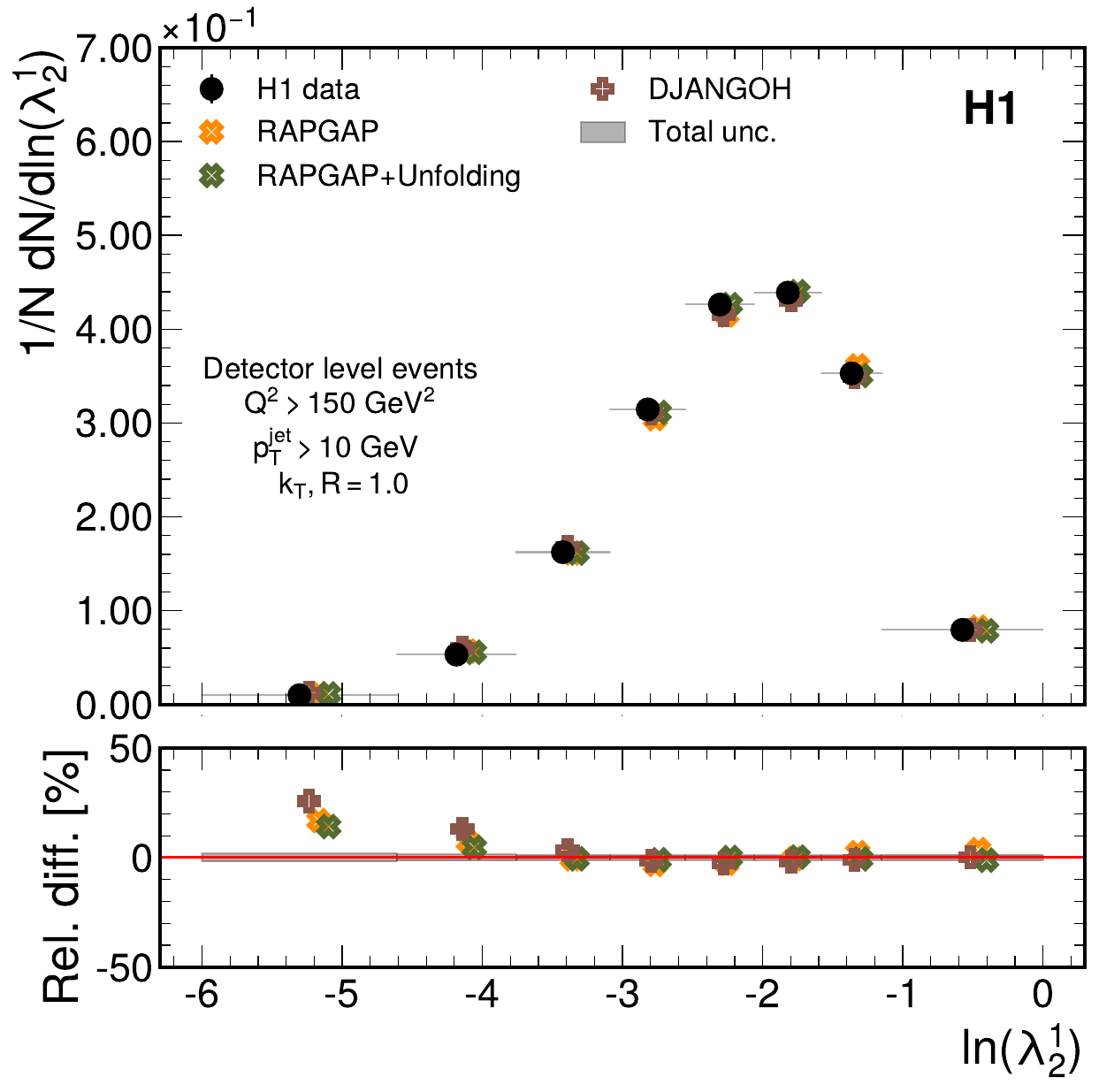} 
    \includegraphics[width=0.32\textwidth]{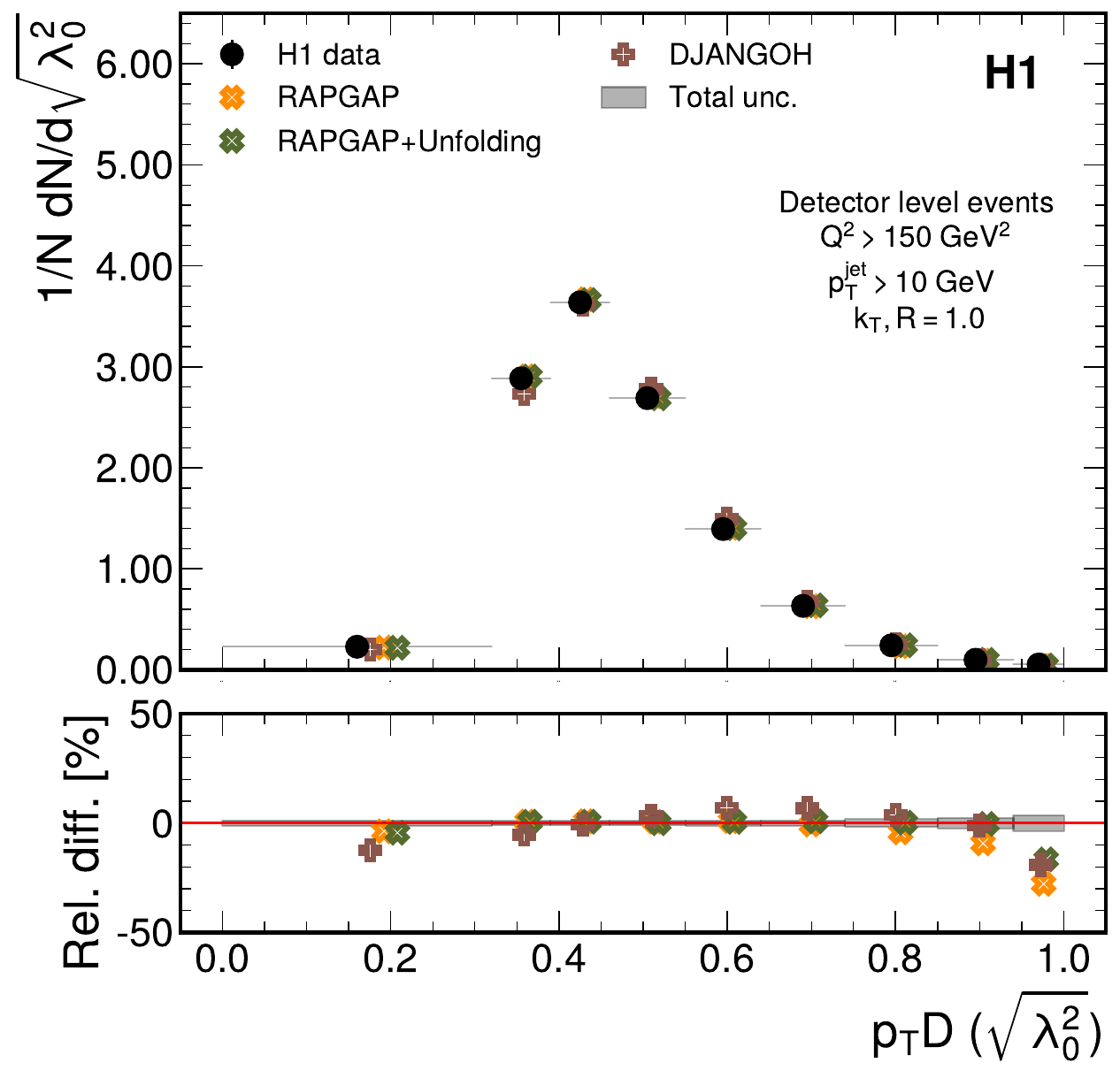}   
    \includegraphics[width=0.32\textwidth]{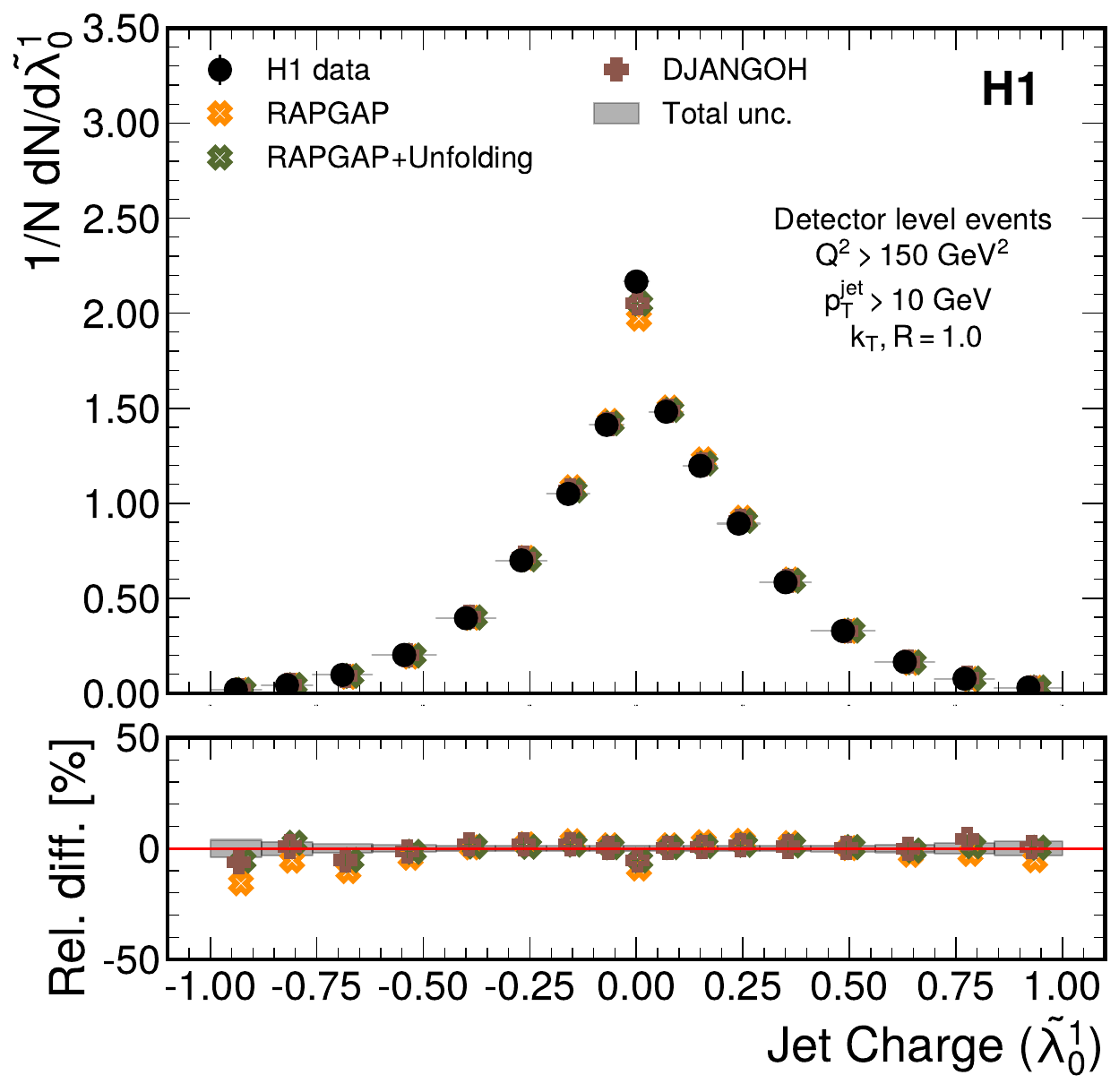}
    \includegraphics[width=0.32\textwidth]{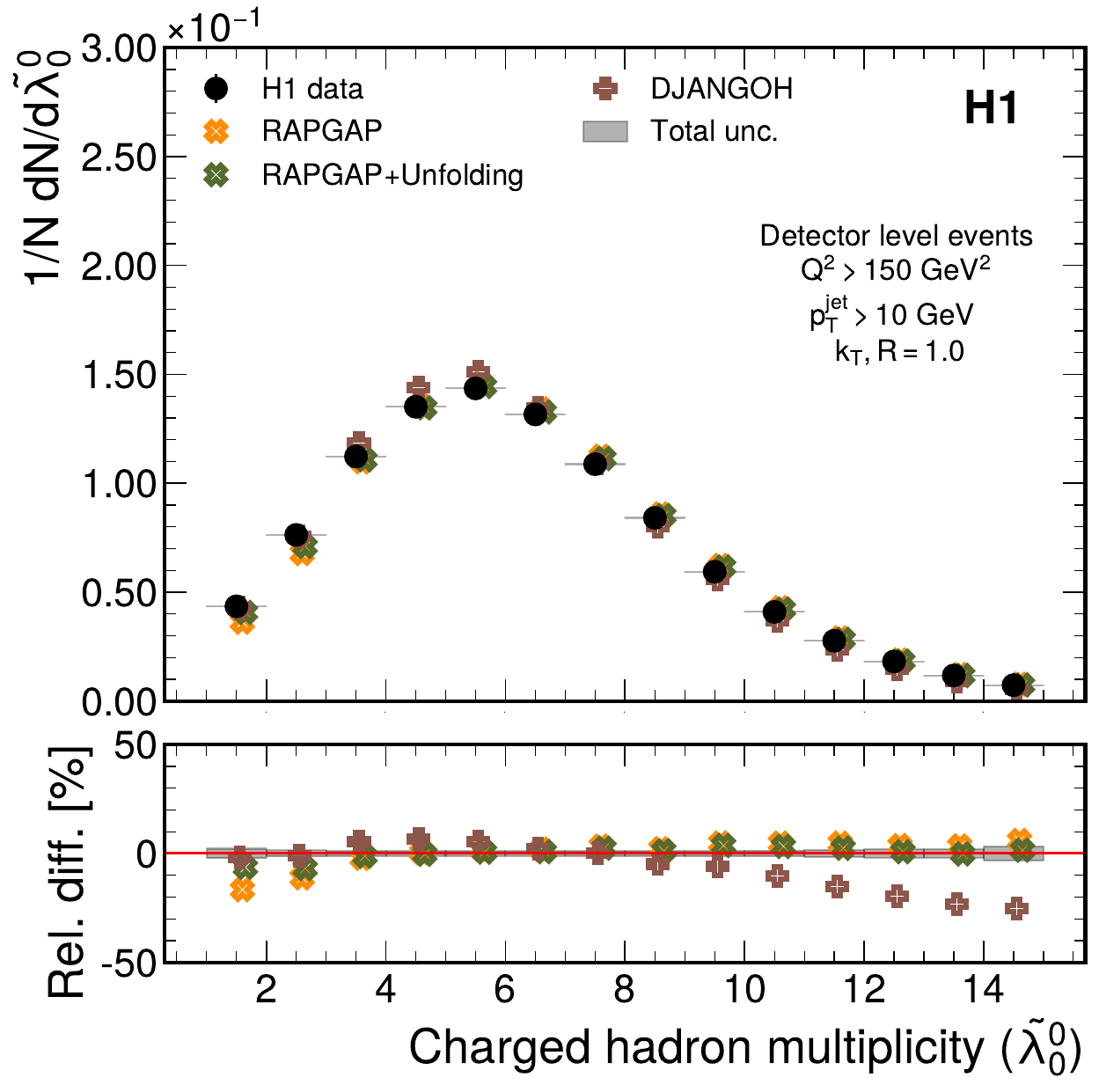}

    \caption{Measured cross sections at reconstruction level, normalized to the inclusive jet production cross section, as a function of the jet angularities. Data are shown as solid dots, horizontal bars indicate the bin ranges. Predictions from \textsc{Djangoh} and \textsc{Rapgap} simulations are shown for comparison, as well as \textsc{Rapgap} predictions with unfolding weights. The predictions are offset horizontally for visual clarity. The relative differences between data and predictions are shown in the bottom panels. Gray bands represent the total systematic uncertainties from detector effects.
      Data statistical uncertainties are shown as vertical bars, while the statistical uncertainties on the simulations are smaller than the marker sizes.}
    \label{fig:dataMC_inclusive}
\end{figure*}

Prior to unfolding, a comparison between the data and simulation is performed using the selection criteria described in Sec.~\ref{sec:h1} and the angularities calculated from the reconstructed detector objects. Both \textsc{Rapgap} and \textsc{Djangoh} simulations are compared to data in Fig.~\ref{fig:dataMC_inclusive}. The \textsc{Rapgap} simulation with unfolding weights applied is also shown.

\section{Binned distribution of unfolded observables at different energy scales}\label{app:q2_int}

\begin{figure*}[htb]
    \centering
    
    \includegraphics[width=0.45\textwidth]{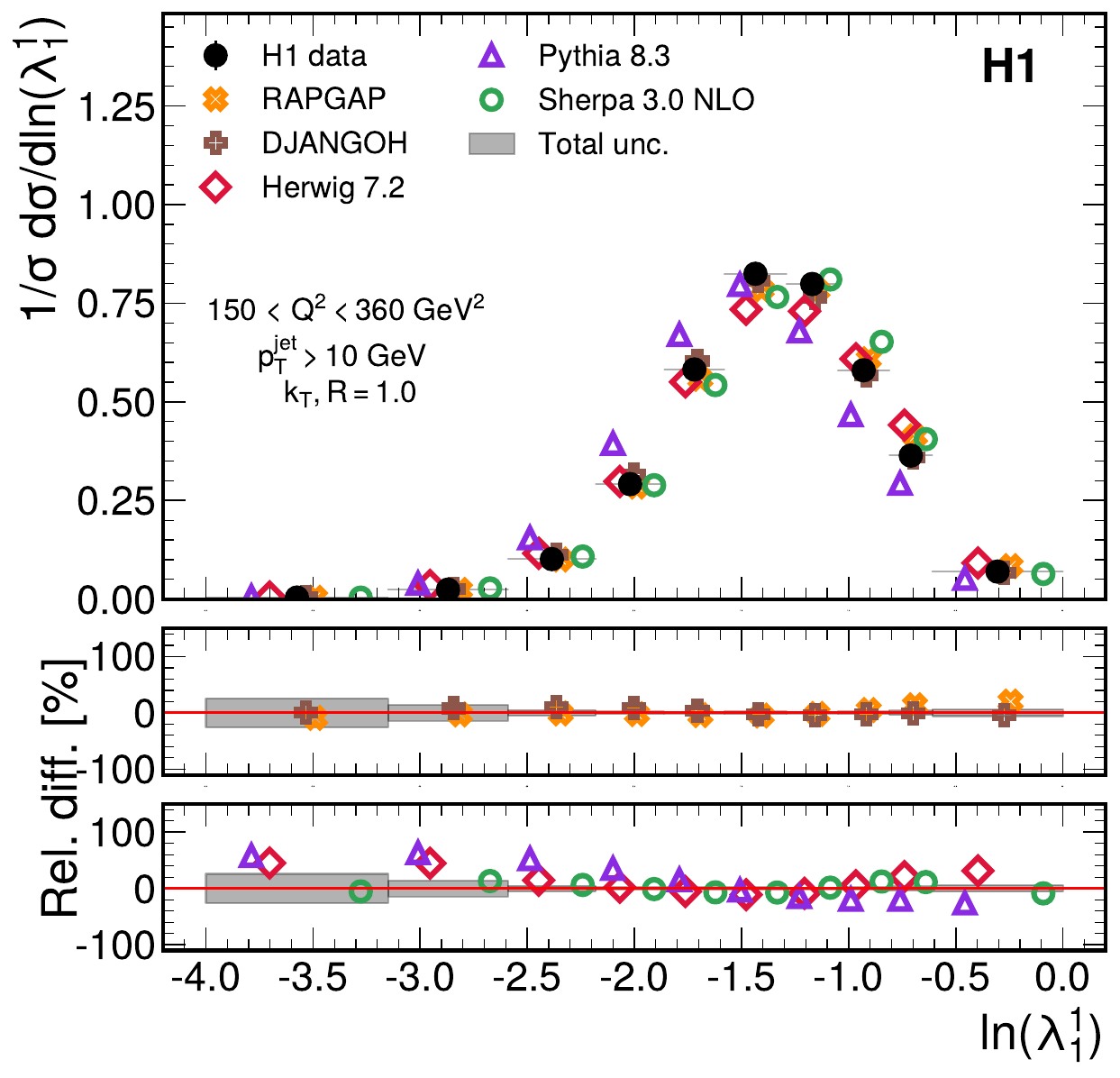}
\includegraphics[width=0.45\textwidth]{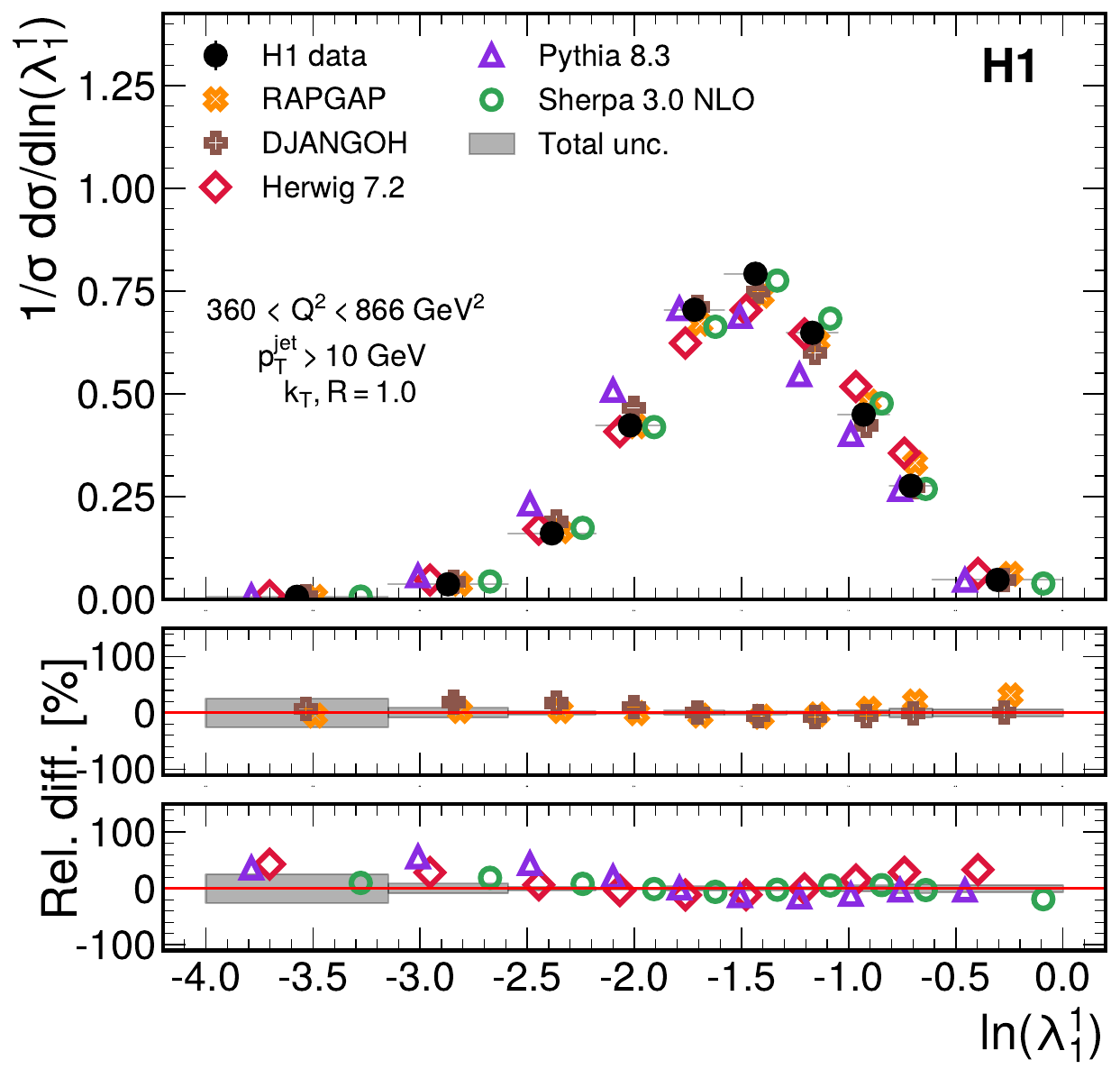}
    \includegraphics[width=0.45\textwidth]{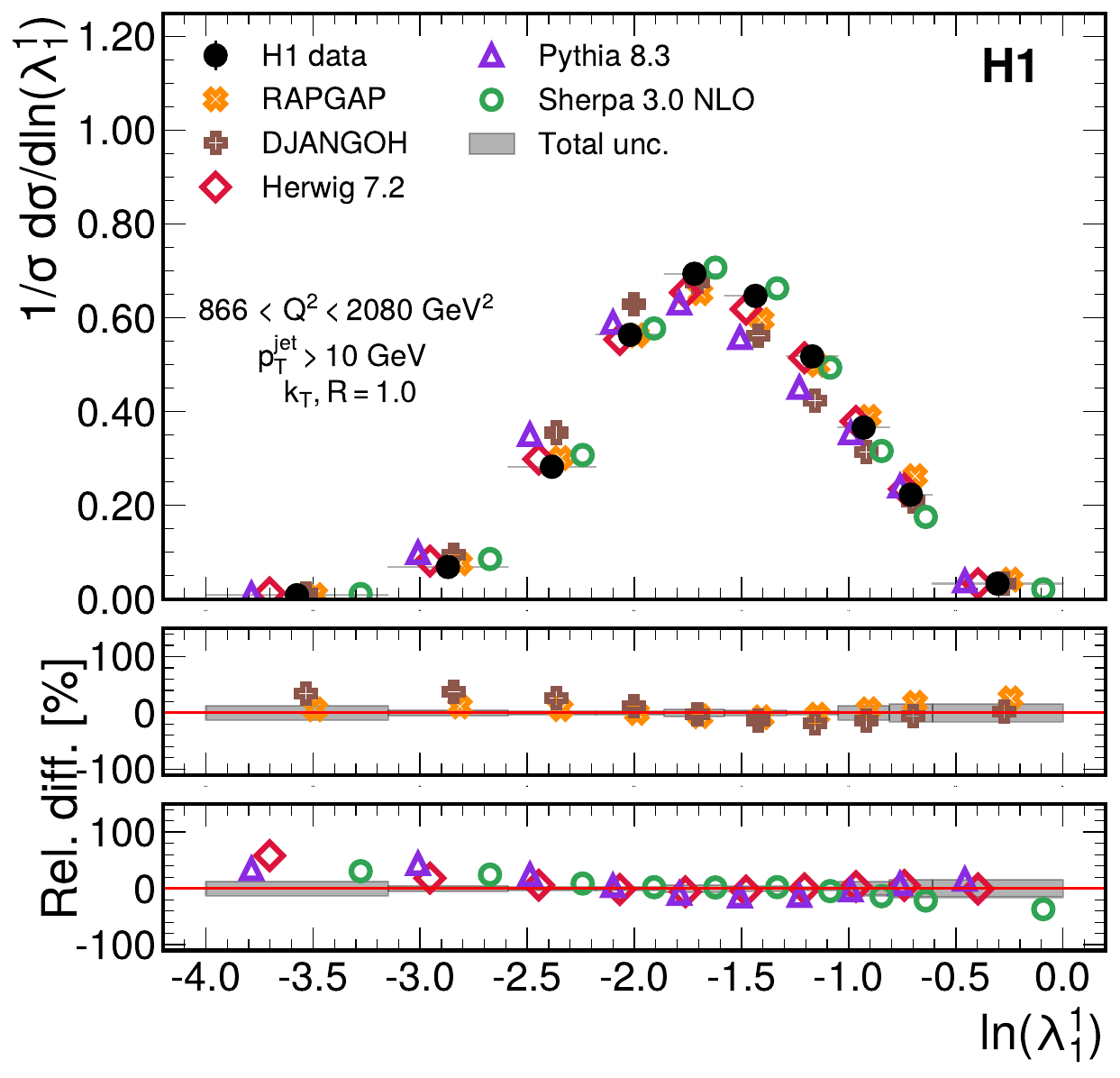}
    \includegraphics[width=0.45\textwidth]{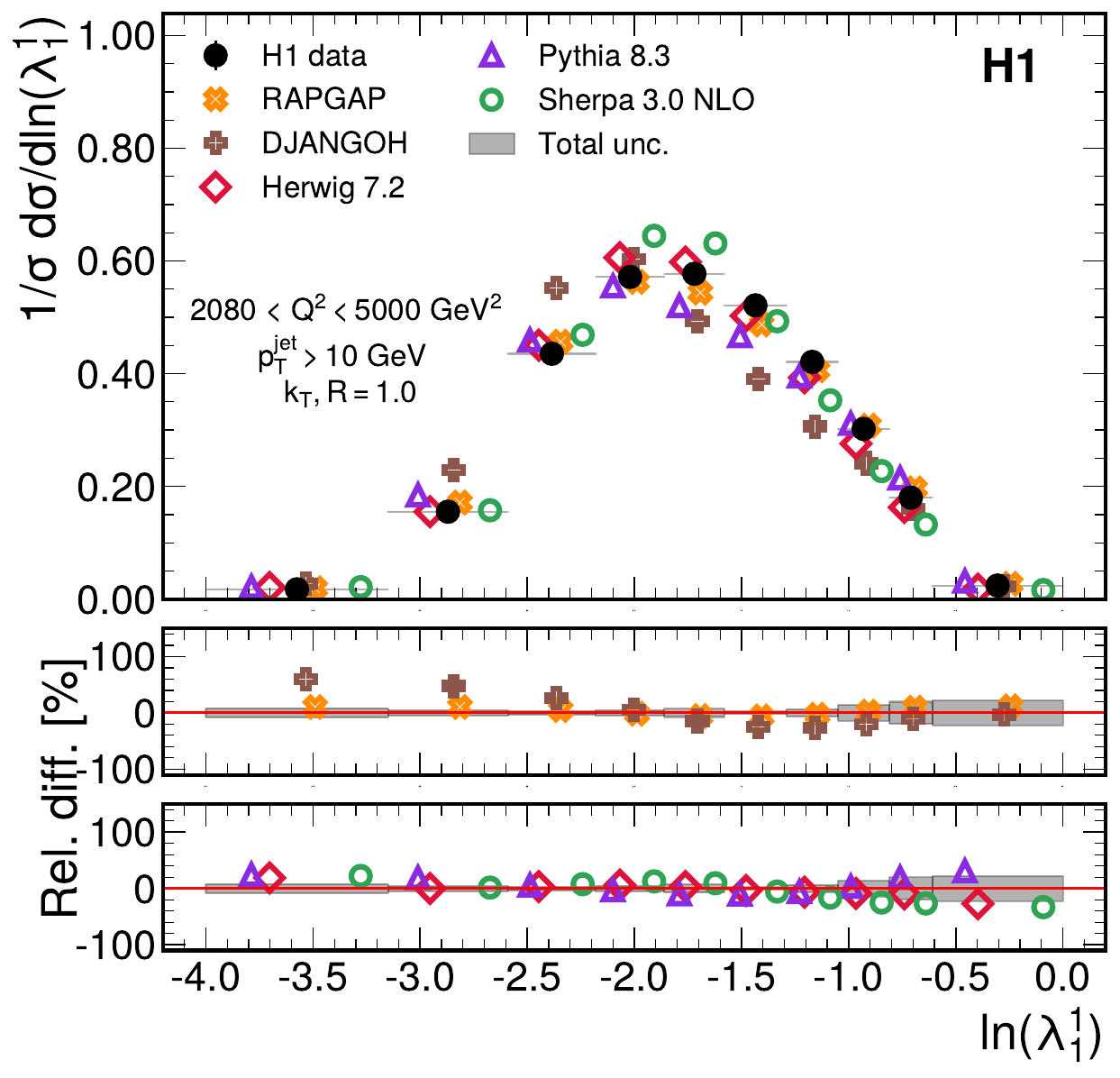}
    
    \caption{Measured cross sections, normalized to the inclusive jet production cross section, for multiple \Qs{} intervals as a function of $\ln(\lambda_1^1)$. Data are shown as solid dots, horizontal bars indicate the bin ranges. Predictions from multiple simulations are shown for comparison, and are offset horizontally for visual clarity. The relative differences between data and predictions are shown in the bottom panels, split between dedicated DIS simulators (middle) and general purpose simulators (bottom). Gray bands represent the total data uncertainties.
        }
    \label{fig:rapgap_data_sys_q2_4}
\end{figure*}

\begin{figure*}[htb]
    \centering
    \includegraphics[width=0.45\textwidth]{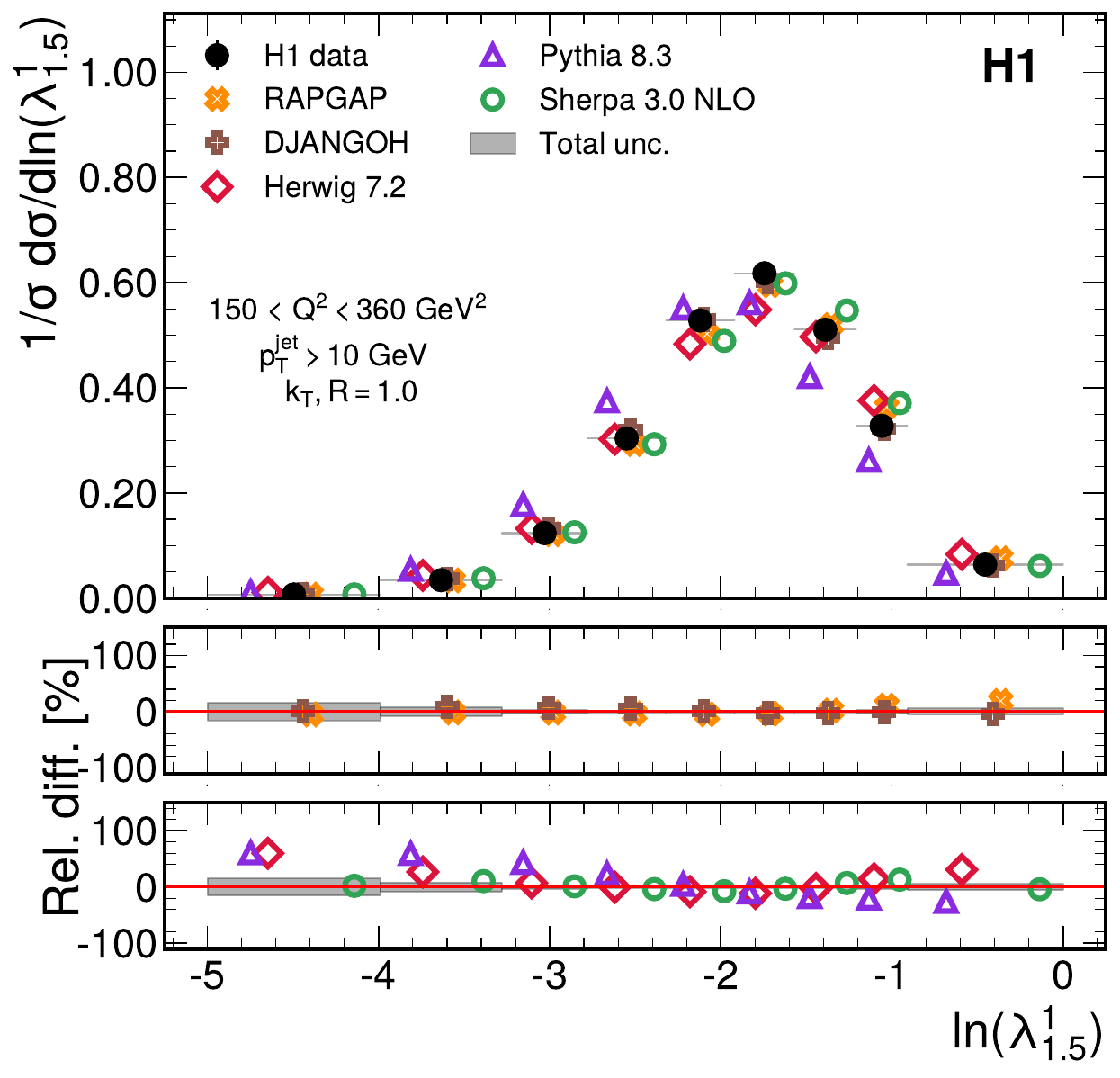}
\includegraphics[width=0.45\textwidth]{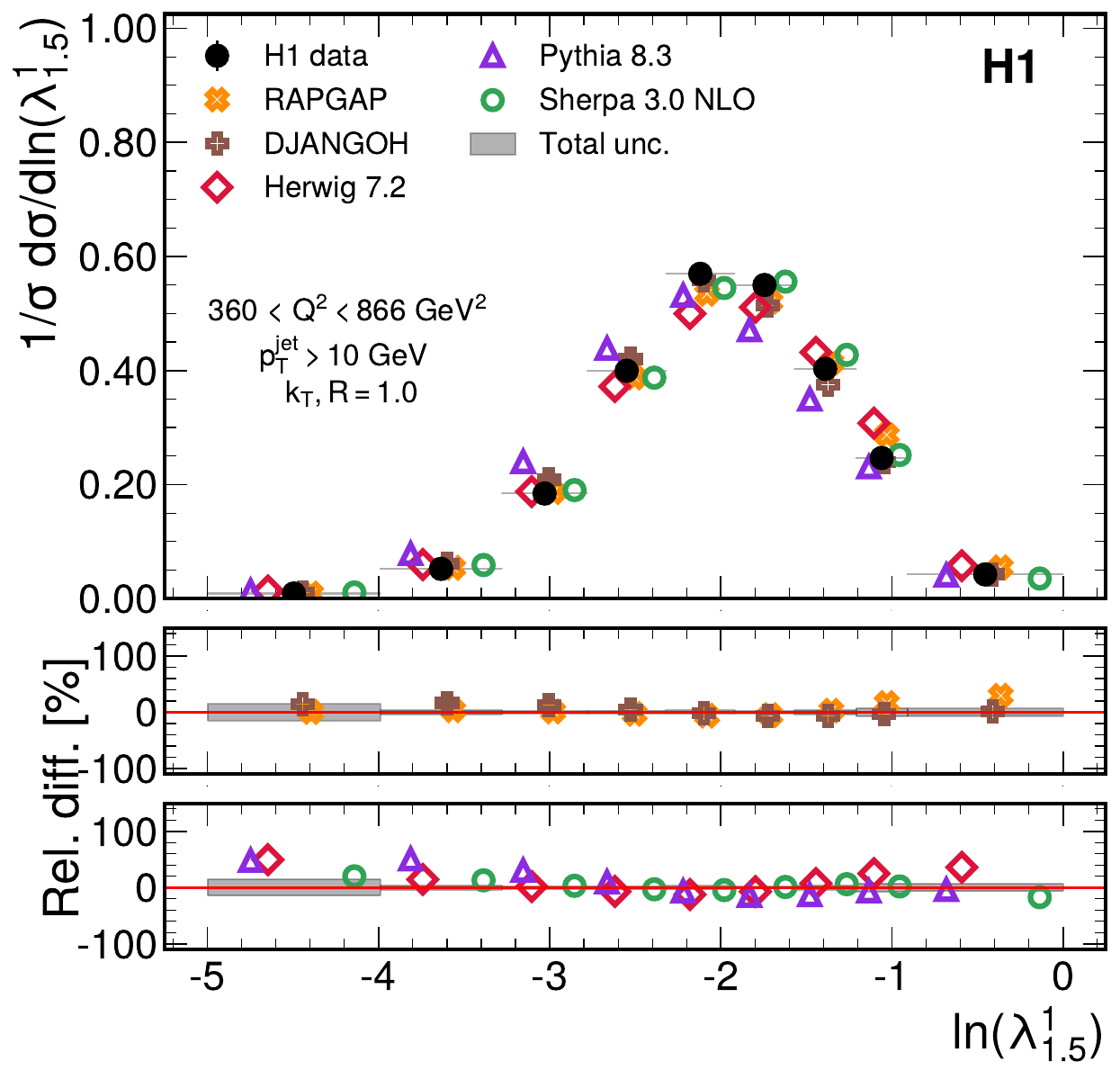}
    \includegraphics[width=0.45\textwidth]{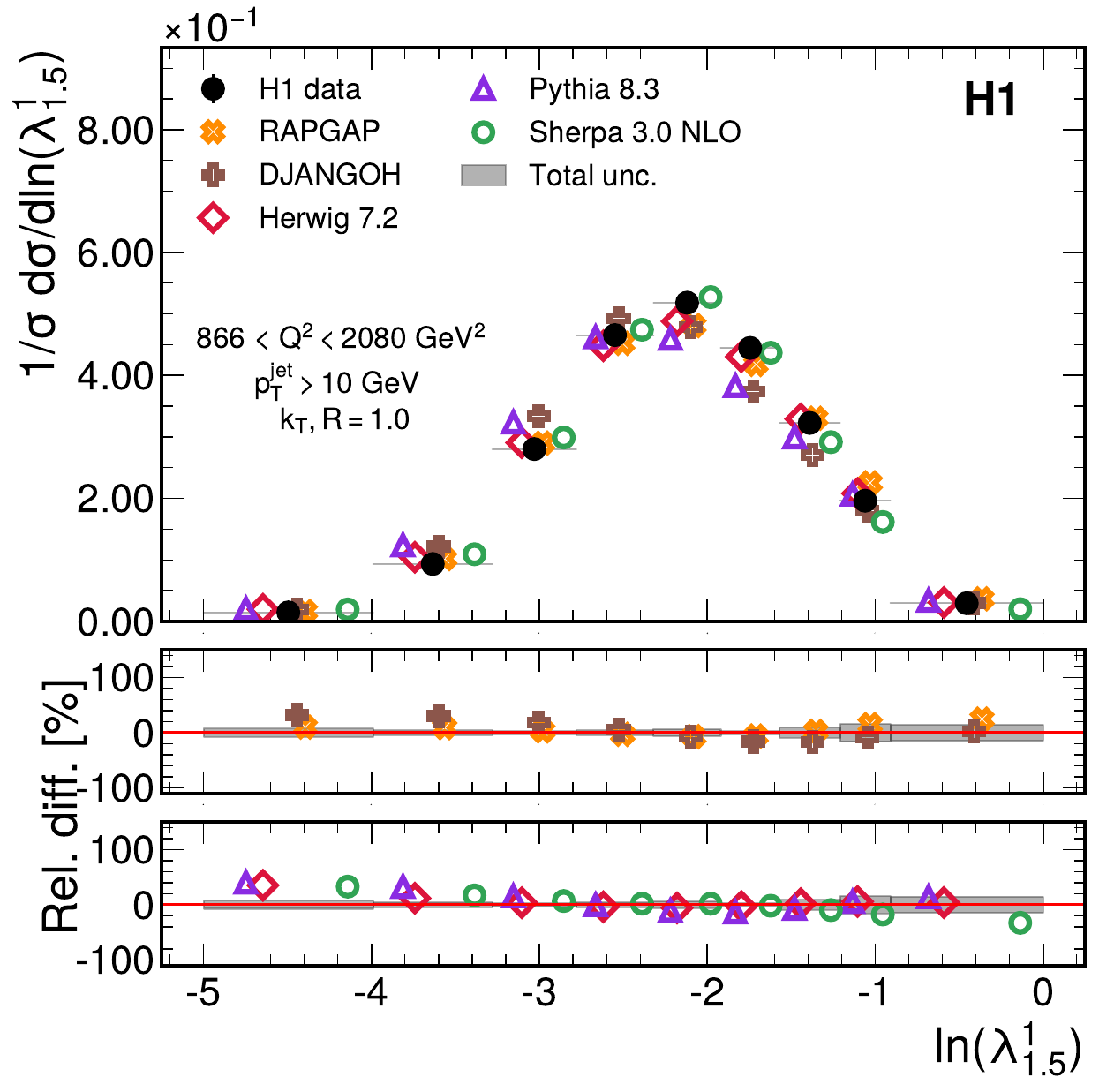}
    \includegraphics[width=0.45\textwidth]{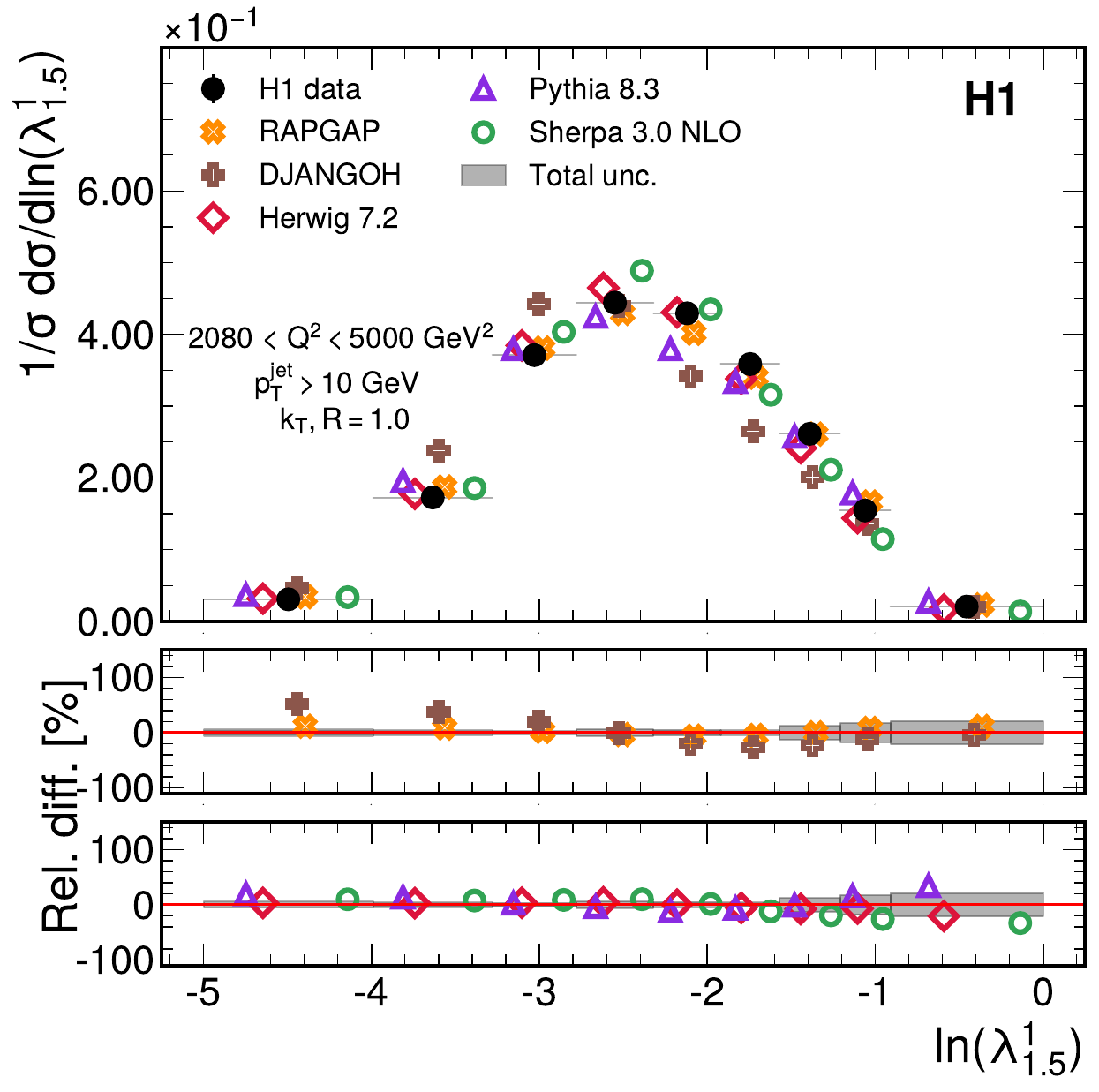}
    \caption{Measured cross sections, normalized to the inclusive jet production cross section, for multiple \Qs{} intervals as a function of $\ln(\lambda_{1.5}^{1})$. Data are shown as solid dots, horizontal bars indicate the bin ranges. 
       Predictions from multiple simulations are shown for comparison, and are offset horizontally for visual clarity. The relative differences between data and predictions are shown in the bottom panels, split between dedicated DIS simulators (middle) and general purpose simulators (bottom). Gray bands represent the total data uncertainties.
}
    \label{fig:rapgap_data_sys_q2_5}
\end{figure*}

\begin{figure*}[htb]
    \centering
    \includegraphics[width=0.45\textwidth]{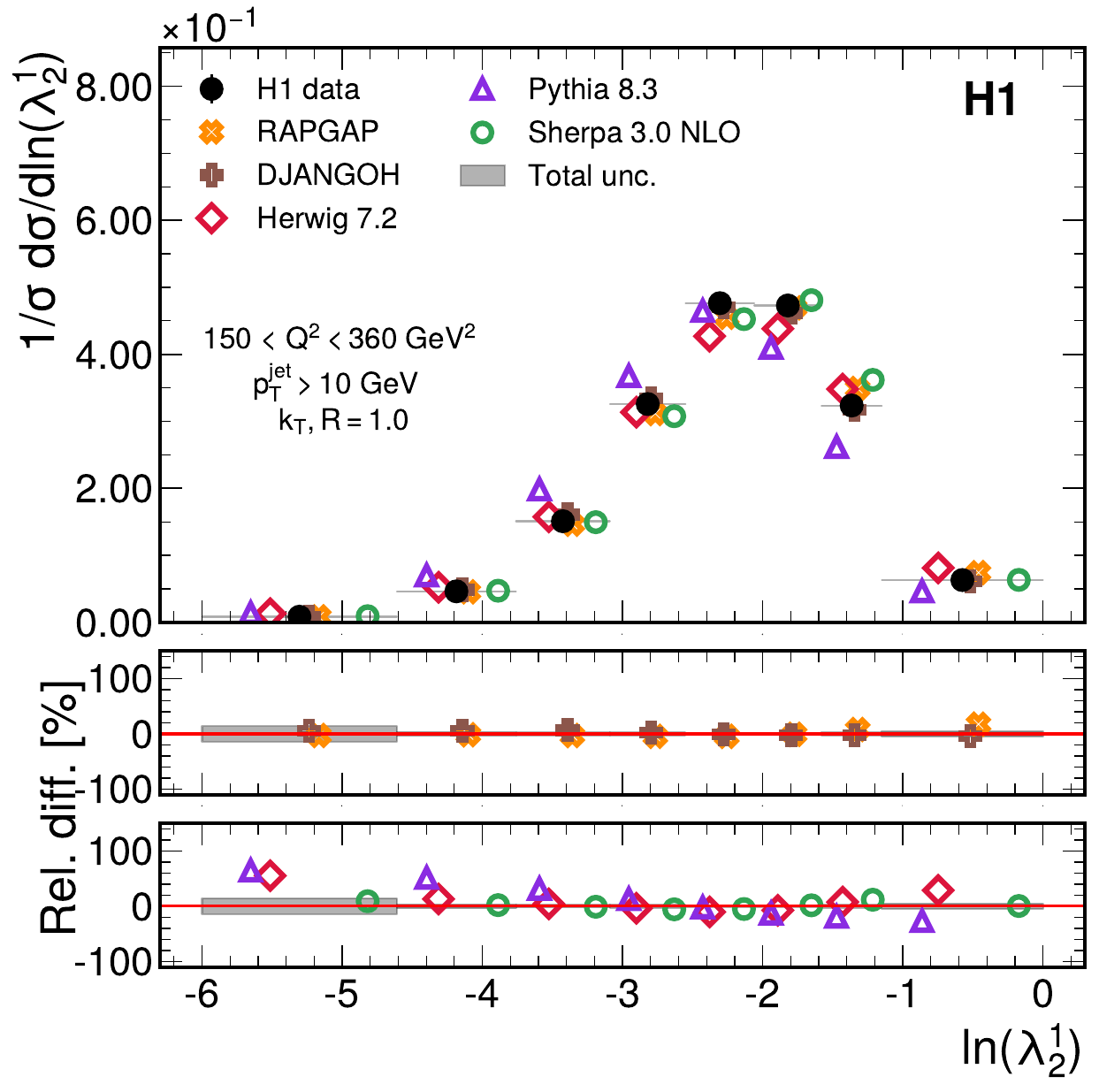}
\includegraphics[width=0.45\textwidth]{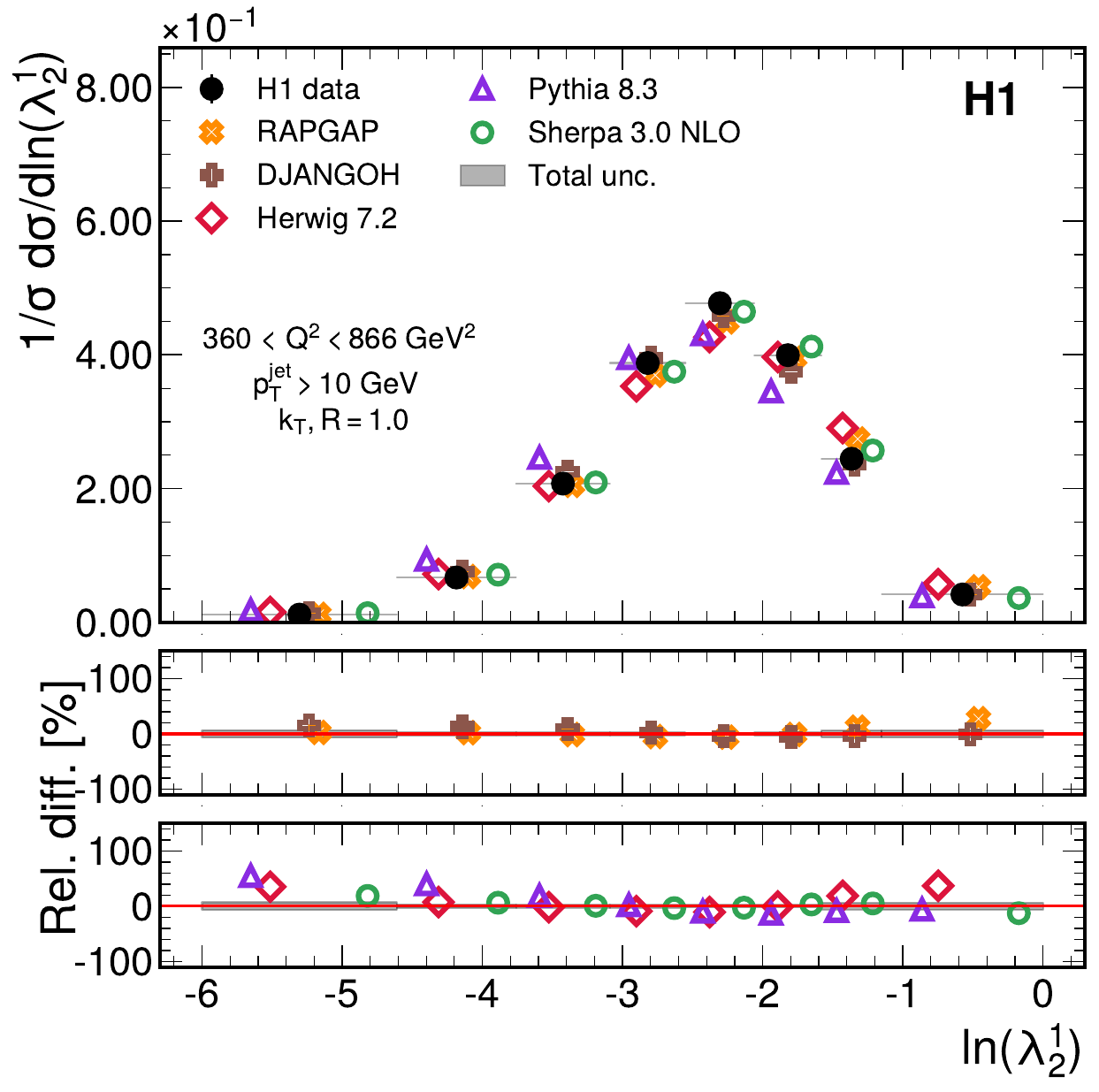}
    \includegraphics[width=0.45\textwidth]{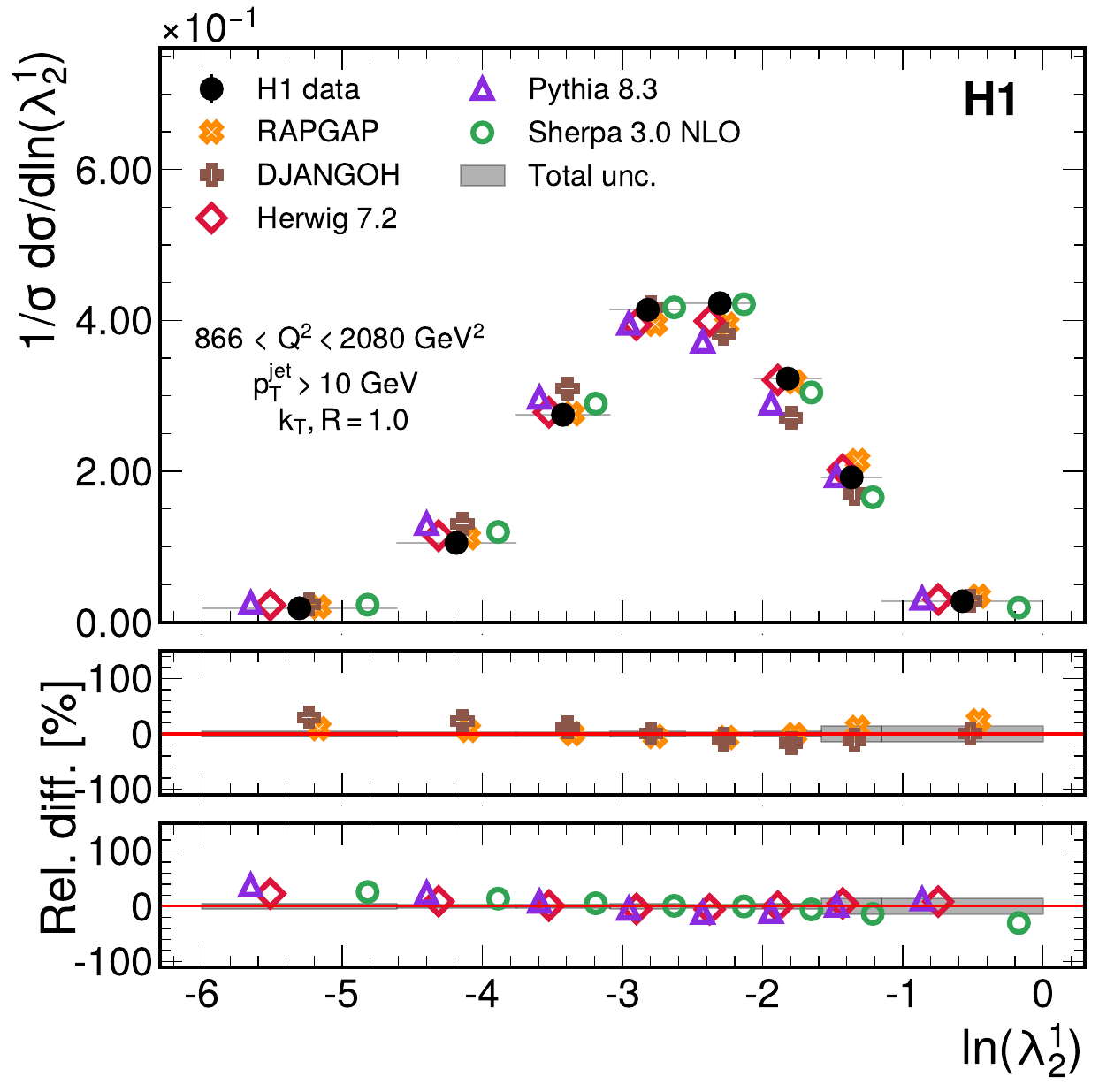}
    \includegraphics[width=0.45\textwidth]{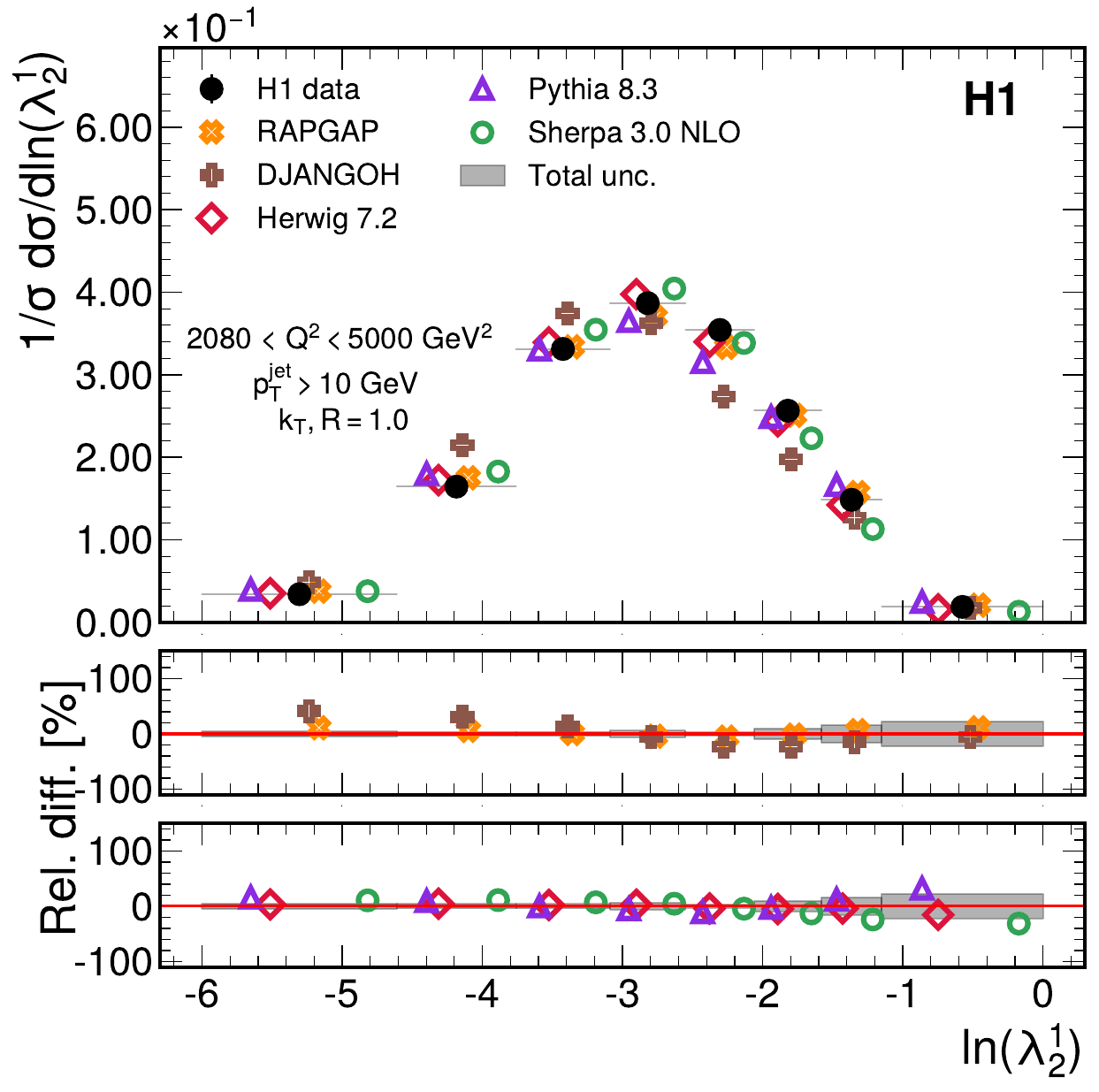}
    
    \caption{Measured cross sections, normalized to the inclusive jet production cross section, for multiple \Qs{} intervals as a function of $\ln(\lambda_2^1)$. Data are shown as solid dots, horizontal bars indicate the bin ranges. 
             Predictions from multiple simulations are shown for comparison, and are offset horizontally for visual clarity. The relative differences between data and predictions are shown in the bottom panels, split between dedicated DIS simulators (middle) and general purpose simulators (bottom). Gray bands represent the total data uncertainties.
}
    \label{fig:rapgap_data_sys_q2_6}
\end{figure*}

\begin{figure*}[htb]
    \centering
    \includegraphics[width=0.45\textwidth]{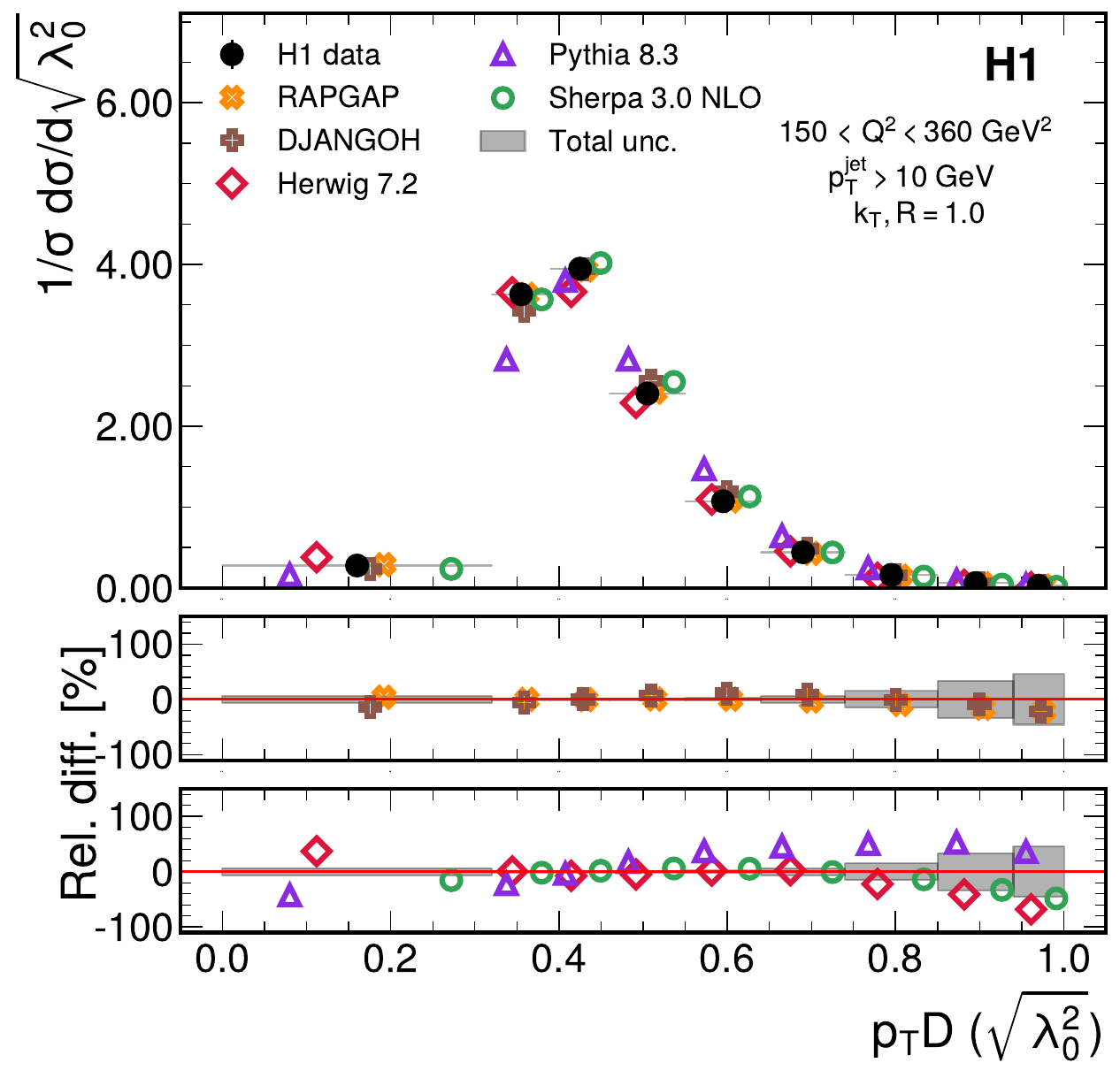}
\includegraphics[width=0.45\textwidth]{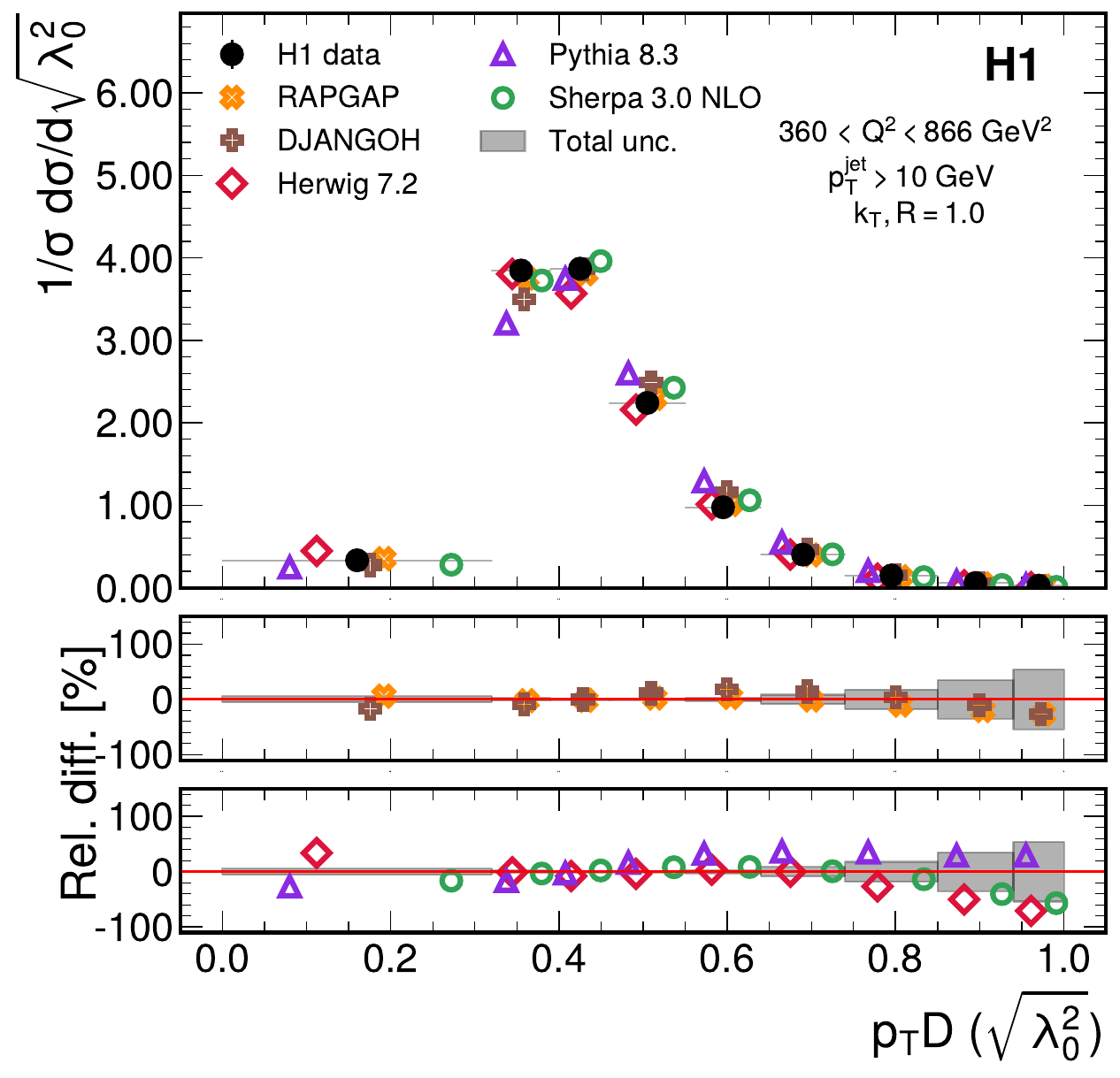}
    \includegraphics[width=0.45\textwidth]{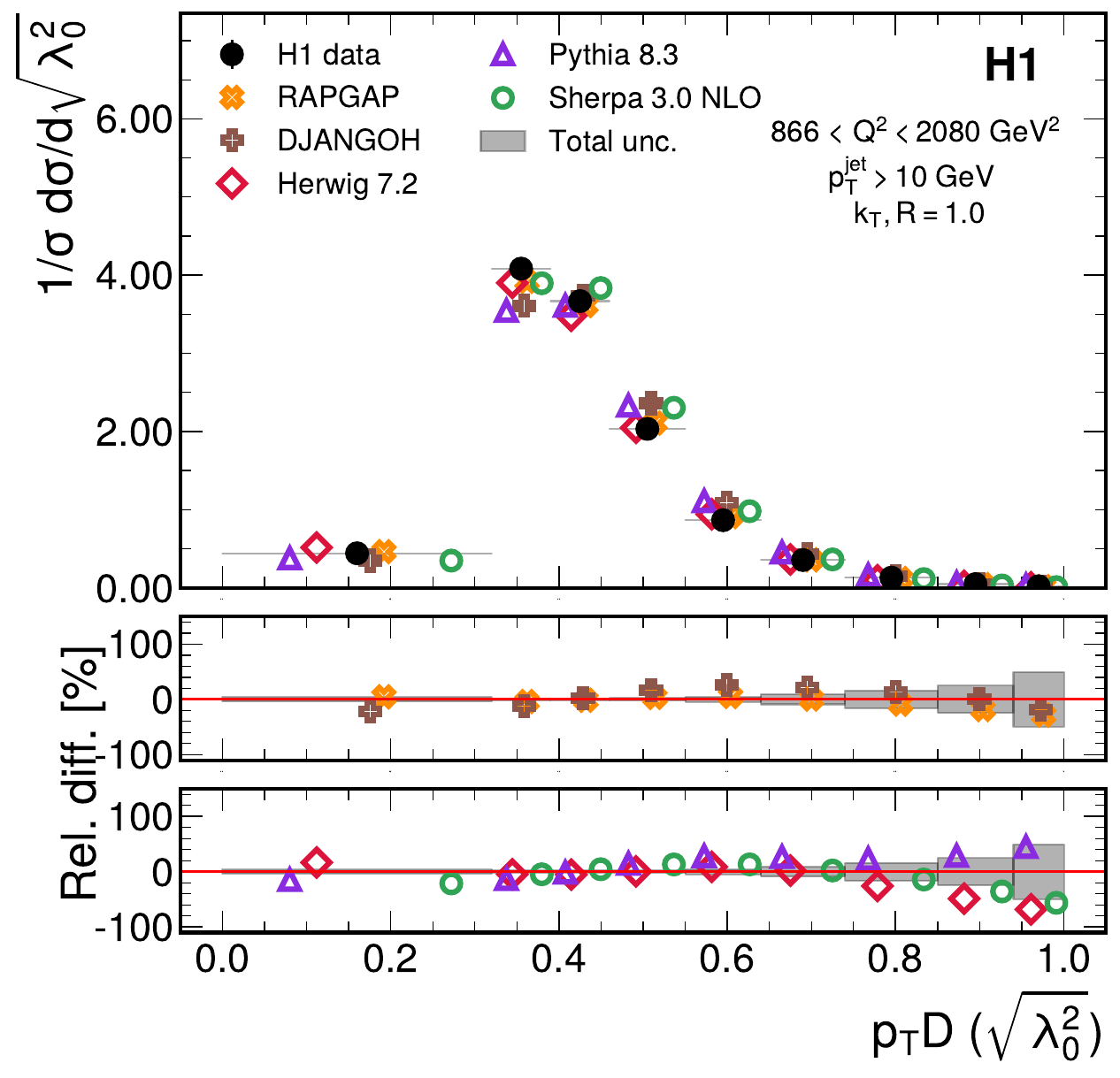}
    \includegraphics[width=0.45\textwidth]{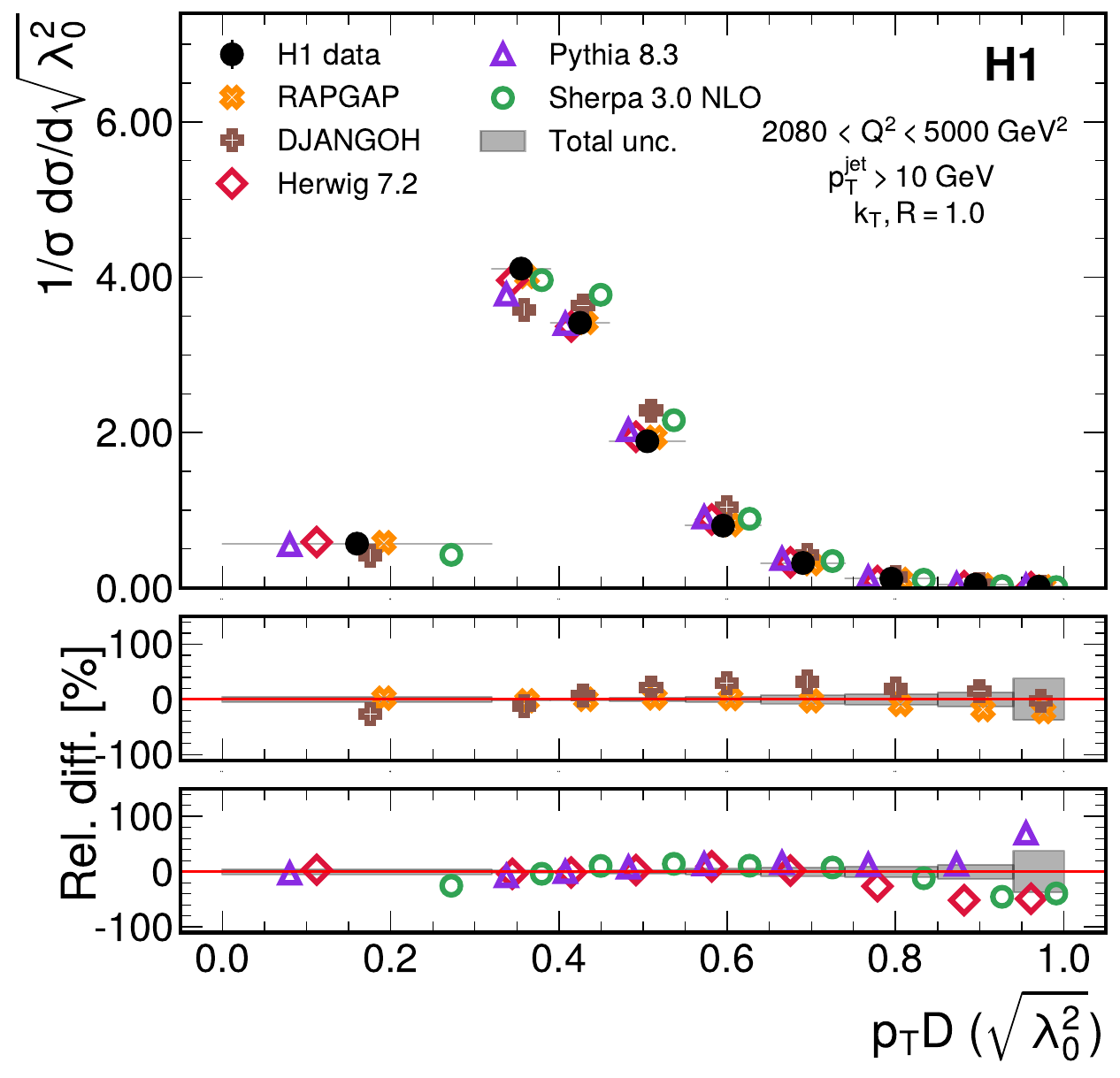}
    
    \caption{Measured cross sections, normalized to the inclusive jet production cross section, for multiple \Qs{} intervals (columns) as a function of $\pt\text{D}$. Data are shown as solid dots, horizontal bars indicate the bin ranges. 
                   Predictions from multiple simulations are shown for comparison, and are offset horizontally for visual clarity. The relative differences between data and predictions are shown in the bottom panels, split between dedicated DIS simulators (middle) and general purpose simulators (bottom). Gray bands represent the total data uncertainties.
}
    \label{fig:rapgap_data_sys_q2_3}
\end{figure*}

\begin{figure*}[htb]
    \centering
    \includegraphics[width=0.450\textwidth]{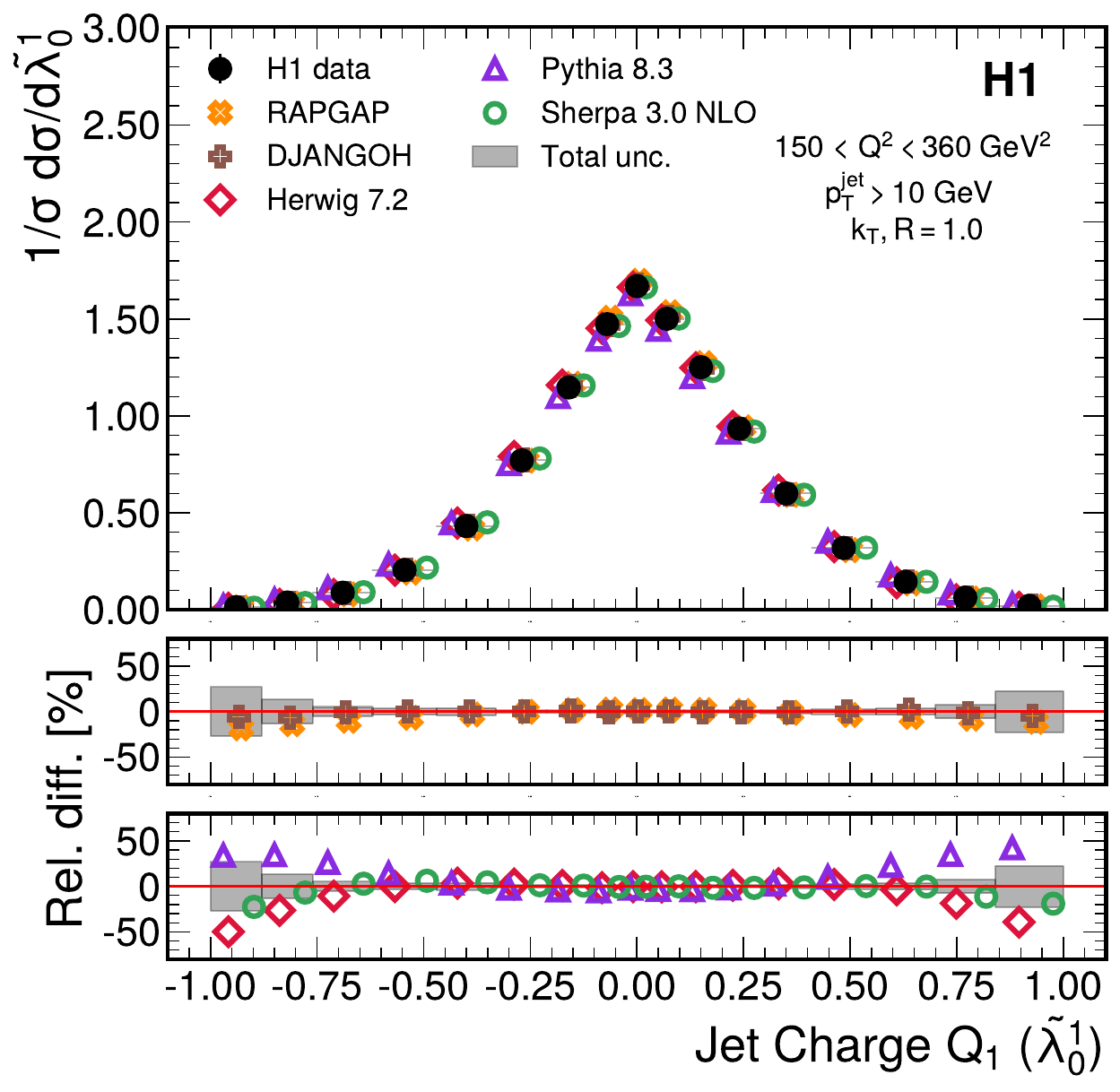}
\includegraphics[width=0.450\textwidth]{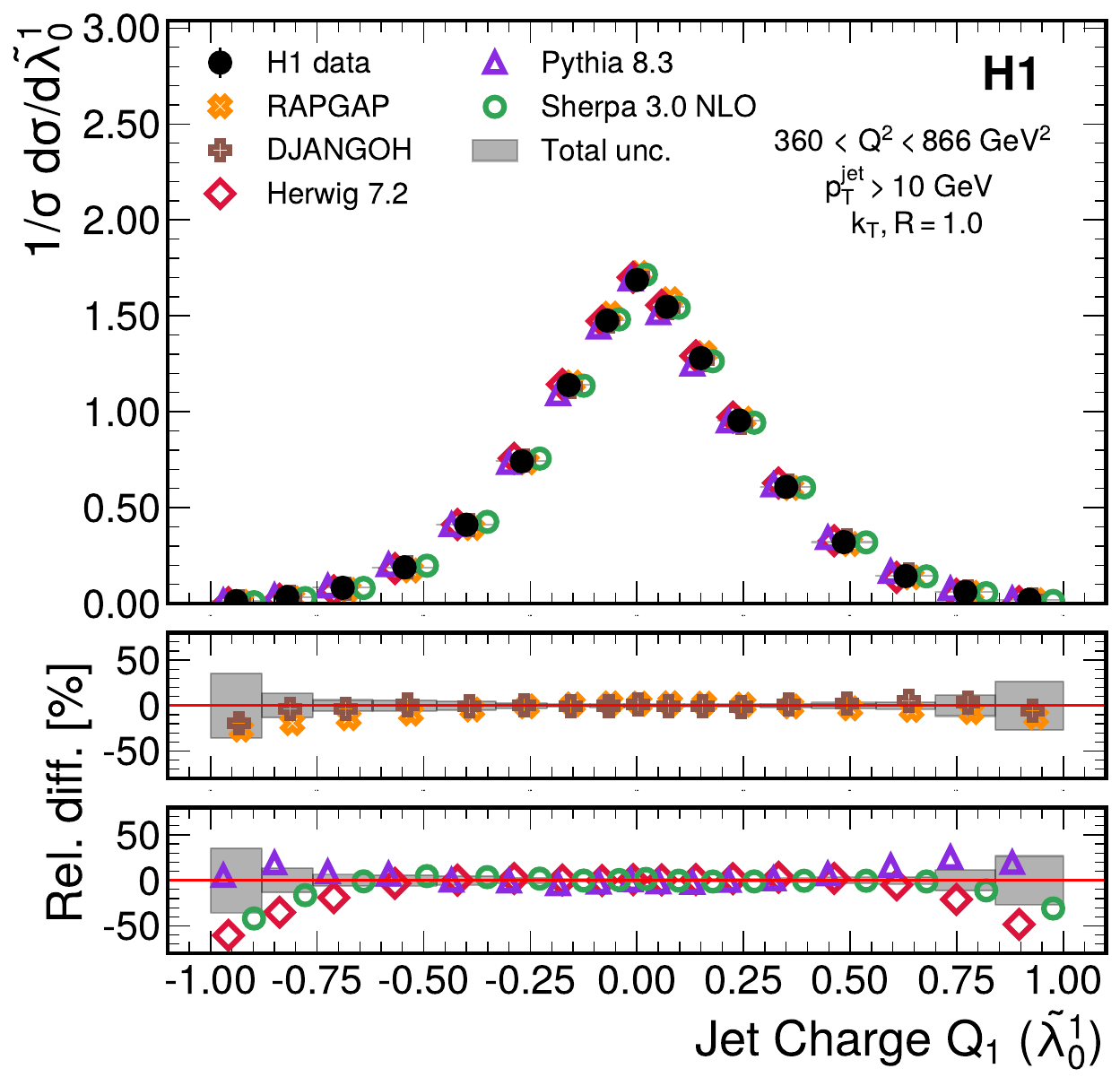}
    \includegraphics[width=0.45\textwidth]{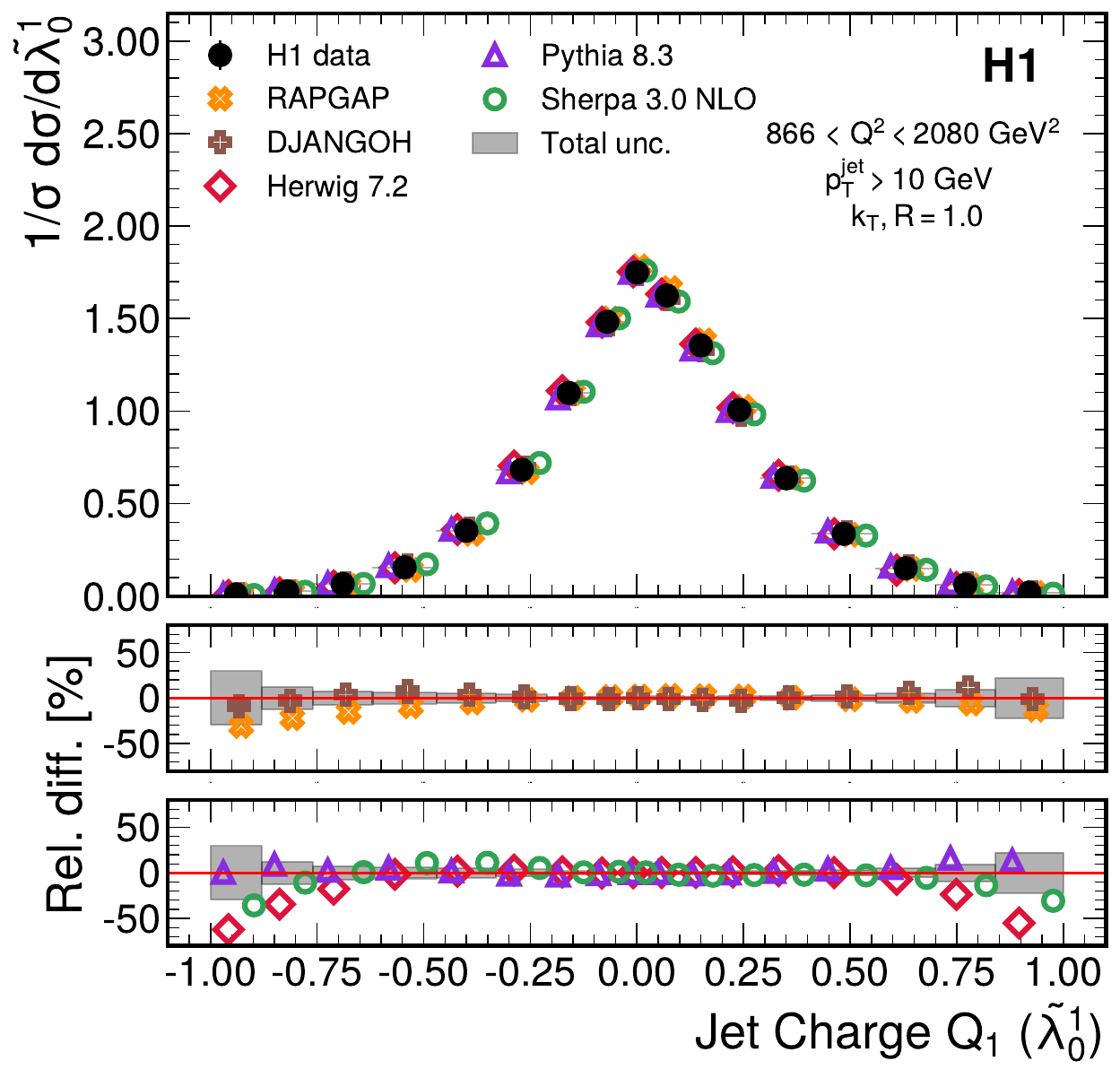}
    \includegraphics[width=0.45\textwidth]{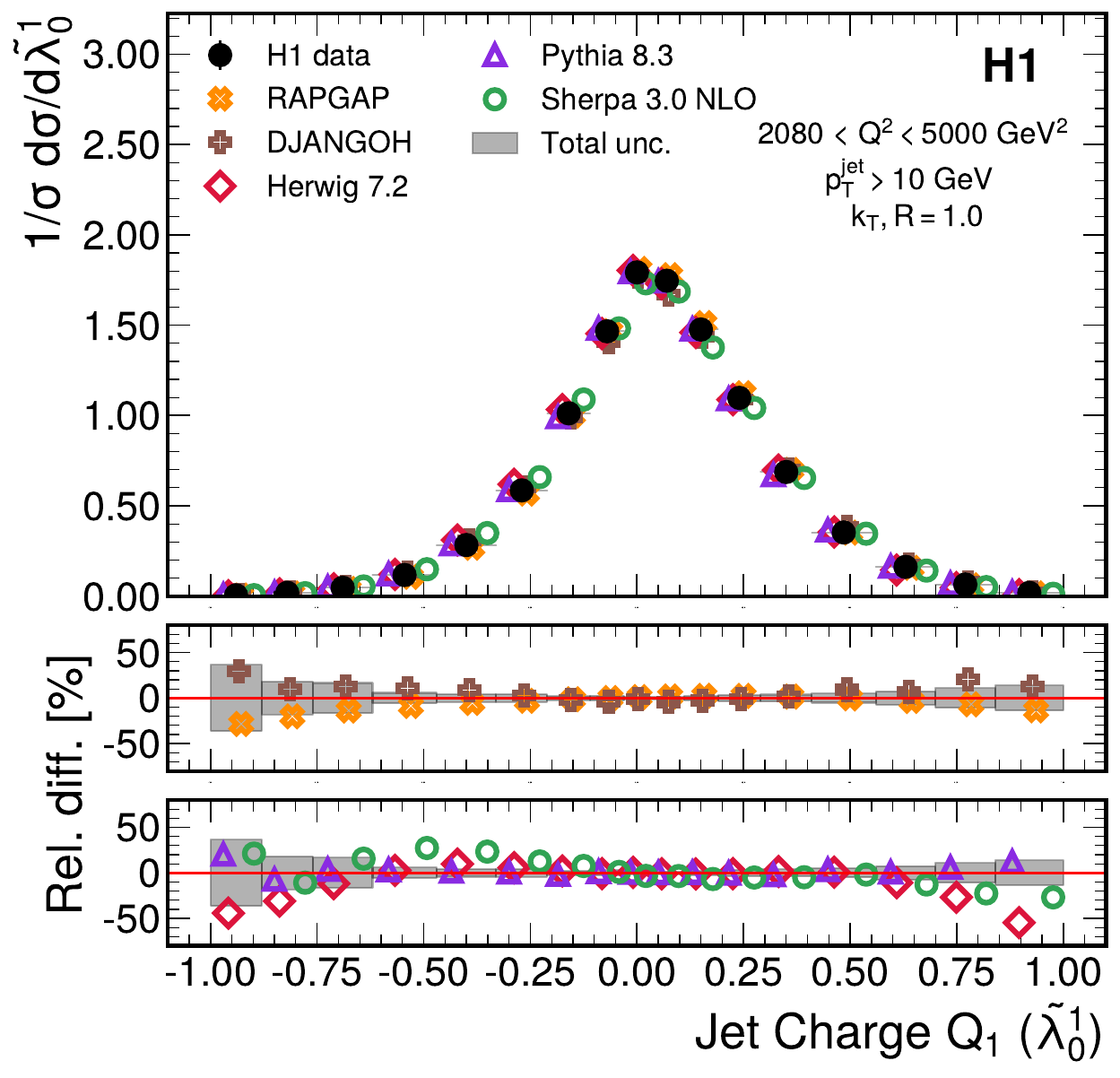}

    \caption{Measured cross sections, normalized to the inclusive jet production cross section, for multiple \Qs{} intervals  as a function of Q$_1$. Data are shown as solid dots, horizontal bars indicate the bin ranges. 
      Predictions from multiple simulations are shown for comparison, and are offset horizontally for visual clarity. The relative differences between data and predictions are shown in the bottom panels, split between dedicated DIS simulators (middle) and general purpose simulators (bottom). Gray bands represent the total data uncertainties.
}
    \label{fig:rapgap_data_sys_q2_1}
\end{figure*}

\begin{figure*}[htb]
    \centering
    
    \includegraphics[width=0.45\textwidth]{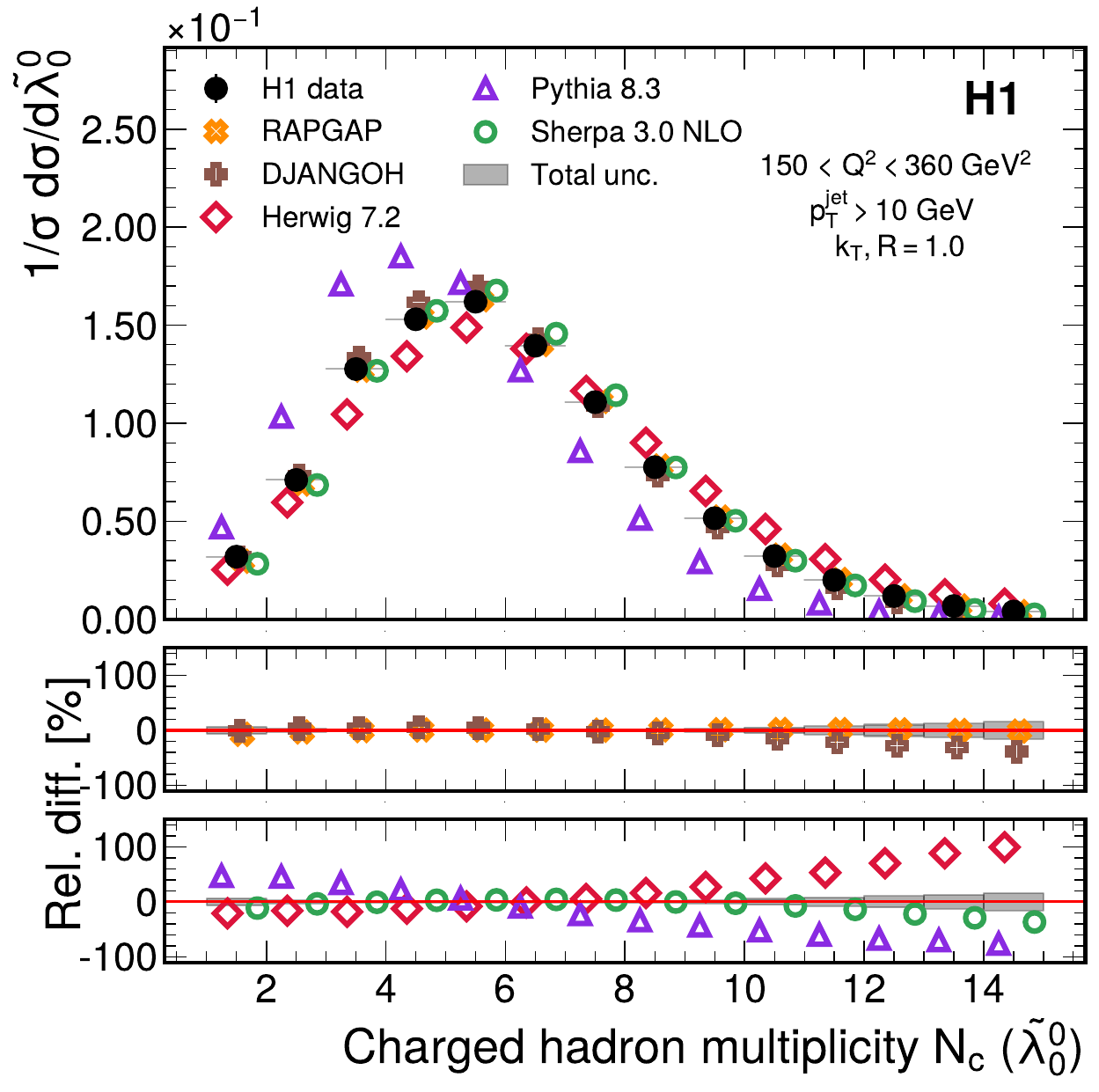}
\includegraphics[width=0.45\textwidth]{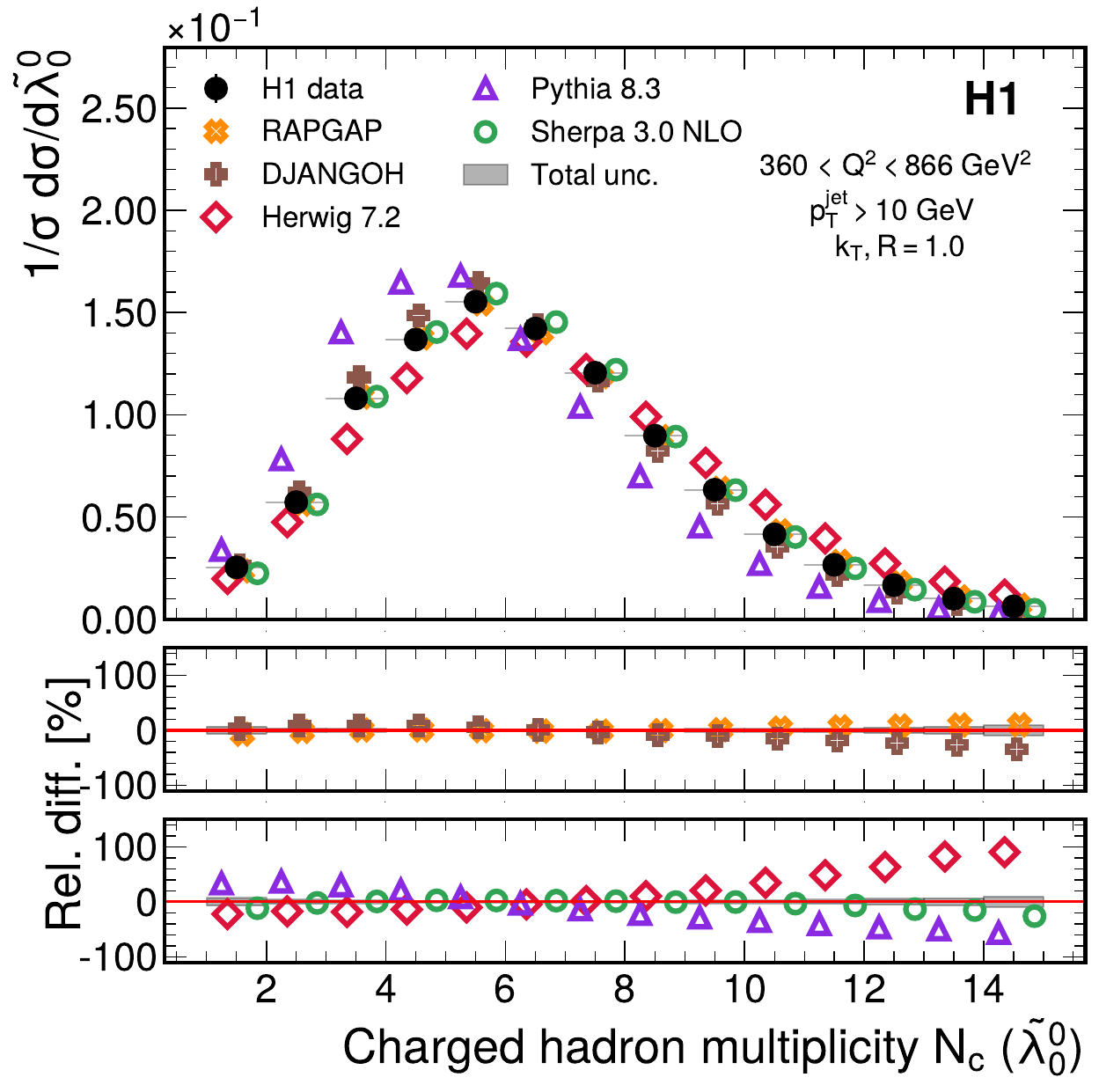}
    \includegraphics[width=0.45\textwidth]{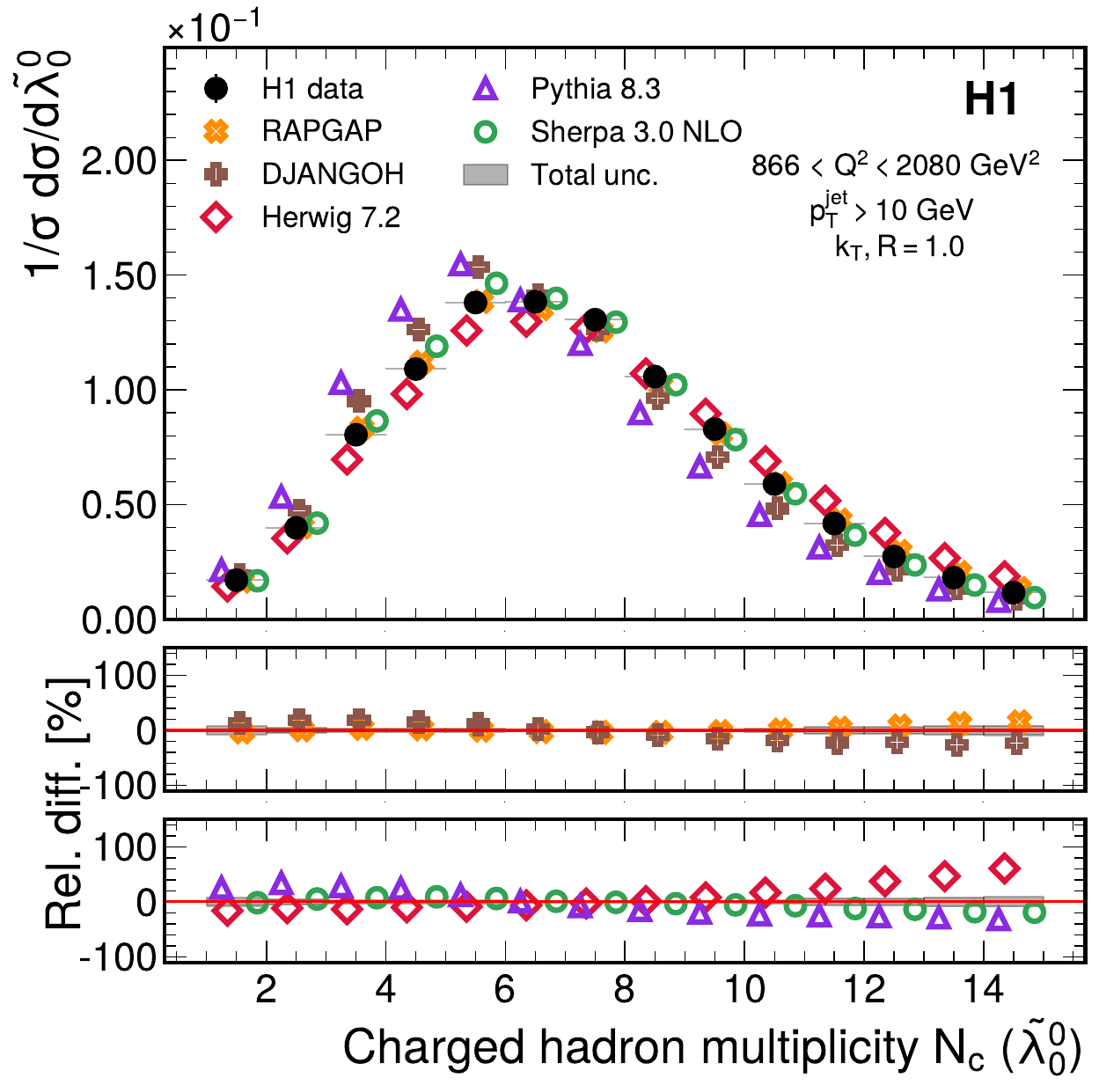}
    \includegraphics[width=0.45\textwidth]{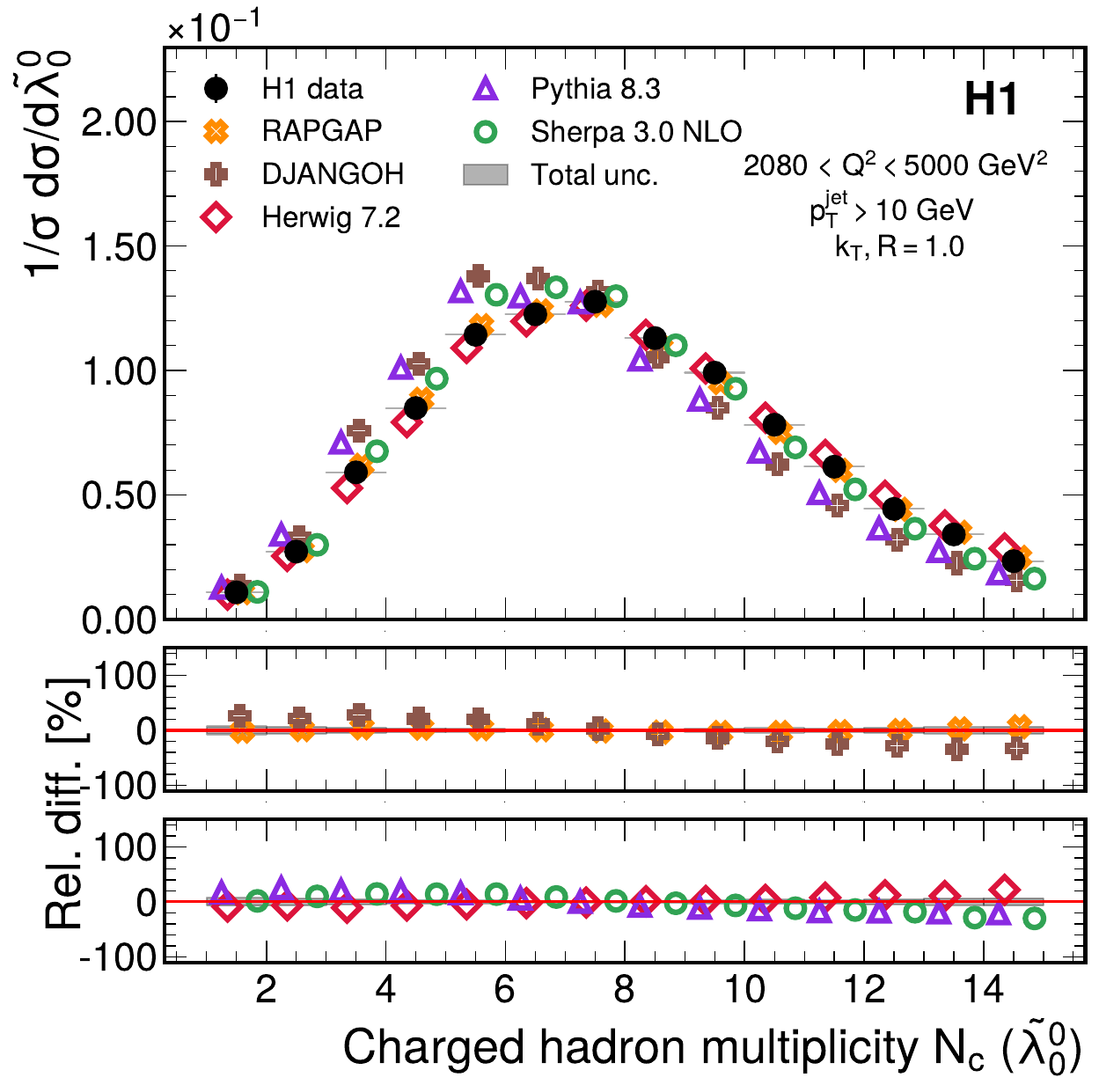}

    \caption{Measured cross sections, normalized to the inclusive jet production cross section, for multiple \Qs{} intervals as a function of N$_c$. Data are shown as solid dots, horizontal bars indicate the bin ranges. 
      Predictions from multiple simulations are shown for comparison, and are offset horizontally for visual clarity. The relative differences between data and predictions are shown in the bottom panels, split between dedicated DIS simulators (middle) and general purpose simulators (bottom). Gray bands represent the total data uncertainties.
}
    \label{fig:rapgap_data_sys_q2_2}
\end{figure*}

In this section, the normalized differential cross section of the jet substructure observables after unfolding are presented in Figs.~\ref{fig:rapgap_data_sys_q2_1}, \ref{fig:rapgap_data_sys_q2_2}, \ref{fig:rapgap_data_sys_q2_3}, \ref{fig:rapgap_data_sys_q2_4}, \ref{fig:rapgap_data_sys_q2_5}, and \ref{fig:rapgap_data_sys_q2_6}. Histograms showing the results are created after detector unfolding for comparison with different generators. The normalized differential cross section for observable $\mathcal{O}$ is calculated as:

\begin{equation}
    \frac{1}{\sigma}\frac{\mathrm{d}\sigma}{\mathrm{d}\mathcal{O}} = \frac{1}{\text{N}}\frac{\Delta\text{N}}{\Delta X},
\end{equation}
with $N$ being the total number of unfolded jets in the considered $Q^2$ range and $\Delta\text{N}$ representing the number of unfolded jets in the respective bin of $\mathcal{O}$. The bin width is denoted $\Delta X$.

\section{Data tables for all presented histograms}\label{app:tables}
Measured values of the normalized differential cross sections and contributions from each uncertainty source to the unfolded values are listed in Tables~\ref{tab:q2150}, \ref{tab:q2360}, \ref{tab:q2866}, \ref{tab:q22080}, and~\ref{tab:q25000}.  

\begin{table*}[h!]
\centering
\caption{Measured values of the normalized unfolded differential cross sections and uncertainties for each systematic source considered in the range $Q^{2}>$~150 GeV$^{2}$.}
\label{tab:q2150}
\begin{adjustbox}{width=1\textwidth}
\begin{tabular}{| c | c | c | c || c | c | c | c | c | c | c | c | c |}
\hline
$\ln(\lambda_1^1)$ & $1/\sigma$ $\mathrm{d}\sigma/\mathrm{d}$$\ln(\lambda_1^1)$ & Stat. (\%) & Tot. (\%) & HFS(jet) (\%) & HFS(other) (\%) & HFS($\phi$) (\%) & Lepton(E) (\%) & Lepton($\phi$) (\%) & Model (\%) & Closure (\%)\\ 
\hline
$[-4.00,-3.15]$ & 0.006 & 3.122 & 21.454 & 11.697 & 6.089 & 10.095 & 9.280 & 8.787 & 3.030 & 1.483\\ 
$[-3.15,-2.59]$ & 0.038 & 2.238 & 10.056 & 5.187 & 4.311 & 3.258 & 4.283 & 3.378 & 2.237 & 2.295\\ 
$[-2.59,-2.18]$ & 0.151 & 1.412 & 4.036 & 0.717 & 0.186 & 0.904 & 1.390 & 1.141 & 1.385 & 2.789\\ 
$[-2.18,-1.86]$ & 0.369 & 0.841 & 2.339 & 0.640 & 0.520 & 0.828 & 0.839 & 0.081 & 0.810 & 1.425\\ 
$[-1.86,-1.58]$ & 0.629 & 0.749 & 3.105 & 1.670 & 1.440 & 0.527 & 1.369 & 1.183 & 0.724 & 0.375\\ 
$[-1.58,-1.29]$ & 0.784 & 0.750 & 2.799 & 1.408 & 1.291 & 0.399 & 1.604 & 0.706 & 0.379 & 0.500\\ 
$[-1.29,-1.05]$ & 0.709 & 0.607 & 1.696 & 0.568 & 0.517 & 0.597 & 0.419 & 0.960 & 0.312 & 0.604\\ 
$[-1.05,-0.81]$ & 0.508 & 1.070 & 3.443 & 1.140 & 1.150 & 1.038 & 1.143 & 0.572 & 0.590 & 2.242\\ 
$[-0.81,-0.61]$ & 0.315 & 1.960 & 6.917 & 2.648 & 2.136 & 1.616 & 2.544 & 1.956 & 0.757 & 4.352\\ 
$[-0.61,0.00]$ & 0.057 & 2.871 & 8.053 & 1.292 & 2.468 & 2.674 & 1.761 & 1.152 & 2.760 & 5.445\\ 
\hline
$\ln(\lambda_{1.5}^1)$ & $1/\sigma$ $\mathrm{d}\sigma/\mathrm{d}$$\ln(\lambda_{1.5}^1)$ & Stat. (\%) & Tot. (\%) & HFS(jet) (\%) & HFS(other) (\%) & HFS($\phi$) (\%) & Lepton(E) (\%) & Lepton($\phi$) (\%) & Model (\%) & Closure (\%)\\ 
\hline
$[-5.00,-3.99]$ & 0.010 & 2.489 & 14.137 & 7.035 & 5.404 & 5.577 & 5.227 & 4.817 & 1.854 & 5.467\\ 
$[-3.99,-3.28]$ & 0.052 & 1.755 & 5.929 & 2.495 & 1.510 & 1.007 & 2.365 & 0.570 & 1.516 & 3.787\\ 
$[-3.28,-2.78]$ & 0.167 & 1.100 & 3.832 & 1.063 & 1.073 & 1.068 & 0.590 & 1.042 & 1.074 & 2.731\\ 
$[-2.78,-2.32]$ & 0.354 & 0.759 & 2.256 & 0.941 & 1.305 & 0.532 & 0.312 & 0.875 & 0.637 & 0.609\\ 
$[-2.32,-1.92]$ & 0.535 & 0.735 & 2.873 & 1.565 & 1.378 & 0.209 & 1.514 & 0.498 & 0.655 & 0.598\\ 
$[-1.92,-1.57]$ & 0.569 & 0.624 & 1.690 & 0.572 & 0.516 & 0.552 & 0.972 & 0.293 & 0.392 & 0.621\\ 
$[-1.57,-1.21]$ & 0.449 & 0.789 & 2.226 & 0.499 & 0.944 & 0.765 & 0.206 & 1.156 & 0.789 & 0.781\\ 
$[-1.21,-0.91]$ & 0.283 & 1.659 & 5.811 & 2.316 & 1.985 & 1.358 & 2.248 & 1.588 & 0.256 & 3.496\\ 
$[-0.91,0.00]$ & 0.052 & 2.678 & 7.593 & 1.015 & 2.240 & 2.524 & 1.613 & 1.072 & 2.542 & 5.278\\ 
\hline
$\ln(\lambda_2^1)$ & $1/\sigma$ $\mathrm{d}\sigma/\mathrm{d}$$\ln(\lambda_2^1)$ & Stat. (\%) & Tot. (\%) & HFS(jet) (\%) & HFS(other) (\%) & HFS($\phi$) (\%) & Lepton(E) (\%) & Lepton($\phi$) (\%) & Model (\%) & Closure (\%)\\ 
\hline
$[-6.00,-4.61]$ & 0.012 & 2.039 & 11.029 & 4.839 & 4.291 & 3.877 & 3.747 & 3.223 & 1.913 & 5.705\\ 
$[-4.61,-3.76]$ & 0.064 & 1.384 & 5.820 & 0.773 & 1.058 & 1.184 & 0.450 & 1.376 & 1.380 & 4.984\\ 
$[-3.76,-3.09]$ & 0.188 & 0.897 & 2.421 & 0.724 & 0.575 & 0.871 & 0.885 & 0.099 & 0.722 & 1.460\\ 
$[-3.09,-2.55]$ & 0.355 & 0.716 & 2.745 & 1.263 & 1.808 & 0.303 & 0.865 & 0.839 & 0.643 & 0.449\\ 
$[-2.55,-2.06]$ & 0.466 & 0.622 & 2.002 & 1.002 & 0.340 & 0.453 & 1.158 & 0.622 & 0.464 & 0.596\\ 
$[-2.06,-1.58]$ & 0.427 & 0.597 & 1.880 & 0.565 & 0.195 & 0.597 & 0.587 & 1.170 & 0.588 & 0.638\\ 
$[-1.58,-1.15]$ & 0.279 & 1.346 & 4.385 & 1.814 & 1.388 & 1.135 & 1.620 & 0.989 & 1.003 & 2.511\\ 
$[-1.15,0.00]$ & 0.051 & 2.485 & 6.921 & 0.944 & 1.929 & 2.328 & 1.459 & 1.289 & 2.427 & 4.692\\ 
\hline
$\sqrt{\lambda_0^2}$ & $1/\sigma$ $\mathrm{d}\sigma/\mathrm{d}$$\sqrt{\lambda_0^2}$ & Stat. (\%) & Tot. (\%) & HFS(jet) (\%) & HFS(other) (\%) & HFS($\phi$) (\%) & Lepton(E) (\%) & Lepton($\phi$) (\%) & Model (\%) & Closure (\%)\\ 
\hline
$[0.00,0.32]$ & 0.322 & 1.336 & 5.555 & 1.335 & 0.810 & 1.963 & 1.212 & 4.160 & 1.815 & 0.844\\ 
$[0.32,0.39]$ & 3.761 & 0.599 & 1.486 & 0.272 & 0.183 & 0.694 & 0.224 & 0.859 & 0.296 & 0.621\\ 
$[0.39,0.46]$ & 3.875 & 0.353 & 2.429 & 0.427 & 0.295 & 0.308 & 0.197 & 0.329 & 1.931 & 1.239\\ 
$[0.46,0.55]$ & 2.296 & 0.553 & 1.531 & 0.501 & 0.552 & 0.544 & 0.550 & 0.168 & 0.456 & 0.805\\ 
$[0.55,0.64]$ & 1.011 & 1.062 & 2.868 & 1.042 & 1.042 & 0.866 & 0.920 & 0.685 & 1.628 & 0.459\\ 
$[0.64,0.74]$ & 0.415 & 1.563 & 7.673 & 2.566 & 2.178 & 3.974 & 1.441 & 4.361 & 2.434 & 1.514\\ 
$[0.74,0.85]$ & 0.152 & 2.551 & 15.804 & 6.838 & 2.984 & 8.837 & 3.763 & 9.109 & 2.539 & 2.436\\ 
$[0.85,0.94]$ & 0.057 & 4.153 & 32.704 & 14.773 & 9.364 & 16.999 & 13.020 & 16.045 & 3.663 & 4.124\\ 
$[0.94,1.00]$ & 0.024 & 5.688 & 49.586 & 21.621 & 13.182 & 24.299 & 22.638 & 21.964 & 11.637 & 8.026\\ 
\hline
$\tilde{\lambda}_0^1$ & $1/\sigma$ $\mathrm{d}\sigma/\mathrm{d}$$\tilde{\lambda}_0^1$ & Stat. (\%) & Tot. (\%) & HFS(jet) (\%) & HFS(other) (\%) & HFS($\phi$) (\%) & Lepton(E) (\%) & Lepton($\phi$) (\%) & Model (\%) & Closure (\%)\\ 
\hline
$[-1.00,-0.88]$ & 0.012 & 4.945 & 30.010 & 15.368 & 6.793 & 13.754 & 9.789 & 16.375 & 4.940 & 4.041\\ 
$[-0.88,-0.76]$ & 0.034 & 3.307 & 13.156 & 6.265 & 1.598 & 5.088 & 2.438 & 8.461 & 3.302 & 2.451\\ 
$[-0.76,-0.62]$ & 0.082 & 2.142 & 5.746 & 1.401 & 2.138 & 2.053 & 0.339 & 3.181 & 2.075 & 1.771\\ 
$[-0.62,-0.47]$ & 0.191 & 1.325 & 4.767 & 0.979 & 1.606 & 1.440 & 3.015 & 1.211 & 1.323 & 1.748\\ 
$[-0.47,-0.33]$ & 0.411 & 0.786 & 4.584 & 1.544 & 1.585 & 2.288 & 2.898 & 0.990 & 0.809 & 0.478\\ 
$[-0.33,-0.21]$ & 0.745 & 0.553 & 2.226 & 0.796 & 0.428 & 0.996 & 1.438 & 0.579 & 0.542 & 0.375\\ 
$[-0.21,-0.11]$ & 1.134 & 0.488 & 1.242 & 0.478 & 0.487 & 0.488 & 0.393 & 0.344 & 0.368 & 0.438\\ 
$[-0.11,-0.03]$ & 1.474 & 0.433 & 1.435 & 0.388 & 0.086 & 0.240 & 0.411 & 0.427 & 0.430 & 1.058\\ 
$[-0.03,0.03]$ & 1.688 & 0.337 & 1.630 & 0.585 & 0.887 & 0.796 & 0.460 & 0.227 & 0.699 & 0.167\\ 
$[0.03,0.11]$ & 1.537 & 0.367 & 1.270 & 0.430 & 0.277 & 0.499 & 0.628 & 0.231 & 0.663 & 0.285\\ 
$[0.11,0.19]$ & 1.280 & 0.388 & 1.041 & 0.131 & 0.332 & 0.265 & 0.530 & 0.359 & 0.356 & 0.445\\ 
$[0.19,0.29]$ & 0.955 & 0.491 & 1.101 & 0.491 & 0.287 & 0.447 & 0.240 & 0.395 & 0.239 & 0.421\\ 
$[0.29,0.41]$ & 0.610 & 0.716 & 1.635 & 0.344 & 0.718 & 0.291 & 0.694 & 0.127 & 0.711 & 0.661\\ 
$[0.41,0.56]$ & 0.323 & 1.068 & 2.522 & 0.865 & 1.094 & 1.000 & 0.780 & 1.050 & 0.375 & 0.651\\ 
$[0.56,0.70]$ & 0.145 & 1.563 & 4.131 & 1.508 & 1.537 & 1.246 & 0.858 & 0.959 & 1.294 & 2.259\\ 
$[0.70,0.84]$ & 0.062 & 2.357 & 8.393 & 1.441 & 1.244 & 4.414 & 4.344 & 3.894 & 2.301 & 1.567\\ 
$[0.84,1.00]$ & 0.020 & 3.775 & 23.469 & 6.677 & 6.895 & 13.110 & 11.793 & 10.396 & 3.408 & 3.717\\ 
\hline
$\tilde{\lambda}_0^0$ & $1/\sigma$ $\mathrm{d}\sigma/\mathrm{d}$$\tilde{\lambda}_0^0$ & Stat. (\%) & Tot. (\%) & HFS(jet) (\%) & HFS(other) (\%) & HFS($\phi$) (\%) & Lepton(E) (\%) & Lepton($\phi$) (\%) & Model (\%) & Closure (\%)\\ 
\hline
$[1.00,2.00]$ & 0.028 & 1.839 & 7.749 & 1.503 & 1.802 & 4.571 & 1.504 & 3.082 & 3.129 & 2.952\\ 
$[2.00,3.00]$ & 0.062 & 1.173 & 4.658 & 1.169 & 1.167 & 1.596 & 0.996 & 1.064 & 2.173 & 2.864\\ 
$[3.00,4.00]$ & 0.114 & 0.798 & 4.048 & 0.541 & 0.761 & 0.685 & 0.191 & 0.598 & 2.080 & 3.113\\ 
$[4.00,5.00]$ & 0.141 & 0.575 & 2.199 & 0.156 & 0.571 & 0.524 & 0.345 & 0.402 & 0.897 & 1.671\\ 
$[5.00,6.00]$ & 0.156 & 0.384 & 0.964 & 0.378 & 0.362 & 0.328 & 0.356 & 0.208 & 0.339 & 0.337\\ 
$[6.00,7.00]$ & 0.139 & 0.341 & 1.024 & 0.313 & 0.343 & 0.340 & 0.341 & 0.228 & 0.242 & 0.611\\ 
$[7.00,8.00]$ & 0.116 & 0.459 & 2.159 & 0.389 & 0.214 & 0.347 & 0.425 & 0.445 & 0.980 & 1.672\\ 
$[8.00,9.00]$ & 0.085 & 0.645 & 3.522 & 0.609 & 0.482 & 0.337 & 0.559 & 0.322 & 1.808 & 2.754\\ 
$[9.00,10.00]$ & 0.060 & 0.877 & 5.289 & 0.734 & 0.850 & 0.373 & 0.499 & 0.588 & 2.958 & 4.056\\ 
$[10.00,11.00]$ & 0.040 & 1.137 & 5.345 & 1.070 & 0.333 & 1.125 & 0.879 & 1.928 & 2.262 & 3.893\\ 
$[11.00,12.00]$ & 0.026 & 1.430 & 4.706 & 1.219 & 0.203 & 1.173 & 0.609 & 2.796 & 0.741 & 2.908\\ 
$[12.00,13.00]$ & 0.016 & 1.738 & 6.529 & 1.732 & 0.799 & 2.490 & 1.509 & 4.521 & 1.737 & 2.009\\ 
$[13.00,14.00]$ & 0.010 & 2.075 & 6.515 & 1.945 & 0.713 & 2.390 & 1.280 & 4.835 & 1.669 & 0.575\\ 
$[14.00,15.00]$ & 0.006 & 2.390 & 8.898 & 2.375 & 1.335 & 3.504 & 1.947 & 6.307 & 2.316 & 2.197\\ 
\hline
\end{tabular}
\end{adjustbox}
\end{table*}

\begin{table*}[h!]
\centering
\caption{Measured values of the normalized unfolded differential cross sections and uncertainties for each systematic source considered in the range 150 $ < Q^2 < 360 $~GeV$^2$ .}
\label{tab:q2360}
\begin{adjustbox}{width=1\textwidth}
\begin{tabular}{| c | c | c | c || c | c | c | c | c | c | c | c | c |}
\hline
$\ln(\lambda_1^1)$ & $1/\sigma$ $\mathrm{d}\sigma/\mathrm{d}$$\ln(\lambda_1^1)$ & Stat. (\%) & Tot. (\%) & HFS(jet) (\%) & HFS(other) (\%) & HFS($\phi$) (\%) & Lepton(E) (\%) & Lepton($\phi$) (\%) & Model (\%) & Closure (\%)\\ 
\hline
$[-4.00,-3.15]$ & 0.004 & 3.21 & 26.30 & 11.99 & 8.20 & 14.77 & 11.96 & 9.95 & 2.53 & 1.91\\ 
$[-3.15,-2.59]$ & 0.024 & 2.11 & 14.28 & 5.73 & 6.77 & 7.64 & 5.56 & 4.86 & 2.00 & 1.97\\ 
$[-2.59,-2.18]$ & 0.101 & 1.41 & 5.23 & 1.93 & 0.84 & 3.28 & 2.83 & 0.86 & 1.03 & 0.65\\ 
$[-2.18,-1.86]$ & 0.292 & 0.96 & 3.10 & 0.96 & 0.95 & 0.27 & 0.50 & 0.88 & 2.24 & 0.85\\ 
$[-1.86,-1.58]$ & 0.582 & 0.77 & 2.51 & 0.81 & 0.90 & 0.70 & 0.64 & 0.37 & 1.72 & 0.49\\ 
$[-1.58,-1.29]$ & 0.825 & 0.69 & 2.24 & 1.05 & 1.08 & 0.11 & 1.03 & 0.68 & 0.45 & 0.74\\ 
$[-1.29,-1.05]$ & 0.799 & 0.55 & 2.24 & 0.15 & 0.53 & 0.43 & 0.67 & 0.57 & 1.35 & 1.27\\ 
$[-1.05,-0.81]$ & 0.580 & 0.94 & 2.44 & 0.76 & 0.75 & 0.72 & 0.90 & 0.92 & 0.94 & 0.93\\ 
$[-0.81,-0.61]$ & 0.364 & 1.83 & 4.66 & 1.31 & 1.34 & 1.78 & 1.02 & 1.78 & 1.79 & 2.06\\ 
$[-0.61,0.00]$ & 0.070 & 2.85 & 6.12 & 0.62 & 2.70 & 2.81 & 2.78 & 1.28 & 1.02 & 1.86\\ 
\hline
$\ln(\lambda_{1.5}^1)$ & $1/\sigma$ $\mathrm{d}\sigma/\mathrm{d}$$\ln(\lambda_{1.5}^1)$ & Stat. (\%) & Tot. (\%) & HFS(jet) (\%) & HFS(other) (\%) & HFS($\phi$) (\%) & Lepton(E) (\%) & Lepton($\phi$) (\%) & Model (\%) & Closure (\%)\\ 
\hline
$[-5.00,-3.99]$ & 0.007 & 2.52 & 17.46 & 6.58 & 7.02 & 8.97 & 6.08 & 4.75 & 4.17 & 6.96\\ 
$[-3.99,-3.28]$ & 0.034 & 1.67 & 8.62 & 3.28 & 2.74 & 4.86 & 3.82 & 2.47 & 1.66 & 2.48\\ 
$[-3.28,-2.78]$ & 0.124 & 1.16 & 3.41 & 0.64 & 0.78 & 1.69 & 1.79 & 1.14 & 0.71 & 1.17\\ 
$[-2.78,-2.32]$ & 0.304 & 0.85 & 3.11 & 0.42 & 0.86 & 0.84 & 0.81 & 0.23 & 2.56 & 0.18\\ 
$[-2.32,-1.92]$ & 0.528 & 0.72 & 1.92 & 0.98 & 0.98 & 0.49 & 0.46 & 0.49 & 0.19 & 0.72\\ 
$[-1.92,-1.57]$ & 0.618 & 0.57 & 1.58 & 0.65 & 0.41 & 0.36 & 1.01 & 0.19 & 0.36 & 0.51\\ 
$[-1.57,-1.21]$ & 0.510 & 0.69 & 1.76 & 0.67 & 0.45 & 0.33 & 0.65 & 0.46 & 0.89 & 0.65\\ 
$[-1.21,-0.91]$ & 0.328 & 1.51 & 3.72 & 0.99 & 0.99 & 1.50 & 0.84 & 1.51 & 1.47 & 1.49\\ 
$[-0.91,0.00]$ & 0.064 & 2.63 & 5.76 & 1.01 & 2.44 & 2.62 & 2.57 & 1.30 & 1.14 & 1.67\\ 
\hline
$\ln(\lambda_2^1)$ & $1/\sigma$ $\mathrm{d}\sigma/\mathrm{d}$$\ln(\lambda_2^1)$ & Stat. (\%) & Tot. (\%) & HFS(jet) (\%) & HFS(other) (\%) & HFS($\phi$) (\%) & Lepton(E) (\%) & Lepton($\phi$) (\%) & Model (\%) & Closure (\%)\\ 
\hline
$[-6.00,-4.61]$ & 0.008 & 2.08 & 14.61 & 5.60 & 7.21 & 7.61 & 5.49 & 4.29 & 0.64 & 4.35\\ 
$[-4.61,-3.76]$ & 0.046 & 1.38 & 6.10 & 0.14 & 0.97 & 2.28 & 1.90 & 1.29 & 1.16 & 4.75\\ 
$[-3.76,-3.09]$ & 0.151 & 0.99 & 3.46 & 0.99 & 0.99 & 0.43 & 0.84 & 0.94 & 2.53 & 0.94\\ 
$[-3.09,-2.55]$ & 0.326 & 0.78 & 3.05 & 0.83 & 1.86 & 0.71 & 0.68 & 0.44 & 1.73 & 0.62\\ 
$[-2.55,-2.06]$ & 0.476 & 0.60 & 1.56 & 0.62 & 0.54 & 0.42 & 0.58 & 0.54 & 0.60 & 0.47\\ 
$[-2.06,-1.58]$ & 0.473 & 0.53 & 1.51 & 0.42 & 0.45 & 0.46 & 0.57 & 0.68 & 0.60 & 0.51\\ 
$[-1.58,-1.15]$ & 0.324 & 1.20 & 2.51 & 0.45 & 0.41 & 1.20 & 0.95 & 1.15 & 0.77 & 0.51\\ 
$[-1.15,0.00]$ & 0.063 & 2.42 & 5.37 & 0.85 & 2.18 & 2.41 & 2.34 & 0.79 & 1.54 & 1.78\\ 
\hline
$\sqrt{\lambda_0^2}$ & $1/\sigma$ $\mathrm{d}\sigma/\mathrm{d}$$\sqrt{\lambda_0^2}$ & Stat. (\%) & Tot. (\%) & HFS(jet) (\%) & HFS(other) (\%) & HFS($\phi$) (\%) & Lepton(E) (\%) & Lepton($\phi$) (\%) & Model (\%) & Closure (\%)\\ 
\hline
$[0.00,0.32]$ & 0.276 & 1.43 & 6.84 & 0.27 & 2.61 & 2.66 & 1.26 & 4.93 & 1.39 & 1.72\\ 
$[0.32,0.39]$ & 3.632 & 0.60 & 1.56 & 0.21 & 0.60 & 0.75 & 0.22 & 0.53 & 0.42 & 0.78\\ 
$[0.39,0.46]$ & 3.949 & 0.32 & 2.04 & 0.26 & 0.42 & 0.32 & 0.28 & 0.29 & 1.39 & 1.28\\ 
$[0.46,0.55]$ & 2.403 & 0.53 & 1.42 & 0.49 & 0.08 & 0.53 & 0.52 & 0.44 & 0.53 & 0.68\\ 
$[0.55,0.64]$ & 1.071 & 0.97 & 3.02 & 0.97 & 0.82 & 0.90 & 0.94 & 0.83 & 2.02 & 0.27\\ 
$[0.64,0.74]$ & 0.439 & 1.44 & 7.29 & 2.52 & 2.47 & 3.91 & 0.53 & 3.45 & 3.04 & 1.38\\ 
$[0.74,0.85]$ & 0.160 & 2.34 & 15.51 & 6.65 & 4.22 & 8.77 & 5.10 & 7.89 & 2.02 & 1.99\\ 
$[0.85,0.94]$ & 0.059 & 3.79 & 34.15 & 15.03 & 11.75 & 17.26 & 15.01 & 15.61 & 2.65 & 3.71\\ 
$[0.94,1.00]$ & 0.025 & 5.14 & 48.64 & 20.24 & 15.04 & 22.85 & 23.17 & 19.35 & 12.84 & 10.28\\ 
\hline
$\tilde{\lambda}_0^1$ & $1/\sigma$ $\mathrm{d}\sigma/\mathrm{d}$$\tilde{\lambda}_0^1$ & Stat. (\%) & Tot. (\%) & HFS(jet) (\%) & HFS(other) (\%) & HFS($\phi$) (\%) & Lepton(E) (\%) & Lepton($\phi$) (\%) & Model (\%) & Closure (\%)\\ 
\hline
$[-1.00,-0.88]$ & 0.013 & 4.29 & 27.48 & 12.72 & 8.62 & 12.57 & 9.19 & 15.17 & 4.01 & 3.45\\ 
$[-0.88,-0.76]$ & 0.037 & 2.93 & 13.46 & 6.40 & 0.70 & 6.31 & 2.72 & 8.88 & 0.30 & 2.25\\ 
$[-0.76,-0.62]$ & 0.088 & 1.97 & 5.36 & 1.03 & 1.60 & 0.58 & 1.94 & 3.70 & 1.30 & 1.31\\ 
$[-0.62,-0.47]$ & 0.204 & 1.34 & 3.60 & 1.29 & 1.76 & 0.82 & 1.60 & 1.07 & 1.34 & 0.45\\ 
$[-0.47,-0.33]$ & 0.432 & 0.84 & 3.90 & 1.22 & 1.65 & 1.91 & 2.36 & 0.72 & 0.76 & 0.07\\ 
$[-0.33,-0.21]$ & 0.771 & 0.59 & 1.56 & 0.33 & 0.58 & 0.47 & 0.76 & 0.43 & 0.56 & 0.58\\ 
$[-0.21,-0.11]$ & 1.147 & 0.52 & 1.34 & 0.48 & 0.43 & 0.50 & 0.52 & 0.52 & 0.20 & 0.52\\ 
$[-0.11,-0.03]$ & 1.473 & 0.46 & 1.62 & 0.34 & 0.15 & 0.26 & 0.40 & 0.46 & 0.21 & 1.34\\ 
$[-0.03,0.03]$ & 1.669 & 0.34 & 1.87 & 0.65 & 1.38 & 0.80 & 0.43 & 0.19 & 0.10 & 0.43\\ 
$[0.03,0.11]$ & 1.501 & 0.39 & 1.06 & 0.22 & 0.35 & 0.32 & 0.28 & 0.33 & 0.62 & 0.36\\ 
$[0.11,0.19]$ & 1.251 & 0.41 & 1.15 & 0.37 & 0.38 & 0.38 & 0.25 & 0.36 & 0.15 & 0.71\\ 
$[0.19,0.29]$ & 0.934 & 0.53 & 1.13 & 0.31 & 0.22 & 0.14 & 0.51 & 0.20 & 0.53 & 0.50\\ 
$[0.29,0.41]$ & 0.600 & 0.78 & 2.03 & 0.55 & 0.82 & 0.67 & 0.71 & 0.76 & 0.65 & 0.76\\ 
$[0.41,0.56]$ & 0.319 & 1.16 & 3.03 & 0.42 & 1.72 & 0.78 & 1.14 & 0.85 & 1.08 & 0.96\\ 
$[0.56,0.70]$ & 0.144 & 1.59 & 5.22 & 1.40 & 1.49 & 1.51 & 0.87 & 1.45 & 1.52 & 3.61\\ 
$[0.70,0.84]$ & 0.062 & 2.27 & 8.02 & 1.85 & 1.81 & 3.78 & 4.96 & 2.06 & 2.27 & 2.04\\ 
$[0.84,1.00]$ & 0.019 & 3.55 & 23.22 & 6.19 & 7.92 & 12.31 & 13.36 & 8.59 & 3.27 & 3.33\\ 
\hline
$\tilde{\lambda}_0^0$ & $1/\sigma$ $\mathrm{d}\sigma/\mathrm{d}$$\tilde{\lambda}_0^0$ & Stat. (\%) & Tot. (\%) & HFS(jet) (\%) & HFS(other) (\%) & HFS($\phi$) (\%) & Lepton(E) (\%) & Lepton($\phi$) (\%) & Model (\%) & Closure (\%)\\ 
\hline
$[1.00,2.00]$ & 0.032 & 1.71 & 8.77 & 2.22 & 1.71 & 4.57 & 1.52 & 2.45 & 3.63 & 4.87\\ 
$[2.00,3.00]$ & 0.071 & 1.06 & 5.83 & 1.06 & 0.99 & 1.52 & 1.02 & 0.72 & 2.65 & 4.45\\ 
$[3.00,4.00]$ & 0.128 & 0.69 & 5.11 & 0.41 & 0.59 & 0.39 & 0.69 & 0.64 & 2.35 & 4.31\\ 
$[4.00,5.00]$ & 0.153 & 0.51 & 2.51 & 0.49 & 0.40 & 0.49 & 0.32 & 0.40 & 0.38 & 2.23\\ 
$[5.00,6.00]$ & 0.162 & 0.39 & 1.23 & 0.28 & 0.23 & 0.42 & 0.25 & 0.65 & 0.65 & 0.36\\ 
$[6.00,7.00]$ & 0.140 & 0.39 & 2.51 & 0.28 & 1.06 & 0.35 & 0.38 & 0.59 & 1.42 & 1.53\\ 
$[7.00,8.00]$ & 0.111 & 0.50 & 3.45 & 0.47 & 0.88 & 0.23 & 0.43 & 0.45 & 1.68 & 2.72\\ 
$[8.00,9.00]$ & 0.078 & 0.70 & 5.65 & 0.68 & 0.69 & 1.02 & 0.58 & 0.18 & 2.64 & 4.69\\ 
$[9.00,10.00]$ & 0.052 & 0.98 & 7.63 & 0.74 & 0.21 & 1.78 & 0.73 & 1.40 & 3.15 & 6.40\\ 
$[10.00,11.00]$ & 0.032 & 1.32 & 8.31 & 0.35 & 3.22 & 2.78 & 0.68 & 3.01 & 0.67 & 6.26\\ 
$[11.00,12.00]$ & 0.020 & 1.71 & 8.87 & 0.74 & 4.21 & 3.53 & 0.93 & 4.63 & 1.14 & 4.62\\ 
$[12.00,13.00]$ & 0.012 & 2.16 & 12.07 & 2.60 & 4.36 & 5.34 & 1.20 & 6.93 & 6.05 & 0.85\\ 
$[13.00,14.00]$ & 0.007 & 2.65 & 14.30 & 2.87 & 6.60 & 5.94 & 0.77 & 7.49 & 6.03 & 4.18\\ 
$[14.00,15.00]$ & 0.004 & 3.17 & 19.73 & 4.34 & 6.80 & 7.61 & 2.39 & 9.98 & 11.90 & 3.10\\ 
\hline
\end{tabular}
\end{adjustbox}

\end{table*}

\begin{table*}[h!]
\centering
\caption{Measured values of the normalized unfolded differential cross sections and uncertainties for each systematic source considered in the range 360 $ < Q^2 < 866 $~GeV$^2$ .}
\label{tab:q2866}
\begin{adjustbox}{width=1\textwidth}
\begin{tabular}{| c | c | c | c || c | c | c | c | c | c | c | c | c |}
\hline
$\ln(\lambda_1^1)$ & $1/\sigma$ $\mathrm{d}\sigma/\mathrm{d}$$\ln(\lambda_1^1)$ & Stat. (\%) & Tot. (\%) & HFS(jet) (\%) & HFS(other) (\%) & HFS($\phi$) (\%) & Lepton(E) (\%) & Lepton($\phi$) (\%) & Model (\%) & Closure (\%)\\ 
\hline
$[-4.00,-3.15]$ & 0.006 & 3.66 & 26.20 & 12.04 & 5.86 & 14.89 & 9.92 & 12.29 & 3.90 & 2.67\\ 
$[-3.15,-2.59]$ & 0.036 & 2.31 & 8.89 & 3.86 & 0.81 & 5.54 & 2.33 & 4.48 & 1.41 & 0.12\\ 
$[-2.59,-2.18]$ & 0.160 & 1.48 & 3.34 & 1.41 & 1.38 & 0.60 & 1.40 & 1.13 & 1.05 & 0.62\\ 
$[-2.18,-1.86]$ & 0.423 & 0.92 & 1.90 & 0.62 & 0.78 & 0.87 & 0.52 & 0.07 & 0.41 & 0.77\\ 
$[-1.86,-1.58]$ & 0.704 & 0.80 & 4.36 & 2.04 & 1.51 & 1.08 & 2.04 & 1.38 & 1.61 & 1.44\\ 
$[-1.58,-1.29]$ & 0.792 & 0.82 & 2.80 & 1.27 & 1.21 & 0.19 & 1.72 & 0.76 & 0.69 & 0.16\\ 
$[-1.29,-1.05]$ & 0.648 & 0.72 & 2.23 & 0.44 & 0.56 & 0.70 & 0.61 & 1.31 & 1.05 & 0.53\\ 
$[-1.05,-0.81]$ & 0.449 & 1.37 & 5.70 & 2.60 & 2.74 & 1.21 & 2.76 & 0.90 & 2.01 & 1.52\\ 
$[-0.81,-0.61]$ & 0.276 & 2.39 & 9.72 & 3.68 & 3.15 & 2.02 & 4.59 & 2.39 & 3.55 & 4.67\\ 
$[-0.61,0.00]$ & 0.047 & 3.34 & 11.48 & 1.35 & 1.48 & 2.99 & 3.84 & 2.42 & 3.95 & 8.45\\ 
\hline
$\ln(\lambda_{1.5}^1)$ & $1/\sigma$ $\mathrm{d}\sigma/\mathrm{d}$$\ln(\lambda_{1.5}^1)$ & Stat. (\%) & Tot. (\%) & HFS(jet) (\%) & HFS(other) (\%) & HFS($\phi$) (\%) & Lepton(E) (\%) & Lepton($\phi$) (\%) & Model (\%) & Closure (\%)\\ 
\hline
$[-5.00,-3.99]$ & 0.009 & 2.80 & 14.70 & 6.54 & 3.29 & 8.85 & 4.64 & 7.21 & 0.66 & 1.53\\ 
$[-3.99,-3.28]$ & 0.052 & 1.80 & 5.10 & 0.72 & 1.67 & 2.24 & 1.21 & 0.64 & 1.98 & 2.95\\ 
$[-3.28,-2.78]$ & 0.184 & 1.17 & 2.82 & 0.79 & 0.58 & 1.17 & 1.04 & 1.03 & 0.94 & 1.10\\ 
$[-2.78,-2.32]$ & 0.400 & 0.82 & 2.94 & 1.33 & 1.53 & 0.36 & 1.02 & 0.85 & 0.80 & 1.14\\ 
$[-2.32,-1.92]$ & 0.570 & 0.78 & 3.12 & 1.65 & 1.18 & 0.69 & 1.89 & 0.62 & 0.76 & 0.16\\ 
$[-1.92,-1.57]$ & 0.550 & 0.70 & 1.75 & 0.50 & 0.62 & 0.65 & 0.59 & 0.46 & 0.70 & 0.69\\ 
$[-1.57,-1.21]$ & 0.403 & 1.01 & 3.89 & 1.88 & 1.92 & 0.82 & 1.78 & 1.33 & 0.99 & 0.56\\ 
$[-1.21,-0.91]$ & 0.247 & 2.07 & 8.27 & 3.52 & 3.39 & 1.72 & 4.13 & 2.02 & 2.49 & 3.15\\ 
$[-0.91,0.00]$ & 0.042 & 3.15 & 11.23 & 1.71 & 0.63 & 2.80 & 3.91 & 2.37 & 4.28 & 8.11\\ 
\hline
$\ln(\lambda_2^1)$ & $1/\sigma$ $\mathrm{d}\sigma/\mathrm{d}$$\ln(\lambda_2^1)$ & Stat. (\%) & Tot. (\%) & HFS(jet) (\%) & HFS(other) (\%) & HFS($\phi$) (\%) & Lepton(E) (\%) & Lepton($\phi$) (\%) & Model (\%) & Closure (\%)\\ 
\hline
$[-6.00,-4.61]$ & 0.012 & 2.24 & 8.58 & 2.64 & 1.56 & 4.72 & 0.94 & 3.37 & 2.71 & 4.18\\ 
$[-4.61,-3.76]$ & 0.067 & 1.45 & 5.33 & 1.45 & 1.42 & 0.98 & 1.44 & 1.43 & 2.22 & 3.48\\ 
$[-3.76,-3.09]$ & 0.207 & 0.96 & 2.85 & 1.09 & 1.66 & 0.77 & 0.43 & 0.80 & 0.69 & 1.17\\ 
$[-3.09,-2.55]$ & 0.388 & 0.75 & 2.56 & 1.31 & 1.41 & 0.26 & 1.27 & 0.71 & 0.25 & 0.12\\ 
$[-2.55,-2.06]$ & 0.477 & 0.66 & 2.14 & 1.00 & 0.47 & 0.04 & 1.43 & 0.66 & 0.44 & 0.49\\ 
$[-2.06,-1.58]$ & 0.399 & 0.73 & 2.16 & 0.81 & 0.74 & 0.63 & 0.32 & 1.50 & 0.23 & 0.36\\ 
$[-1.58,-1.15]$ & 0.245 & 1.71 & 6.83 & 3.11 & 2.73 & 1.38 & 3.40 & 1.41 & 2.15 & 2.56\\ 
$[-1.15,0.00]$ & 0.042 & 2.96 & 10.01 & 1.77 & 1.64 & 2.64 & 3.76 & 2.32 & 3.35 & 6.92\\ 
\hline
$\sqrt{\lambda_0^2}$ & $1/\sigma$ $\mathrm{d}\sigma/\mathrm{d}$$\sqrt{\lambda_0^2}$ & Stat. (\%) & Tot. (\%) & HFS(jet) (\%) & HFS(other) (\%) & HFS($\phi$) (\%) & Lepton(E) (\%) & Lepton($\phi$) (\%) & Model (\%) & Closure (\%)\\ 
\hline
$[0.00,0.32]$ & 0.334 & 1.49 & 7.95 & 1.46 & 1.01 & 1.96 & 1.49 & 4.32 & 5.18 & 2.51\\ 
$[0.32,0.39]$ & 3.847 & 0.77 & 2.79 & 0.30 & 1.27 & 0.60 & 0.57 & 1.02 & 1.59 & 1.11\\ 
$[0.39,0.46]$ & 3.870 & 0.47 & 3.31 & 0.59 & 0.46 & 0.22 & 0.41 & 0.35 & 2.76 & 1.49\\ 
$[0.46,0.55]$ & 2.242 & 0.72 & 2.73 & 0.72 & 0.37 & 0.67 & 0.62 & 0.42 & 1.86 & 1.35\\ 
$[0.55,0.64]$ & 0.976 & 1.43 & 3.35 & 1.40 & 0.61 & 1.24 & 0.93 & 1.44 & 1.04 & 1.13\\ 
$[0.64,0.74]$ & 0.402 & 2.10 & 8.88 & 2.68 & 2.59 & 4.39 & 2.09 & 5.65 & 1.62 & 1.50\\ 
$[0.74,0.85]$ & 0.149 & 3.30 & 18.46 & 7.80 & 2.46 & 10.30 & 3.85 & 11.20 & 3.19 & 2.55\\ 
$[0.85,0.94]$ & 0.058 & 5.18 & 35.41 & 16.05 & 6.62 & 19.16 & 14.35 & 17.76 & 3.30 & 5.12\\ 
$[0.94,1.00]$ & 0.025 & 6.93 & 56.26 & 24.31 & 11.43 & 28.18 & 26.18 & 25.75 & 13.08 & 9.04\\ 
\hline
$\tilde{\lambda}_0^1$ & $1/\sigma$ $\mathrm{d}\sigma/\mathrm{d}$$\tilde{\lambda}_0^1$ & Stat. (\%) & Tot. (\%) & HFS(jet) (\%) & HFS(other) (\%) & HFS($\phi$) (\%) & Lepton(E) (\%) & Lepton($\phi$) (\%) & Model (\%) & Closure (\%)\\ 
\hline
$[-1.00,-0.88]$ & 0.012 & 6.40 & 36.35 & 19.74 & 5.44 & 15.96 & 14.10 & 18.59 & 5.55 & 5.60\\ 
$[-0.88,-0.76]$ & 0.034 & 4.39 & 14.90 & 7.00 & 2.66 & 3.76 & 3.77 & 8.75 & 3.55 & 5.38\\ 
$[-0.76,-0.62]$ & 0.083 & 2.79 & 6.82 & 2.52 & 2.73 & 2.57 & 2.44 & 1.83 & 1.46 & 2.61\\ 
$[-0.62,-0.47]$ & 0.189 & 1.62 & 6.08 & 0.92 & 0.60 & 3.06 & 4.31 & 1.62 & 0.80 & 1.39\\ 
$[-0.47,-0.33]$ & 0.412 & 0.92 & 5.68 & 1.47 & 0.81 & 2.89 & 3.21 & 0.98 & 2.96 & 0.52\\ 
$[-0.33,-0.21]$ & 0.742 & 0.65 & 2.90 & 0.99 & 0.49 & 1.45 & 1.76 & 0.92 & 0.54 & 0.65\\ 
$[-0.21,-0.11]$ & 1.139 & 0.56 & 1.74 & 0.50 & 0.14 & 0.52 & 0.34 & 0.61 & 0.39 & 1.24\\ 
$[-0.11,-0.03]$ & 1.476 & 0.50 & 1.77 & 0.28 & 1.11 & 0.37 & 0.39 & 0.49 & 0.16 & 1.02\\ 
$[-0.03,0.03]$ & 1.688 & 0.39 & 2.37 & 0.73 & 0.26 & 1.12 & 0.62 & 0.62 & 1.64 & 0.39\\ 
$[0.03,0.11]$ & 1.547 & 0.41 & 1.56 & 0.51 & 0.27 & 0.74 & 0.74 & 0.15 & 0.82 & 0.37\\ 
$[0.11,0.19]$ & 1.280 & 0.46 & 1.44 & 0.32 & 0.42 & 0.33 & 0.54 & 0.36 & 0.94 & 0.41\\ 
$[0.19,0.29]$ & 0.954 & 0.60 & 1.43 & 0.60 & 0.50 & 0.60 & 0.17 & 0.50 & 0.30 & 0.60\\ 
$[0.29,0.41]$ & 0.608 & 0.87 & 2.45 & 0.25 & 1.32 & 0.17 & 0.84 & 0.51 & 0.83 & 1.32\\ 
$[0.41,0.56]$ & 0.321 & 1.27 & 3.55 & 1.14 & 1.77 & 1.27 & 0.17 & 1.25 & 0.74 & 1.68\\ 
$[0.56,0.70]$ & 0.144 & 1.92 & 4.78 & 1.67 & 1.70 & 1.34 & 0.93 & 1.28 & 1.83 & 2.42\\ 
$[0.70,0.84]$ & 0.061 & 2.94 & 11.96 & 1.81 & 3.82 & 6.46 & 5.16 & 6.03 & 2.86 & 1.90\\ 
$[0.84,1.00]$ & 0.020 & 4.60 & 27.36 & 7.15 & 6.11 & 16.75 & 13.17 & 12.42 & 3.33 & 4.39\\ 
\hline
$\tilde{\lambda}_0^0$ & $1/\sigma$ $\mathrm{d}\sigma/\mathrm{d}$$\tilde{\lambda}_0^0$ & Stat. (\%) & Tot. (\%) & HFS(jet) (\%) & HFS(other) (\%) & HFS($\phi$) (\%) & Lepton(E) (\%) & Lepton($\phi$) (\%) & Model (\%) & Closure (\%)\\ 
\hline
$[1.00,2.00]$ & 0.025 & 2.38 & 8.55 & 1.27 & 1.55 & 4.67 & 2.32 & 3.44 & 4.13 & 2.70\\ 
$[2.00,3.00]$ & 0.057 & 1.54 & 5.82 & 1.52 & 1.23 & 1.59 & 1.18 & 1.95 & 3.18 & 3.13\\ 
$[3.00,4.00]$ & 0.108 & 1.03 & 5.05 & 1.04 & 1.02 & 1.02 & 0.73 & 0.83 & 3.00 & 3.32\\ 
$[4.00,5.00]$ & 0.137 & 0.69 & 3.81 & 1.10 & 0.57 & 0.49 & 0.83 & 0.67 & 2.41 & 2.31\\ 
$[5.00,6.00]$ & 0.155 & 0.45 & 1.50 & 0.25 & 0.21 & 0.43 & 0.37 & 0.45 & 0.98 & 0.67\\ 
$[6.00,7.00]$ & 0.142 & 0.42 & 1.11 & 0.42 & 0.42 & 0.41 & 0.41 & 0.42 & 0.13 & 0.41\\ 
$[7.00,8.00]$ & 0.121 & 0.58 & 2.29 & 0.63 & 0.58 & 0.52 & 0.74 & 0.40 & 0.99 & 1.50\\ 
$[8.00,9.00]$ & 0.090 & 0.78 & 3.51 & 0.29 & 1.19 & 0.76 & 0.55 & 0.77 & 1.98 & 2.20\\ 
$[9.00,10.00]$ & 0.063 & 1.01 & 7.86 & 0.91 & 0.99 & 1.00 & 1.00 & 0.97 & 5.39 & 5.19\\ 
$[10.00,11.00]$ & 0.042 & 1.27 & 8.34 & 0.51 & 1.25 & 0.88 & 0.75 & 0.77 & 5.74 & 5.59\\ 
$[11.00,12.00]$ & 0.027 & 1.56 & 10.70 & 1.08 & 1.51 & 0.77 & 1.28 & 2.36 & 7.04 & 7.16\\ 
$[12.00,13.00]$ & 0.017 & 1.88 & 10.20 & 1.77 & 1.59 & 1.44 & 1.85 & 3.81 & 5.20 & 6.91\\ 
$[13.00,14.00]$ & 0.010 & 2.20 & 10.86 & 2.20 & 1.16 & 2.36 & 2.19 & 5.04 & 4.28 & 7.27\\ 
$[14.00,15.00]$ & 0.006 & 2.53 & 9.72 & 2.26 & 3.92 & 4.04 & 2.08 & 6.59 & 1.31 & 1.29\\ 
\hline
\end{tabular}
\end{adjustbox}

\end{table*}

\begin{table*}[h!]
\centering
\caption{Measured values of the normalized unfolded differential cross sections and uncertainties for each systematic source considered in the range 866 $ < Q^2 < 2080 $~GeV$^2$ .}
\label{tab:q22080}
\begin{adjustbox}{width=1\textwidth}
\begin{tabular}{| c | c | c | c || c | c | c | c | c | c | c | c | c |}
\hline
$\ln(\lambda_1^1)$ & $1/\sigma$ $\mathrm{d}\sigma/\mathrm{d}$$\ln(\lambda_1^1)$ & Stat. (\%) & Tot. (\%) & HFS(jet) (\%) & HFS(other) (\%) & HFS($\phi$) (\%) & Lepton(E) (\%) & Lepton($\phi$) (\%) & Model (\%) & Closure (\%)\\ 
\hline
$[-4.00,-3.15]$ & 0.008 & 3.17 & 12.73 & 6.62 & 3.15 & 3.86 & 2.56 & 8.49 & 1.32 & 1.75\\ 
$[-3.15,-2.59]$ & 0.069 & 2.29 & 5.70 & 1.77 & 1.00 & 2.12 & 2.19 & 2.34 & 2.17 & 1.91\\ 
$[-2.59,-2.18]$ & 0.282 & 1.54 & 3.75 & 1.53 & 0.23 & 1.04 & 1.12 & 1.54 & 1.53 & 1.49\\ 
$[-2.18,-1.86]$ & 0.564 & 0.97 & 3.40 & 1.43 & 0.82 & 1.00 & 2.19 & 0.78 & 0.74 & 0.97\\ 
$[-1.86,-1.58]$ & 0.694 & 0.93 & 6.42 & 3.12 & 2.45 & 1.45 & 3.80 & 2.24 & 1.76 & 0.21\\ 
$[-1.58,-1.29]$ & 0.647 & 1.01 & 4.69 & 2.28 & 2.44 & 0.87 & 2.38 & 0.45 & 1.62 & 0.76\\ 
$[-1.29,-1.05]$ & 0.518 & 1.14 & 3.58 & 0.24 & 0.42 & 1.01 & 1.82 & 1.08 & 0.37 & 2.37\\ 
$[-1.05,-0.81]$ & 0.366 & 1.89 & 12.67 & 6.03 & 4.70 & 3.28 & 7.96 & 4.13 & 1.95 & 1.89\\ 
$[-0.81,-0.61]$ & 0.223 & 2.71 & 18.11 & 8.65 & 4.46 & 2.77 & 11.39 & 1.34 & 8.08 & 4.63\\ 
$[-0.61,0.00]$ & 0.033 & 3.37 & 25.76 & 7.48 & 3.33 & 2.06 & 12.11 & 3.16 & 15.70 & 13.34\\ 
\hline
$\ln(\lambda_{1.5}^1)$ & $1/\sigma$ $\mathrm{d}\sigma/\mathrm{d}$$\ln(\lambda_{1.5}^1)$ & Stat. (\%) & Tot. (\%) & HFS(jet) (\%) & HFS(other) (\%) & HFS($\phi$) (\%) & Lepton(E) (\%) & Lepton($\phi$) (\%) & Model (\%) & Closure (\%)\\ 
\hline
$[-5.00,-3.99]$ & 0.015 & 2.57 & 7.51 & 3.85 & 1.06 & 1.11 & 2.39 & 4.88 & 1.16 & 1.35\\ 
$[-3.99,-3.28]$ & 0.093 & 1.91 & 4.98 & 1.84 & 1.49 & 1.58 & 1.65 & 1.73 & 1.91 & 1.91\\ 
$[-3.28,-2.78]$ & 0.280 & 1.27 & 3.09 & 0.84 & 1.12 & 0.61 & 1.16 & 1.03 & 1.27 & 1.27\\ 
$[-2.78,-2.32]$ & 0.466 & 0.86 & 4.28 & 2.05 & 1.25 & 1.13 & 2.83 & 1.32 & 0.24 & 0.86\\ 
$[-2.32,-1.92]$ & 0.518 & 0.93 & 6.26 & 2.99 & 3.22 & 0.87 & 3.45 & 1.69 & 1.87 & 0.08\\ 
$[-1.92,-1.57]$ & 0.445 & 1.00 & 2.87 & 0.86 & 0.96 & 0.56 & 1.00 & 0.90 & 1.19 & 1.43\\ 
$[-1.57,-1.21]$ & 0.323 & 1.57 & 9.10 & 4.05 & 3.27 & 2.55 & 5.71 & 3.42 & 1.28 & 0.82\\ 
$[-1.21,-0.91]$ & 0.196 & 2.50 & 17.04 & 8.48 & 4.90 & 3.61 & 10.98 & 3.00 & 6.18 & 2.71\\ 
$[-0.91,0.00]$ & 0.029 & 3.24 & 24.81 & 7.17 & 2.94 & 2.01 & 11.59 & 3.08 & 15.29 & 12.79\\ 
\hline
$\ln(\lambda_2^1)$ & $1/\sigma$ $\mathrm{d}\sigma/\mathrm{d}$$\ln(\lambda_2^1)$ & Stat. (\%) & Tot. (\%) & HFS(jet) (\%) & HFS(other) (\%) & HFS($\phi$) (\%) & Lepton(E) (\%) & Lepton($\phi$) (\%) & Model (\%) & Closure (\%)\\ 
\hline
$[-6.00,-4.61]$ & 0.019 & 2.19 & 5.17 & 1.87 & 1.60 & 2.01 & 2.18 & 2.38 & 0.92 & 0.74\\ 
$[-4.61,-3.76]$ & 0.105 & 1.63 & 3.69 & 1.52 & 1.38 & 0.39 & 0.57 & 1.61 & 1.50 & 1.20\\ 
$[-3.76,-3.09]$ & 0.275 & 1.05 & 3.56 & 1.28 & 1.04 & 1.28 & 2.26 & 0.88 & 0.55 & 1.00\\ 
$[-3.09,-2.55]$ & 0.414 & 0.78 & 4.96 & 2.35 & 2.14 & 1.00 & 3.04 & 1.30 & 1.16 & 0.75\\ 
$[-2.55,-2.06]$ & 0.423 & 0.84 & 3.24 & 1.68 & 1.52 & 0.81 & 1.79 & 0.16 & 0.21 & 0.84\\ 
$[-2.06,-1.58]$ & 0.323 & 1.23 & 5.28 & 1.83 & 1.70 & 1.71 & 3.13 & 2.17 & 1.17 & 1.19\\ 
$[-1.58,-1.15]$ & 0.192 & 2.25 & 14.74 & 7.50 & 4.34 & 3.66 & 9.80 & 3.60 & 3.49 & 1.57\\ 
$[-1.15,0.00]$ & 0.028 & 3.11 & 23.68 & 6.71 & 2.45 & 1.90 & 11.05 & 2.99 & 14.73 & 12.19\\ 
\hline
$\sqrt{\lambda_0^2}$ & $1/\sigma$ $\mathrm{d}\sigma/\mathrm{d}$$\sqrt{\lambda_0^2}$ & Stat. (\%) & Tot. (\%) & HFS(jet) (\%) & HFS(other) (\%) & HFS($\phi$) (\%) & Lepton(E) (\%) & Lepton($\phi$) (\%) & Model (\%) & Closure (\%)\\ 
\hline
$[0.00,0.32]$ & 0.442 & 1.70 & 7.05 & 0.40 & 1.70 & 1.46 & 1.08 & 2.08 & 5.66 & 2.05\\ 
$[0.32,0.39]$ & 4.084 & 0.81 & 2.13 & 0.91 & 0.27 & 0.55 & 0.36 & 1.24 & 0.80 & 0.60\\ 
$[0.39,0.46]$ & 3.667 & 0.53 & 3.84 & 1.40 & 0.79 & 0.30 & 0.27 & 0.30 & 3.17 & 1.27\\ 
$[0.46,0.55]$ & 2.034 & 0.92 & 3.28 & 0.90 & 0.66 & 0.92 & 0.82 & 0.50 & 2.05 & 1.64\\ 
$[0.55,0.64]$ & 0.865 & 1.63 & 4.74 & 1.27 & 1.75 & 0.58 & 1.40 & 2.99 & 1.29 & 1.50\\ 
$[0.64,0.74]$ & 0.355 & 2.33 & 9.31 & 3.79 & 2.28 & 3.26 & 2.33 & 6.30 & 1.44 & 1.93\\ 
$[0.74,0.85]$ & 0.130 & 3.58 & 17.10 & 7.45 & 3.57 & 6.31 & 3.54 & 11.13 & 5.11 & 3.02\\ 
$[0.85,0.94]$ & 0.047 & 5.40 & 25.01 & 13.09 & 4.73 & 11.20 & 2.97 & 15.86 & 2.13 & 3.50\\ 
$[0.94,1.00]$ & 0.020 & 7.46 & 50.24 & 24.52 & 18.27 & 23.18 & 14.86 & 26.30 & 5.82 & 7.01\\ 
\hline
$\tilde{\lambda}_0^1$ & $1/\sigma$ $\mathrm{d}\sigma/\mathrm{d}$$\tilde{\lambda}_0^1$ & Stat. (\%) & Tot. (\%) & HFS(jet) (\%) & HFS(other) (\%) & HFS($\phi$) (\%) & Lepton(E) (\%) & Lepton($\phi$) (\%) & Model (\%) & Closure (\%)\\ 
\hline
$[-1.00,-0.88]$ & 0.009 & 7.62 & 30.62 & 20.47 & 6.72 & 11.66 & 4.92 & 13.67 & 5.56 & 6.11\\ 
$[-0.88,-0.76]$ & 0.026 & 5.02 & 12.57 & 6.63 & 5.01 & 4.38 & 2.95 & 5.38 & 2.45 & 0.89\\ 
$[-0.76,-0.62]$ & 0.065 & 3.22 & 8.18 & 2.64 & 3.22 & 2.11 & 4.84 & 1.24 & 0.86 & 3.00\\ 
$[-0.62,-0.47]$ & 0.156 & 1.93 & 6.91 & 1.80 & 1.15 & 3.14 & 3.88 & 1.92 & 1.19 & 3.08\\ 
$[-0.47,-0.33]$ & 0.353 & 1.28 & 5.49 & 1.40 & 1.34 & 3.30 & 3.11 & 1.53 & 0.70 & 1.16\\ 
$[-0.33,-0.21]$ & 0.683 & 0.92 & 4.00 & 1.50 & 0.43 & 2.05 & 2.11 & 1.70 & 0.91 & 0.63\\ 
$[-0.21,-0.11]$ & 1.098 & 0.70 & 1.64 & 0.51 & 0.66 & 0.56 & 0.38 & 0.62 & 0.43 & 0.68\\ 
$[-0.11,-0.03]$ & 1.483 & 0.52 & 1.60 & 0.47 & 0.81 & 0.37 & 0.14 & 0.62 & 0.78 & 0.49\\ 
$[-0.03,0.03]$ & 1.749 & 0.41 & 1.36 & 0.40 & 0.44 & 0.35 & 0.41 & 0.32 & 0.88 & 0.41\\ 
$[0.03,0.11]$ & 1.624 & 0.40 & 2.27 & 0.58 & 0.81 & 0.96 & 0.99 & 0.19 & 1.37 & 0.40\\ 
$[0.11,0.19]$ & 1.355 & 0.50 & 1.61 & 0.02 & 0.48 & 0.65 & 0.92 & 0.42 & 0.66 & 0.48\\ 
$[0.19,0.29]$ & 1.006 & 0.71 & 2.58 & 0.34 & 0.53 & 0.36 & 1.17 & 0.55 & 1.51 & 1.29\\ 
$[0.29,0.41]$ & 0.637 & 1.06 & 3.04 & 1.06 & 0.90 & 1.03 & 0.45 & 1.02 & 1.78 & 0.84\\ 
$[0.41,0.56]$ & 0.336 & 1.60 & 5.09 & 1.56 & 0.75 & 1.30 & 0.36 & 1.13 & 4.11 & 0.55\\ 
$[0.56,0.70]$ & 0.151 & 2.41 & 5.74 & 1.68 & 2.09 & 1.38 & 0.97 & 2.96 & 1.76 & 2.28\\ 
$[0.70,0.84]$ & 0.061 & 3.53 & 10.31 & 2.42 & 2.63 & 3.96 & 2.06 & 6.62 & 2.34 & 3.44\\ 
$[0.84,1.00]$ & 0.019 & 5.29 & 23.38 & 8.50 & 6.77 & 10.03 & 2.90 & 15.40 & 5.16 & 5.29\\ 
\hline
$\tilde{\lambda}_0^0$ & $1/\sigma$ $\mathrm{d}\sigma/\mathrm{d}$$\tilde{\lambda}_0^0$ & Stat. (\%) & Tot. (\%) & HFS(jet) (\%) & HFS(other) (\%) & HFS($\phi$) (\%) & Lepton(E) (\%) & Lepton($\phi$) (\%) & Model (\%) & Closure (\%)\\ 
\hline
$[1.00,2.00]$ & 0.017 & 2.60 & 7.82 & 1.97 & 2.55 & 1.73 & 1.66 & 5.41 & 2.44 & 1.74\\ 
$[2.00,3.00]$ & 0.040 & 1.93 & 5.65 & 1.25 & 1.47 & 1.55 & 0.68 & 3.82 & 1.87 & 1.89\\ 
$[3.00,4.00]$ & 0.080 & 1.43 & 3.82 & 1.38 & 1.27 & 1.23 & 1.20 & 0.53 & 1.37 & 1.97\\ 
$[4.00,5.00]$ & 0.109 & 1.05 & 2.89 & 0.79 & 0.86 & 0.84 & 1.11 & 1.05 & 0.82 & 1.47\\ 
$[5.00,6.00]$ & 0.138 & 0.73 & 2.57 & 1.31 & 1.41 & 1.00 & 0.63 & 0.59 & 0.61 & 0.50\\ 
$[6.00,7.00]$ & 0.138 & 0.57 & 3.96 & 0.88 & 0.66 & 0.43 & 0.43 & 0.54 & 3.00 & 2.11\\ 
$[7.00,8.00]$ & 0.131 & 0.61 & 2.06 & 0.93 & 0.56 & 0.52 & 0.35 & 0.60 & 1.28 & 0.54\\ 
$[8.00,9.00]$ & 0.106 & 0.75 & 2.48 & 0.67 & 0.70 & 0.66 & 0.73 & 0.02 & 0.74 & 1.76\\ 
$[9.00,10.00]$ & 0.083 & 1.02 & 2.72 & 0.46 & 0.71 & 0.55 & 0.63 & 1.00 & 1.01 & 1.71\\ 
$[10.00,11.00]$ & 0.059 & 1.40 & 4.37 & 0.81 & 1.35 & 0.90 & 1.23 & 1.21 & 1.14 & 3.10\\ 
$[11.00,12.00]$ & 0.042 & 1.84 & 6.56 & 3.36 & 2.10 & 1.45 & 3.87 & 1.71 & 0.80 & 1.82\\ 
$[12.00,13.00]$ & 0.027 & 2.29 & 10.73 & 3.15 & 1.93 & 1.91 & 3.34 & 2.04 & 6.06 & 6.38\\ 
$[13.00,14.00]$ & 0.018 & 2.74 & 9.16 & 4.24 & 2.76 & 2.18 & 4.62 & 2.56 & 3.14 & 2.89\\ 
$[14.00,15.00]$ & 0.012 & 3.13 & 17.35 & 4.30 & 2.84 & 2.94 & 5.12 & 2.04 & 10.96 & 10.28\\ 
\hline
\end{tabular}

\end{adjustbox}

\end{table*}

\begin{table*}[h!]
\centering
\caption{Measured values of the normalized unfolded differential cross sections and uncertainties for each systematic source considered in the range $2080 < Q^2 < 5000 $~GeV$^2$ .}
\label{tab:q25000}
\begin{adjustbox}{width=1\textwidth}
\begin{tabular}{| c | c | c | c || c | c | c | c | c | c | c | c | c |}
\hline
$\ln(\lambda_1^1)$ & $1/\sigma$ $\mathrm{d}\sigma/\mathrm{d}$$\ln(\lambda_1^1)$ & Stat. (\%) & Tot. (\%) & HFS(jet) (\%) & HFS(other) (\%) & HFS($\phi$) (\%) & Lepton(E) (\%) & Lepton($\phi$) (\%) & Model (\%) & Closure (\%)\\ 
\hline
$[-4.00,-3.15]$ & 0.017 & 3.45 & 13.80 & 2.86 & 5.17 & 2.29 & 2.84 & 3.00 & 6.13 & 9.14\\ 
$[-3.15,-2.59]$ & 0.155 & 2.41 & 8.28 & 2.25 & 0.96 & 1.16 & 2.20 & 2.01 & 3.17 & 6.03\\ 
$[-2.59,-2.18]$ & 0.435 & 1.54 & 4.67 & 1.19 & 1.53 & 1.11 & 1.42 & 1.53 & 0.75 & 3.09\\ 
$[-2.18,-1.86]$ & 0.572 & 1.15 & 5.07 & 2.13 & 2.55 & 0.96 & 3.19 & 0.71 & 0.65 & 1.15\\ 
$[-1.86,-1.58]$ & 0.577 & 1.10 & 11.81 & 3.96 & 2.66 & 1.44 & 4.26 & 2.98 & 5.89 & 7.18\\ 
$[-1.58,-1.29]$ & 0.521 & 1.24 & 11.20 & 1.10 & 0.67 & 0.90 & 0.61 & 0.42 & 5.74 & 9.37\\ 
$[-1.29,-1.05]$ & 0.421 & 1.84 & 9.32 & 1.92 & 2.43 & 2.90 & 4.27 & 1.63 & 0.87 & 6.62\\ 
$[-1.05,-0.81]$ & 0.302 & 2.80 & 15.61 & 6.72 & 3.85 & 4.05 & 9.79 & 3.92 & 6.90 & 0.93\\ 
$[-0.81,-0.61]$ & 0.180 & 3.78 & 27.77 & 11.51 & 2.51 & 4.50 & 14.21 & 3.83 & 16.64 & 10.21\\ 
$[-0.61,0.00]$ & 0.024 & 4.40 & 35.10 & 12.68 & 4.40 & 3.44 & 16.23 & 3.47 & 21.64 & 16.65\\ 
\hline
$\ln(\lambda_{1.5}^1)$ & $1/\sigma$ $\mathrm{d}\sigma/\mathrm{d}$$\ln(\lambda_{1.5}^1)$ & Stat. (\%) & Tot. (\%) & HFS(jet) (\%) & HFS(other) (\%) & HFS($\phi$) (\%) & Lepton(E) (\%) & Lepton($\phi$) (\%) & Model (\%) & Closure (\%)\\ 
\hline
$[-5.00,-3.99]$ & 0.031 & 2.86 & 7.77 & 2.78 & 1.62 & 1.08 & 2.11 & 2.47 & 1.67 & 5.24\\ 
$[-3.99,-3.28]$ & 0.173 & 2.01 & 5.50 & 1.98 & 1.59 & 0.62 & 1.29 & 1.91 & 0.29 & 3.73\\ 
$[-3.28,-2.78]$ & 0.371 & 1.35 & 3.94 & 0.66 & 1.14 & 1.31 & 2.13 & 1.19 & 1.18 & 1.70\\ 
$[-2.78,-2.32]$ & 0.444 & 1.06 & 6.40 & 3.01 & 2.85 & 1.42 & 3.76 & 1.74 & 1.81 & 0.38\\ 
$[-2.32,-1.92]$ & 0.429 & 1.08 & 11.82 & 3.01 & 1.72 & 0.86 & 2.88 & 2.17 & 6.25 & 8.59\\ 
$[-1.92,-1.57]$ & 0.359 & 1.50 & 8.89 & 1.01 & 1.03 & 2.40 & 1.86 & 1.07 & 2.91 & 7.48\\ 
$[-1.57,-1.21]$ & 0.261 & 2.48 & 13.07 & 5.20 & 5.10 & 3.98 & 8.31 & 3.77 & 2.70 & 2.30\\ 
$[-1.21,-0.91]$ & 0.155 & 3.59 & 25.45 & 10.12 & 0.45 & 3.92 & 13.03 & 3.39 & 15.70 & 9.43\\ 
$[-0.91,0.00]$ & 0.020 & 4.30 & 33.92 & 12.06 & 4.25 & 3.97 & 15.71 & 3.69 & 20.90 & 15.98\\ 
\hline
$\ln(\lambda_2^1)$ & $1/\sigma$ $\mathrm{d}\sigma/\mathrm{d}$$\ln(\lambda_2^1)$ & Stat. (\%) & Tot. (\%) & HFS(jet) (\%) & HFS(other) (\%) & HFS($\phi$) (\%) & Lepton(E) (\%) & Lepton($\phi$) (\%) & Model (\%) & Closure (\%)\\ 
\hline
$[-6.00,-4.61]$ & 0.034 & 2.49 & 5.25 & 2.35 & 1.09 & 0.75 & 1.72 & 1.99 & 1.83 & 1.98\\ 
$[-4.61,-3.76]$ & 0.164 & 1.76 & 4.80 & 1.72 & 1.54 & 1.10 & 1.19 & 1.76 & 1.72 & 2.43\\ 
$[-3.76,-3.09]$ & 0.331 & 1.19 & 3.85 & 1.58 & 0.30 & 1.42 & 2.68 & 0.46 & 1.10 & 0.43\\ 
$[-3.09,-2.55]$ & 0.387 & 0.94 & 6.57 & 2.92 & 2.95 & 0.93 & 3.41 & 1.87 & 2.38 & 1.83\\ 
$[-2.55,-2.06]$ & 0.354 & 1.11 & 10.15 & 1.16 & 1.10 & 0.93 & 0.71 & 0.39 & 5.36 & 8.31\\ 
$[-2.06,-1.58]$ & 0.256 & 2.05 & 9.49 & 3.57 & 2.46 & 3.78 & 6.25 & 3.04 & 1.70 & 1.23\\ 
$[-1.58,-1.15]$ & 0.148 & 3.34 & 21.48 & 8.40 & 3.67 & 3.45 & 11.36 & 2.84 & 12.83 & 7.25\\ 
$[-1.15,0.00]$ & 0.019 & 4.21 & 33.32 & 12.50 & 4.21 & 5.42 & 16.01 & 2.96 & 19.92 & 15.07\\ 
\hline
$\sqrt{\lambda_0^2}$ & $1/\sigma$ $\mathrm{d}\sigma/\mathrm{d}$$\sqrt{\lambda_0^2}$ & Stat. (\%) & Tot. (\%) & HFS(jet) (\%) & HFS(other) (\%) & HFS($\phi$) (\%) & Lepton(E) (\%) & Lepton($\phi$) (\%) & Model (\%) & Closure (\%)\\ 
\hline
$[0.00,0.32]$ & 0.569 & 2.28 & 5.37 & 0.14 & 2.24 & 2.15 & 1.64 & 1.99 & 1.51 & 2.23\\ 
$[0.32,0.39]$ & 4.111 & 0.93 & 3.18 & 0.31 & 0.65 & 0.92 & 0.70 & 0.59 & 1.40 & 2.26\\ 
$[0.39,0.46]$ & 3.413 & 0.64 & 1.46 & 0.40 & 0.59 & 0.64 & 0.43 & 0.54 & 0.49 & 0.30\\ 
$[0.46,0.55]$ & 1.888 & 1.33 & 3.73 & 1.28 & 1.82 & 1.29 & 1.29 & 0.97 & 1.11 & 1.32\\ 
$[0.55,0.64]$ & 0.801 & 2.36 & 5.81 & 2.33 & 1.92 & 2.36 & 2.33 & 1.27 & 2.29 & 1.10\\ 
$[0.64,0.74]$ & 0.320 & 3.39 & 10.89 & 3.27 & 3.32 & 3.37 & 2.63 & 2.82 & 4.26 & 6.41\\ 
$[0.74,0.85]$ & 0.117 & 4.68 & 10.32 & 4.26 & 4.54 & 4.62 & 2.86 & 2.41 & 2.49 & 2.08\\ 
$[0.85,0.94]$ & 0.041 & 6.21 & 18.54 & 4.95 & 4.33 & 6.06 & 3.78 & 6.05 & 6.91 & 11.24\\ 
$[0.94,1.00]$ & 0.014 & 7.57 & 38.72 & 16.40 & 25.48 & 4.59 & 7.14 & 18.93 & 6.09 & 7.50\\ 
\hline
$\tilde{\lambda}_0^1$ & $1/\sigma$ $\mathrm{d}\sigma/\mathrm{d}$$\tilde{\lambda}_0^1$ & Stat. (\%) & Tot. (\%) & HFS(jet) (\%) & HFS(other) (\%) & HFS($\phi$) (\%) & Lepton(E) (\%) & Lepton($\phi$) (\%) & Model (\%) & Closure (\%)\\ 
\hline
$[-1.00,-0.88]$ & 0.004 & 8.03 & 70.50 & 21.97 & 12.77 & 9.95 & 6.15 & 21.97 & 53.11 & 28.63\\ 
$[-0.88,-0.76]$ & 0.018 & 5.78 & 20.95 & 1.78 & 9.85 & 7.82 & 10.15 & 5.11 & 10.20 & 3.36\\ 
$[-0.76,-0.62]$ & 0.046 & 4.17 & 17.57 & 3.51 & 14.35 & 1.98 & 4.43 & 4.16 & 4.72 & 3.13\\ 
$[-0.62,-0.47]$ & 0.117 & 2.94 & 6.72 & 1.81 & 2.04 & 1.85 & 2.31 & 2.85 & 2.49 & 2.44\\ 
$[-0.47,-0.33]$ & 0.283 & 1.96 & 4.74 & 1.19 & 2.31 & 1.88 & 1.00 & 1.14 & 1.26 & 2.11\\ 
$[-0.33,-0.21]$ & 0.585 & 1.32 & 5.04 & 2.09 & 0.42 & 1.37 & 2.33 & 1.74 & 2.21 & 1.98\\ 
$[-0.21,-0.11]$ & 1.012 & 0.96 & 2.29 & 0.78 & 0.78 & 0.76 & 0.36 & 0.98 & 0.76 & 0.92\\ 
$[-0.11,-0.03]$ & 1.466 & 0.71 & 2.80 & 0.44 & 2.36 & 0.66 & 0.23 & 0.21 & 0.70 & 0.73\\ 
$[-0.03,0.03]$ & 1.792 & 0.51 & 1.21 & 0.46 & 0.49 & 0.33 & 0.40 & 0.45 & 0.20 & 0.49\\ 
$[0.03,0.11]$ & 1.746 & 0.48 & 3.14 & 0.31 & 0.70 & 0.44 & 0.45 & 0.39 & 1.44 & 2.53\\ 
$[0.11,0.19]$ & 1.476 & 0.59 & 2.32 & 0.82 & 1.44 & 0.59 & 1.25 & 0.36 & 0.50 & 0.04\\ 
$[0.19,0.29]$ & 1.098 & 0.80 & 3.47 & 1.34 & 2.05 & 1.05 & 1.63 & 0.83 & 0.61 & 0.77\\ 
$[0.29,0.41]$ & 0.687 & 1.25 & 4.21 & 1.09 & 3.03 & 0.66 & 1.22 & 0.29 & 1.86 & 0.57\\ 
$[0.41,0.56]$ & 0.352 & 2.13 & 5.36 & 1.95 & 2.07 & 2.11 & 2.08 & 1.76 & 2.01 & 0.44\\ 
$[0.56,0.70]$ & 0.162 & 3.39 & 8.46 & 2.71 & 2.41 & 3.39 & 3.33 & 2.79 & 3.21 & 2.51\\ 
$[0.70,0.84]$ & 0.064 & 4.82 & 13.93 & 4.59 & 4.61 & 4.70 & 4.13 & 3.50 & 3.81 & 7.90\\ 
$[0.84,1.00]$ & 0.018 & 6.50 & 15.66 & 5.17 & 6.27 & 6.47 & 4.42 & 4.19 & 4.59 & 6.08\\ 
\hline
$\tilde{\lambda}_0^0$ & $1/\sigma$ $\mathrm{d}\sigma/\mathrm{d}$$\tilde{\lambda}_0^0$ & Stat. (\%) & Tot. (\%) & HFS(jet) (\%) & HFS(other) (\%) & HFS($\phi$) (\%) & Lepton(E) (\%) & Lepton($\phi$) (\%) & Model (\%) & Closure (\%)\\ 
\hline
$[1.00,2.00]$ & 0.011 & 3.16 & 7.28 & 1.62 & 3.15 & 3.08 & 3.42 & 2.62 & 1.50 & 0.37\\ 
$[2.00,3.00]$ & 0.027 & 2.59 & 6.54 & 1.21 & 2.29 & 2.56 & 2.10 & 3.65 & 1.24 & 1.86\\ 
$[3.00,4.00]$ & 0.059 & 2.08 & 5.60 & 1.76 & 2.83 & 1.79 & 0.82 & 1.97 & 1.99 & 2.05\\ 
$[4.00,5.00]$ & 0.085 & 1.58 & 5.15 & 1.57 & 1.18 & 1.45 & 1.12 & 1.74 & 2.68 & 2.57\\ 
$[5.00,6.00]$ & 0.114 & 1.15 & 2.62 & 1.15 & 0.56 & 0.99 & 1.10 & 0.93 & 0.56 & 0.72\\ 
$[6.00,7.00]$ & 0.123 & 0.79 & 2.06 & 0.70 & 0.76 & 0.72 & 0.79 & 0.42 & 0.79 & 0.77\\ 
$[7.00,8.00]$ & 0.128 & 0.63 & 1.52 & 0.63 & 0.61 & 0.32 & 0.42 & 0.42 & 0.63 & 0.55\\ 
$[8.00,9.00]$ & 0.113 & 0.73 & 1.76 & 0.40 & 0.86 & 0.72 & 0.73 & 0.20 & 0.50 & 0.59\\ 
$[9.00,10.00]$ & 0.099 & 0.98 & 2.81 & 0.32 & 0.98 & 0.97 & 0.98 & 1.44 & 0.98 & 0.98\\ 
$[10.00,11.00]$ & 0.078 & 1.34 & 4.17 & 1.34 & 3.16 & 0.08 & 0.98 & 0.22 & 1.18 & 1.19\\ 
$[11.00,12.00]$ & 0.061 & 1.82 & 3.70 & 1.78 & 1.06 & 1.44 & 0.72 & 0.22 & 1.43 & 1.18\\ 
$[12.00,13.00]$ & 0.044 & 2.31 & 5.12 & 2.27 & 2.28 & 2.05 & 1.32 & 0.04 & 1.10 & 1.81\\ 
$[13.00,14.00]$ & 0.034 & 2.93 & 6.91 & 2.46 & 2.90 & 2.52 & 1.90 & 1.00 & 2.28 & 2.93\\ 
$[14.00,15.00]$ & 0.023 & 3.51 & 7.55 & 2.86 & 0.40 & 3.40 & 2.40 & 2.41 & 2.66 & 2.47\\ 
\hline
\end{tabular}
\end{adjustbox}
\end{table*}
\end{document}